# Advancing Data Justice Research and Practice:

An Integrated Literature Review

Prepared by: The Alan Turing Institute

# Advancing Data Justice Research and Practice:
# An Integrated Literature Review

By David Leslie, Michael Katell, Mhairi Aitken, Jatinder Singh, Morgan Briggs, Rosamund Powell, Cami Rincón, Thompson Chengeta, Abeba Birhane, Antonella Perini, Smera Jayadeva, and Anjali Mazumder

*This report was commissioned by the International Centre of Expertise in Montréal in collaboration with GPAI's Data Governance Working Group, and produced by the Alan Turing Institute. The report reflects the personal opinions of the authors and does not necessarily reflect the views of GPAI and its experts, the OECD, or their respective Members.*

### Acknowledgements

*This research was supported, in part, by a grant from ESRC (ES/T007354/1), Towards Turing 2.0 under the EPSRC Grant EP/W037211/1, and from the public funds that make the Turing's Public Policy Programme possible.*

*The creation of this material would not have been possible without the support and efforts of various partners and collaborators. The authors would like to acknowledge our 12 Policy Pilot Partners— AfroLeadership, CIPESA, CIPIT, WOUGNET, GobLab UAI, ITS Rio, Internet Bolivia, Digital Empowerment Foundation, Digital Natives Academy, Digital Rights Foundation, Open Data China, and EngageMedia—for their extensive contributions and input. The research that each of these partners conducted has contributed so much to the advancement of data justice research and practice and to our understanding of this area. We would like to thank James Wright, Noopur Raval, and Alicia Boyd, and our Advisory Board members, Nii Narku Quaynor, Araba Sey, Judith Okonkwo, Annette Braunack-Mayer, Mohan Dutta, Maru Mora Villalpando, Salima Bah, Os Keyes, Verónica Achá Alvarez, Oluwatoyin Sanni, and Nushin Isabelle Yazdani whose expertise, wisdom, and lived experiences have provided us with a wide range of insights that proved invaluable throughout this research. We would also like to thank those individuals and communities who engaged with our participatory platform on decidim and whose thoughts and opinions on data justice greatly informed the framing of this project. All of these contributions have demonstrated the pressing need for a relocation of data justice and we hope to have emphasised this throughout our research outputs. Finally, we would like to acknowledge the tireless efforts of our colleagues at the International Centre of Expertise in Montréal and GPAI's Data Governance Working Group. We are grateful, in particular, for the unbending support of Ed Teather, Sophie Fallaha, Jacques Rajotte, and Noémie Gervais from CEIMIA, and for the indefatigable dedication of Alison Gillwald, Dewey Murdick, Jeni Tennison, Maja Bogataj Jančič, and all other members of the Data Governance Working Group.*

**Cite this work as:** Leslie, D., Katell, M., Aitken, M., Singh, J., Briggs, M., Powell, R., Rincón, C., Chengeta, T., Birhane, A., Perini, A., Jayadeva, S., and Mazumder, A. (2022). Advancing data justice research and practice: an integrated literature review. The Alan Turing Institute in collaboration with The Global Partnership on AI.

# Contents

# Introduction:
# Relocating Data Justice Research and Practice

The *Advancing Data Justice Research and Practice (ADJRP)* project aims to widen the lens of current thinking around data justice and to provide actionable resources that will help policymakers, practitioners, and impacted communities gain a broader understanding of what equitable, freedom-promoting, and rights-sustaining data collection, governance, and use should look like in increasingly dynamic and global data innovation ecosystems. In this integrated literature review[1] and annotated bibliography we hope to lay the conceptual groundwork needed to support this aspiration.

The endeavour to broaden current visions of what data justice is (and what it could become) involves not only building on the considerable insights that have accrued since the inception of the field less than a decade ago. It also involves identifying where the study of data justice has—thus far—fallen short of engaging with and integrating the perspectives and wisdom of those significantly impacted by the subject matter it broaches. It involves distinguishing where limited fields of vision in the current academic literature, gaps in disciplinarily anchored understandings, and listening deficits in scholarship and policymaking, have cramped the analytical and normative scope of its concerns, conclusions, and proposed solutions.

## The Plan of Work

This introduction motivates the broadening of data justice that is undertaken by the literature review which follows. First, we address how certain limitations of the current study of data justice drive the need for a re-orientation, indeed a re-location, of data justice research and practice. We map out the strengths and shortcomings of the contemporary state of the art and then elaborate on the challenges faced by our own effort to broaden the data justice perspective in the decolonial context. We then lay out three trajectories of re-orientation, what we will term the 'where', the 'when', and the 'who' of data justice. Finally, this introductory section addresses the question, what is data justice? Here, we provide a brief history of the data justice literature and outline results collected via our *decidim* participatory platform survey where we received feedback on what data justice means to different groups. Together, these sections explore current interpretations of the term 'data justice' both within academia and beyond.

Key to the re-orientation of data justice prioritised throughout this literature review is the idea that data justice is contextually determined. Consequently, rather than answer the question, what is data justice, directly, the next section presents six pillars of data justice research and practice. These guiding priorities of power, equity, access, identity, participation, and knowledge are set out as resources for reflection to be taken up by those working to advance data justice globally and remain central throughout this review.

---

[1] Toracco, 2005; Snyder, 2019

The body of the literature review itself covers seven thematic areas. For each theme, the ADJRP team has systematically collected and analysed key texts in order to tell the critical empirical story of how existing social structures and power dynamics present challenges to data justice and related justice fields. In each case, this critical empirical story is also supplemented by the transformational story of how activists, policymakers, and academics are challenging longstanding structures of inequity to advance social justice in data innovation ecosystems and adjacent areas of technological practice. Throughout, key themes and key gaps are summarised at the top of each section to highlight important ideas and areas in need of improvement. Reflection questions for academic researchers, policymakers, developers, and impacted communities are also provided at the end of each theme, offering a means of relating the thematic areas to specific stakeholder perspectives.

Finally, this literature review contains an appendix, which provides details of the 12 organisations who are 'Policy Pilot Partners' (PPPs) for this project.[2] These organisations have been working together with the ADJRP team to evaluate sets of data justice guidelines designed to assist policymakers, developers, and impacted communities. All of the PPPs have also provided input via the *decidim* survey, which has been used to shape the six pillars of data justice presented in this review. This literature review is also accompanied by a separate document that contains an annotated bibliography of relevant works which have contributed to this review. It provides summaries of the key themes discussed in each text. Additionally, this companion document contains a table of organisations conducting data justice and data justice adjacent activism across the globe.

## The State of the Art and its Discontents

Before the advent of data justice research several years ago, prevailing approaches to data ethics and governance tended to frame issues surrounding the societal impacts of datafication and the increasing pervasiveness of data-intensive technologies almost exclusively in terms of data protection, individual rights, privacy, efficiency, and security.[3] They likewise tended largely to focus on building technical solutions to potential harms rather than on interrogating the social structures, human choices, and sociotechnical practices that lie behind the myriad predicaments arising out of an ever more datafied society (Figure 1).

---

[2] A longer list of global organisations working to advance data justice can be found in the long-form version of this literature review. This list serves as a living repository of current organisational efforts to combat injustice globally. It provides specific examples of the critical work being done by these organisations to operationalise each of the six pillars of data justice and directs readers to further information about this important work.

[3] Dencik, Hintz, & Cable, 2016

*Figure 1: A depiction of the datafied society where omnipresent sensors and networked mobile devices exponentially multiply sites of data extraction, measurement, and analysis.*

The first wave of data justice scholarship—emerging in the pathbreaking work undertaken by the Data Justice Lab at Cardiff University and the Global Data Justice project at the Tilburg Institute for Law, Technology, and Society—sought to move beyond these limitations by situating the ethical challenges posed by datafication in the wider context of social justice concerns. This meant that data justice research could overcome tendencies in the field of data ethics and governance to dwell in subject-centred abstractions about individual privacy, negative liberty, and algorithmic fairness by becoming more responsive to the real-world conditions of power asymmetries, inequality, discrimination, and exploitation that have increasingly come to define the 'data-society nexus'.[4] It also meant that globally impacting issues surrounding equitable access to representation through data as well as interests in the just distribution of the benefits of data use and the actualisation of social freedom could be brought to bear in considerations of the social consequences of ubiquitous datafication.[5]

Despite the major gains in understanding and insight generated by this first iteration of data justice research, some have pointed to significant limitations. For instance, the initial focus of data justice research on surveillance, informational capitalism, and the 'political economy of data'[6] has been seen to lead to an overly information-centric and economistic narrowing of the way it approaches critical and ethical questions.[7] In other words, an emphasis on the extractive ways that private companies collect, analyse, exchange, and monetise personal information or on the surveillant manner in which governmental actors marshal datasets to sort, rank, and make predictions about datafied citizens and subjects has served a valuable purpose in illuminating certain power dynamics, but this has also limited, and potentially skewed, the data justice perspective. Such a focus on the political and economic forces surrounding datafication has run the risk of obscuring the underlying sources of data injustice. It has risked masking deeper socio-culturally- and historically-entrenched structures of domination that are rooted in discriminatory or racialised logics of coloniality, imperialism, cultural hegemony, and administrative control.[8] The endeavour to advance data justice research and practice therefore faces the challenge to broaden its critical approach to interrogating the social, historical, cultural, political, and economic forces behind manifestations of discrimination and inequity in contemporary ecologies of data collection, governance, and use. It must work towards building an understanding of how the longer-term path dependencies created by patterns and legacies of inequality, discrimination, and privilege get drawn into contemporary data work and data innovation lifecycles.

---

[4] Dencik, Hintz, Redden, & Treré, 2019

[5] Taylor, 2017

[6] Ibid.

[7] Hoffman, 2021

Some have also stressed the problematic tendency of discussions about the ethical issues around data governance and data-intensive technologies to be dominated by Western perspectives, interests, and values.[9] The first wave of data justice research was predominantly anchored in Anglo-European academic framings of data justice—both in terms of how its problem space was defined (i.e. what issues and challenges it confronted[10] and where these were seen to arise)[11] and in terms of the possible normative and practical responses that could be offered to rectify the range of harms inflicted by planetary-scale datafication. Notwithstanding recent calls for new, globally oriented, and intercultural approaches to data justice,[12] this initial Western bias has led to a deficient representation of non-Western values, insights, and interests within the existing literature. This is a critical deficit. Current approaches to data justice have not yet effectively centred non-Western visions of ethical and just ways of working, acting, and interconnecting with people and the planet that are rooted, for instance, in relational notions of personhood and community—visions arising across non-Western systems of belief ranging from Ubuntu,[13] Buddhism,[14] and Confucianism[15] to various expressions of Indigenous values.[16] Insofar as the principles and priorities of data justice are to ascertain a sufficiently broad reach, they need to align with the forms of life, ways of being, and living contexts of all individuals and communities impacted by the global propagation of datafication and essential digital infrastructures. For this reason, the inclusion of non-Western framings of the ontologies, meanings, and values that might shape and underwrite possible data governance futures is a crucial precondition of advancing data justice research and practice.

---

Widening the approach to data justice along these geospatial lines is also needed to address the way that data justice research and practice confronts global digital divides as well as gaps between the interests and concerns of high-income countries and those of low-and-middle-income countries.[17] The reality of the globalisation of data markets and data flows is that the fair, equitable, and inclusive participation of individuals, communities, and countries has not yet come anywhere near being achieved. Over the course of the last two decades of rapid digitisation, the disproportionate distribution of benefits and harms has largely been determined by a fraught combination. On the one hand, the overwhelming technological capacities and material means of transnational tech corporations and Global North geopolitical actors has enabled them to asymmetrically wield 'network power'[18] while, at the same time, engage in virtually unimpeded data capture and rent-seeking behaviour.[19] On the other hand, sociohistorical legacies of economic inequality and 'slow violence'[20] have all-too-often disadvantaged and marginalised the individuals, organisations, and communities which comprise low-and-middle-income countries. This has rendered such countries and their peoples vulnerable to predatory or extractive data innovation agendas. The inequitable effects of these imbalances have only been exacerbated by the high entry costs of engaging in data-intensive research and innovation (in terms of both technical capabilities and resources) and by the centralisation of the critical data and compute infrastructures needed for information processing at scale. Data justice research and practice therefore faces the challenge of redressing these patterns of economic and sociotechnical disparity. It must reconceptualise the regulation and governance of data work and counterbalance the unequal power dynamics that condition data production by prioritising universal participatory parity and considerations of local contexts and values. Moving in this direction will allow data justice research and practice to foster the collective rights of marginalised and vulnerable groups and to bring all impacted stakeholders to the table as rights-holders and standards-setters for the global digital political economy of tomorrow.

---

[17] Of course, 'digital divides' are not exclusively, or even primarily, an international problem. Data justice research must also confront existing digital inequalities *within* high-income countries—which especially affect Indigenous, marginalised, and vulnerable social groups.

[18] Following Cohen (2019): 'Under background conditions of vastly unequal geopolitical power, [the equivalence of corporate or state policy and mandated standards] sets up the two interlocking dynamics that produce policy hegemony. On one hand, a dominant network enjoys network power— which David Grewal defines as the self-reinforcing power of a dominant network and Manuel Castells explains as a power that is "exercised not by exclusion from the networks, but by the imposition of the rules of inclusion"— simply by virtue of its dominance. On the other, if a particular hub within a dominant network exercises disproportionate control over the content of the standard, then networked organization will amplify that hub's authority to set policy and legally mandated standardization will amplify it still further. When network- and- standard- based legal- institutional arrangements are instituted under background conditions of vastly unequal geopolitical power, network power translates into policy hegemony' (p. 220). See also: Castells, 2011; Grewal, 2008

[19] Birch, 2020; Birch & Cochrane, 2021

[20] Nixon (2011) uses the term 'slow violence' to describe the gradual, and often invisible, forms of harm that happen 'gradually and out of sight, a violence of delayed destruction that is dispersed across time and space, an attritional violence that is typically not viewed as violence at all' (p. 2). This kind of subtle violence, he argues, targets the vulnerabilities of the disempowered, impoverished, and vulnerable peoples of the "Global South" who are subject to the opportunism of global market capitalism, leading to the destruction of local ecosystems, involuntary displacement, and social conflict.

# The Challenge of Relocating Data Justice in the Decolonial Context

The imperative to broaden the perspective of data justice research and practice emerges unequivocally from the need to redress the current limitations and deficiencies noted above. However, any effort to widen the data justice lens that is initiated from within a predominantly Western academic context runs the risk of preserving the very binary constructions of the relationship between the "Global North" and the "Global South", the "core" and the "periphery", the "centre" and the "margins", it seeks to critically interrogate and transform.

To unpack this issue, we need to explore how two kinds of deficiencies in reflexivity (i.e. shortcomings in the self-awareness of the limitations of one's own standpoint and perspective) could lead to Global North or Western academic biases. First, by simply assuming a path of advancement from the core "Global North" perspective (as a beginning point) to an end point in what has been historically referred to as the "periphery", we would potentially be leaving the core-periphery relationship itself undisturbed. Not only would we be failing to sufficiently acknowledge that this supposed Archimedean point in the "Global North" is also hindered with multiple sites of marginalisation and myriad digital divides, but we would likewise be in danger of fabricating a picture of the "periphery", in our own image. This would entrench the legacies of cultural imperialism and ethnocentrism that need to be criticised and destroyed.[21] We must therefore explore alternative paths to broadening the data justice lens that de-prioritise the Northern perspective. We must endeavour to develop a comparative and intercultural approach that raises the independent insights and knowledges of non-Anglo-European thinking to a level of discursive, epistemic, and normative parity with predominant Western points of view.

Second, adopting the imagery of "advancement" that is assumed as the motivation for a project entitled *Advancing Data Justice Research and Practice* runs the risk of reverting to misleading grand narratives of progress. These narratives of "progress" have all-too-often identified Europe as the index and reference point of 'universal human history'[22] and tied humanity's destiny to the forward march of Western rationality.[23] Exceptionalist narratives of this kind have all-too-often prompted the self-understanding of European modernity to slide from a commitment to the humanist and secular ideals of the 'Western enlightenment'[24] into an erroneous assertion of Western cultural superiority. On the latter view (often expressed through the 'transitional narratives'[25] of political and economic modernisation), the fate of humanity itself hangs on a linear drive to global conversion, i.e. on the gradual realisation of a European mission to liberate the rest of humanity from its supposed state of cultural, political, and socioeconomic immaturity.[26] Any attempt to advance data justice research and practice must proceed with a vigilant awareness of tendencies to tell this kind of misleading triumphalist story.

---

[21] Ess, 2008

[22] Chakrabarty, 2000; Mignolo, 2011

[23] Appadurai, 1996; Chakrabarty, 2000

[24] Notably, as Chakrabarty and others point out, there are indispensable legacies of critical and emancipatory thinking that are downstream from the Enlightenment's vision of an equitable and open society typified by universal rights, fundamental freedoms, equality under the law, democracy, and social justice. See fn. 25.

[25] Chakrabarty, 2000

[26] Jean and John Comaroff describe this as a 'European mission to emancipate humankind from its uncivil prehistory, from a life driven by bare necessity, from the thrall of miracle and wonder, enchantment and entropy'.
Comaroff & Comaroff, 2012, p. 2.

With such pitfalls of ethnocentrism and implicit Northern biases in mind, this literature review undertakes an elemental relocation of data justice with the aim of averting the dangers that emerge from both of these deficiencies of reflexivity. Following Homi Bhabha, such an elemental relocation requires data justice research and practice to be knocked off the pedestal of its own geospatial, temporal, and vocational self-centre. Instead, data justice research and practice must start from 'an expanded and ex-centric site of experience and empowerment'—one that foregrounds, amplifies, and interconnects the visions of those who have historically been situated at the margins.'[27] Such a basic relocation of data justice involves an active acknowledgement that, as Leela Gandhi writes, 'what counts as "marginal" in relation to the West has often been central and foundational in the non-West'.[28] Data justice research and practice must accordingly engage in a non-hierarchical and de-polarising way with the 'theoretical self-sufficiency' of non-Western knowledge systems.[29]

Crucially, however, this need for a relocation of data justice does not entail a wholesale rejection or dismissal of the Western intellectual traditions that have helped to form its principal elements and emergent architecture. As Dipesh Chakrabarty argues, the heritage of European thinking is, at once, 'indispensable' and 'inadequate'.[30] It is indispensable in the sense that key components of political and sociocultural modernity, which have roots in the 'universal and secular vision of the human'[31] championed by the European Enlightenment, have provided integral critical, analytical, and normative leverage for the pursuit of justice, the battle against oppression and inequity, and the advancement of human freedom in both national and post-colonial/de-colonial contexts. It is inadequate in the sense that a paramount source of knowledge production across most of the Enlightenment heritage of Western modernity has been that of a non-situated, "neutral", and "objective" subject who claims for itself privileged access to universality. Such a disembodied and de-contextualised knower is supposed to be capable of neutrally and objectively perceiving, representing, and grasping the world through the exercise of its cognitive agency alone.[32] Corollary modes of Western knowing and scientific reason have been predicated on this "view from nowhere" version of cognition and understanding—one that is based on the dubious assertion that rationality is decoupled from the unique social, historical, linguistic, and cultural contexts that unavoidably condition the knower and hence knowledge itself. Such a defective concept of rationality has had a range of damaging downstream impacts.

---

[27] Bhabha, 1994

[28] Gandhi, 1998

[29] Ibid.

[30] Chakrabarty, 2000

[31] Ibid.

[32] For representative internal critiques of this Cartesian starting point and its evolution across Western modernity see Davidson, 2001; Dewey, 1960; Habermas, 1992; Lewis, 1929; Rorty, 1982. It is important to note here that "Western epistemology" is not a homogenous tradition, and that strains critical of the metaphysics of the self and Cartesian subjectivity have been present ab initio.

Over the last several decades, critics of Eurocentric hegemony, Western patriarchal structures, technoscientific hubris, and the 'colonial matrix of power'[33] have pointed out that this privileging of disembodied rationality has led to high levels of self-misrecognition and unclarity. For instance, as Charles Mills points out, it has led central strains of Western thinking about subjectivity to disown the socially situated character of knowledge. It has led them, on Mills' account, to disregard the importance of certain contextually-determined resistances that are linked to one's social attributes and group membership and that act as determinants of 'the kind of experiences one is likely to have and the kinds of concepts one is accordingly likely to develop'.[34] Black feminist[35] and Chicana/Latina feminist[36] thinkers have also emphasised the importance of the sociocultural location of the knower and the formative character of the frictions arising therein. They have stressed that both knowledge and the subjects who produce and bear it are forged in a crucible of oppression, opposition, and dialogue. To decouple speaking and interacting subjects from the racial, gender, ethnic, religious, and socioeconomic hierarchies of the prevailing social order and from the way that these power structures intersect with identity and cognitive agency is to be fatally ignorant of the social, cultural, and historical conditions of possibility of knowledge itself. Other feminist scholars such as Donna Haraway and Sandra Harding take this awareness of the social location of the knower as a starting point to call into question claims to scientific neutrality, objectivity, and impartiality that 'play the God trick'[37] by assuming an absolute point of view which fails to recognise the situated and incarnate character of knowledge. They stress the importance of displacing what D'Ignazio and Klein have more recently called the misguided 'valorization of the neutrality ideal',[38] and, instead, argue for the centring of 'subjugated standpoints' which 'promise more adequate, sustained, objective, transforming accounts of the world'.[39]

Latin American decolonial theorists[40] have taken this critical emphasis on situated knowledge a step further. On their account, 'the cultural complex known as European modernity/rationality'[41] places the perspective of the knowing subject at a 'zero-point'[42] that 'hides and conceals itself as being beyond a particular point of view'.[43] It is a 'point of view that represents itself as being without a point of view... this "god-eye view" that always hides its local and particular perspective under an abstract universalism'.[44] However, this type of subjectivity conceals the power dimension of what, in fact, ends up being the scaffolding for the geopolitical and 'body-political'[45] exercise of Eurocentric control. Here, a Eurocentrism that claims universality for the abstract subject (which it has itself created) becomes the pervasive mode of knowing by relying on 'a confusion between abstract universality and the concrete world hegemony derived from Europe's position as

---

[33] Originally uses by Quijano as the 'patrón colonial de poder' the colonial matrix of power (as later framed by Mignolo, 2011) involves as four interrelated domains: control of the economy, of authority, of gender and sexuality, and of knowledge and subjectivity.

[34] Mills, 1998

[35] Representative works of these Black feminist thinkers include: Collins, 1990, 1998, 2004; hooks, 1984, 1987, 2000; Lorde, 1984, 2009, 2018.

[36] Representative works of these Chicana/Latina feminist thinkers include: Alarcón, 1983, 1991; Anzaldúa, 1990, 1999; Anzaldúa and Moraga, 1999; Lugones, 1994, 2007, 2011.

[37] Haraway, 1988

[38] D'Ignazio & Klein, 2020, p. 82

[39] Haraway, 1988, p. 584

[40] Such as Castro-Gómez, 2021a, 2021b; de Sousa Santos, 2014, 2018; Dussel, 1985, 2012; Grosfoguel, 2003, 2007, 2011, 2013; Grosfoguel et al., 2007; Mignolo, 2002, 2007, 2011; Quijano, 2000, 2007

[41] Quijano, 2007

[42] Castro-Gómez, 2021b

[43] Grosfoguel, 2011, p. 5

[44] Grosfoguel, 2011

[45] Grosfoguel, 2011. He uses the term "body-politics of knowledge," following (Fanon, 1967) and (Anzaldúa, 1987).

centre'.[46] For this reason, decolonial theorists argue that the epistemology of Euro-American modernity is tied to the 'coloniality of power',[47] 'epistemic injustice',[48] and the conquest for 'Cognitive Empire',[49] and that it has allowed the infliction of 'epistemicide'[50] on non-Western ways of knowing and understanding. Such an 'epistemological arrogance'[51] has also made it possible for Western modernity/rationality to render other exteriorised cultures and subjectivities not only inferior, objectified, and invisible but 'incapable of achieving a universal consciousness'[52] and excluded from participating as communicative peers and interlocutors in the production of knowledge and the shaping of human values, purposes, and goals.

Bearing in mind that the motivation for relocating data justice research and practice is *both* critical *and* constructive, we should note here that these decolonial, feminist, and critical theorists do not engage in an outright rejection of reason *per se*. Their interrogations of the inadequacy of dominant forms of Western epistemology do not merely seek to critically disassemble the European heritage of disembodied rationality and self-proclaimed universality. In actuality, a vast majority of the feminist, decolonial, and critical theorists, who analytically confront the flaws of Western epistemology, explicitly reject epistemic and cultural relativism.[53] They embrace instead contextually situated, interculturally inclusive, and 'pluriversal'[54] concepts of "objectivity" and "truth" that are anchored in real life social practices of open, inclusive, and ever-revisable communication, in caring efforts to listen to all affected voices,[55] in 'transversal' interactions from 'lived experience to lived experience'[56] and from 'periphery to periphery',[57] and in knowledge practices that incorporate 'the possibility of fallibility, self-correction, and improvement'.[58] Indeed, decolonial theorist Walter Mignolo argues that "objectivity" and "truth" are shaped, supported, and sustained by a more basic set of ethical demands: they start *from life*, in all its burdens, uncertainties, fragilities, and pluralities of perspective and experience. On this view, ways of knowing are *rooted in* the shared pursuit of the fullness, creativity, harmony, and flourishing of human and biospheric life (what Abya Yala Indigenous traditions of Bolivia and Ecuador have called 'living well' or *sumak kawsay* in Quechua, *suma qamaña* in Aymara, or *buen vivir* in Spanish).[59] For Mignolo, the communal and collaborative character of 'humanness' as *a collective praxis of good living* links situated epistemology with the demands of collective wellbeing, social justice, and *sumak kawsay*.[60]

---

[46] Dussel, 2000, p. 471; Escobar, 2010, p. 38; Quijano, 2000, p. 549

[47] Quijano, 2007

[48] Medina, 2012

[49] de Sousa Santos, 2018

[50] de Sousa Santos (2018) has called 'epistemicide' 'the destruction of an immense variety of ways of knowing', seeing, and experiencing the world in colonised societies and sociabilities because of the displacement of these epistemic and experiential modalities by 'dominant criteria of valid knowledge in Western modernity'.

[51] de Sousa Santos, 2018; Bustamante, 2019

[52] Grosfoguel, 2010. It is helpful to acknowledge that this critique of the dichotomisation between Western rationality and the objectified and invisibilised Other is shared by many of the critics mentioned here, beyond Latin American decolonial theorists, such as Gloria Anzaldúa.

[53] For instance: Haraway, 1988; Harding, 1992, 1995, 2008, 2015

[54] For an elaboration of the concept of 'plurivarsality', see the section on 'Pluriverse and Post-Development Theory' below.

[55] Cole, 1998; Collins, 1990, p. 270

[56] For further discussion of the concept of transversal politics, see Collins, 1990, p. 245-248. For expansions on the epistemological consequences of the transversal perspective in Collins, see: Cole, 1998; Tong & Botts, 2018.

[57] Dussel, 2012

[58] Mohanty, 1995, p. 115

[59] Huanacuni, 2010; Walsh 2015, 2018

[60] Mignolo, 2018, p. 109

Building on these commitments to situated knowledge(s), interculturality, pluriversality, dialogical inter-connection, and a life-centred ethics, several decolonial thinkers advance arguments for *new forms of universality.* Achille Mbembe, for instance, calls for a re-envisioning of universality that promotes the 'resurgence of humanity' and the sustainment of 'the reservoirs of life'.[61] Such a resurgence accords priority to the plural and place-based striving to share 'one world' through mutuality and reciprocity—a common human project to relationally co-construct a 'Whole world' that is shared by all as a 'common place'.[62] This co-created world is a continuously imagined and re-imagined totality that endeavours to sustain the living plenitude which comprises it and that does not reduce the uniqueness of individual cultural places, histories, standpoints, and identities to the totalising homogeneity of the same. Reflecting on these futures in a critical present that seeks rectification and reparation for historical wrongs suffered by those in the colonial context 'to whom the right to have rights is refused, those who are told not to move, …and those who are turned away, deported, expelled',[63] Mbembe writes: 'The path is clear: on the basis of a critique of the past, we must create a future that is inseparable from the notions of justice, dignity, and the in-common'.[64]

The task of relocating data justice in the decolonial context necessitates a striving to re-envision data innovation ecosystems through these constructive, transformative, and restorative lenses. Responsibly taking up the challenge of broadening the data justice perspective requires an imagining of alternative paths for the shared pursuit of individual, collective, and biospheric flourishing and good living. Such a task involves circumventing tendencies to entrench existing geopolitical and socioeconomic power dynamics, to reinforce Western cultural hegemony and coloniality, to revert to ethnocentric meta-narratives of progress, and to shore up injurious 'metropole-periphery' relationships by privileging the perspectives of the former over the latter. Borrowing from Chakrabarty, our approach to relocating data justice must aim to 'provincialize'[65] its current home in the prevailing Western traditions of sociocultural, political, and economic modernity. It must embrace, instead, a socially located concept of epistemology that is place-based and anchored in open, inclusive, and transversal processes of communication as well as a practically-oriented commitment to interculturality, pluriversality, and biocentric ethical perspectives.

---

[61] Mbembe, 2017
[62] Glissant, 2020
[63] Mbembe, 2017
[64] Ibid.
[65] Chakrabarty, 2000

# Three Trajectories of Relocation

## Where

On its face, the goal of relocating data justice research and practice would seem uncomplicated. It would be reasonable to expect that efforts made to relocate data justice in a literature review would focus primarily on the geographic sites of knowledge production that generate relevant scholarship and policy insights. From this point of view, a relocation of the data justice perspective might simply involve a reorientation of the framing of data justice perspectives from a Global North or Eurocentric focus to one that actively spotlights non-Western and Global Southern scholarship and policy thinking which bring to light the myriad data inequities which arise in the contexts of global inequality and asymmetries of geopolitical power as well as novel paths to societal transformation.

However, even at this geospatial level of broadening the '*where*' of data justice, further intricacies are involved. This is because, in addition to an international or global aspect, the need for geospatial relocation has a closely related domestic or internal dimension. A geospatial broadening that only centres the knowledges and insights of and about marginalised groups, regions, areas, and countries that have ended up on the wrong side of global digital divides remains incomplete, because "digital divides" are not exclusively, or even primarily, an international problem. Landscapes of data injustice track patterns of inequality and discrimination that exist *between* high-income countries/regions and low-and-middle-income countries/regions but also *within* them.

Indeed, patterns of inequality and discrimination that affect Indigenous, marginalised, and vulnerable social groups *within* low-and-middle-income countries/regions are likely only to be compounded and magnified by the wider inequalities that exist between these countries/regions and wealthier ones. Likewise, different vulnerable or discriminated-against groups within these countries might suffer different kinds and degrees of inequities. Each of these domestic and global dimensions of potential data injustice must be scrutinised, understood, and interrelated. Similarly, in high-income countries, where patterns of domestic inequality and discrimination that affect Indigenous, marginalised, and vulnerable social groups have a lack of visibility (perhaps in virtue of an outward focus on global digital divides), data injustices can escape needed detection and examination. This can be seen, for instance, in the case of unstudied harms done in wealthy countries to gig workers whose labour is controlled by data-driven algorithmic systems.

Broadening the 'where' of data justice research and practice therefore involves a *double relocation* that widens the data justice lens bifocally, sharpening its vision at both the domestic and global levels. This stereoscopic trajectory of reorientation gives rise to notable advantages. First, it enables more robust interactions between a multiplicity of voices, experiences, and frameworks.[66] By bringing domestically- and globally-focused interrogations of the social and ethical consequences of datafication processes into proximity, insights about data inequities and harms can be connected and brought into conversation 'from the periphery to the periphery' and from experience to experience rather than being mediated from an authoritative nucleus of interpretation or siloed in specific areas of study. Second, and as a result of this transversality, there can be greater cross-fertilisation of insights from divergent sociocultural sites of experience, introducing novel critical and normative angles into domains of research and practice where they might not have otherwise emerged. For instance, visions of just data ecosystems that spring from biocentric and community-centred perspectives (such as the priority of affirming moral personhood through relationality in Ubuntu[67] or the Abya Yala Indigenous prioritisation of *sumac kawsay* and *Pachamama[68]*) can introduce novel criticism and practical transformation in Western-contexts dominated by market-based ideologies and beliefs in 'possessive individualism'.[69]

---

[66] See footnotes 56 and 57 above.

## When

The task of relocation also involves a 'when'. If the problem space of data justice (i.e. what issues and challenges it confronts, how the range of these issues are delimited, and where they are seen to arise) is limited to the past few decades of planetary scale datafication, longer term patterns of inequity and structural discrimination are likely to be neglected. Widening the historical horizons of data justice makes visible more subtle patterns of injustice and can provide leverage for transformative strategies such as redistribution, reparation, and restitution.

We should look critically at the reasons behind the narrow focus of much relevant scholarship and policy formulation on the era of big data which currently dominates across numerous framings.[70] A near-sighted focus on the era of 'big data'[71] or 'the data revolution',[72] on the 'second machine age',[73] or even on the rise of 'surveillance capitalism',[74] is symptomatic of what we might call *data epochalism*.[75] This is the implicit assumption that there is, for better or worse, something unprecedented or exceptional about our contemporary period of technological change. In particular, data epochalism emerges in the perception that our own epoch of rapidly accelerating datafication, digitisation, and informatisation is unique, unparalleled, and thus worthy of concerted and undivided attention.

While there is undoubtedly virtue to interrogating the degree to which present day digital innovation has brought human society to an 'inflection point',[76] a sense of data epochalism can also lead to modes of information-centrism and tech-centred short-sightedness that impair researchers' visions of the past, present, and future. It can impair understanding of the *past* by concealing longer term sociohistorical patterns of inequity and discrimination that have cascading effects on data innovation ecosystems and that directly and indirectly influence the sociotechnical contexts of data collection and use. It can impair understanding of the *present* by limiting levels of explanation and analysis to areas circumscribed by the narrow set of normative and social justice issues that are seen to surface specifically in current constellations of data innovation practices—rather than embedding these practices in the complex social, cultural, political, and economic histories of injustice, inequality, oppression, racism, discrimination, and coloniality that have inexorably shaped them.[77] And, it can impair visions of the future by creating a false sense of the insuperability of the revolutionary momentum of current technological and scientific change—leaving critics feeling disempowered in the face of a creeping technological determinism.[78] As Morozov puts it, 'the paralyzing influence of epochalism induces passivity and limits our responses to change, for the unfolding trends are perceived to be so monumental and inevitable that all resistance seems futile'.[79]

---

[67] See the section on "Non-Western Values and the Transformation of Data Justice" below.

[68] Cf. fn. 60 above.

[69] de Sousa Santos, 2011

[70] Mayer-Schönberger & Cukier, 2013; Kitchin, 2014; Brynjolfsson & McAfee, 2014; Zuboff, 2015, 2019

[71] Mayer-Schönberger & Cukier, 2013

[72] Kitchin, 2014

[73] Brynjolfsson & McAfee, 2014

[74] Zuboff, 2015, 2019

[75] We draw on Evgeny Mozorov's (2013) notion of how internet "epochalism" led to tech-centrism as well as "technological amnesia and complete indifference to history (especially the history of technological amnesia)" in debates among internet pundits during the meteoric rise of online experience in the 1990s and 2000s. See Morozov 2013, p. 35-39

[76] Crawford & Calo, 2016

Data epochalism can also have a hampering effect on the historical self-understanding of data justice itself. A belief in the singularity of the big data revolution can unduly fetter researchers' notions of the appropriate commencement point and historical arc of relevant data justice concerns. To the contrary, as thinkers from Max Weber, Michel Foucault, and Ian Hacking to Alain Desrosières, Georges Canguilhem, and Theodore Porter have all argued, sociohistorical processes of increasing quantification, measurement, categorisation, datafication, and calculation have a centuries-long history marked by the mobilisation of enumerative and statistical techniques as a means of scientific management, normalisation, and the exercise of state power and administrative control.[80] The development of these methods of data collection into techniques of social control, normalisation, and biopower had massive biopolitical, economic, and sociocultural consequences. This ranged from practices of data-driven slave management in the Antebellum United States, which subjected enslaved men, women, and children to dehumanising experimentation aimed at optimising labour productivity,[81] to the development of the racialised forms of social statistics in the 19th century which came to underwrite theories of eugenics.[82] The dark side of this history is fundamental to the data justice story.

Widening the 'when' of data justice is not only important to those wishing to understand how the past influences the present but also to those whose role is to shape the future, in particular through policymaking where concerns have been raised over the extent to which democratic governments are 'willing to invest in long-term social goods'.[83] The time pressures placed on policymakers can prevent them from looking to complex historical trajectories, while short cycles of government where new political priorities are regularly imposed can prevent them from tackling long term issues.[84] It is therefore essential to widen the temporal limits of evidence available to policymakers so that historically entrenched injustices can be understood, and long-term transformative policies substantiated.

Broadening the 'when' of data justice requires researchers and practitioners to acknowledge the deep history of datafication and to integrate such understandings with analyses of present practice. In Ian Hacking's words, 'the world knows both revolution and evolution'.[85] Both must be appropriately scrutinised.

---

[77] A good critique of this sort of impaired vision can be found in Morozov's review of Shannon Zuboff's *The Age of Surveillance Capitalism*. See Morozov, 2021.

[78] See discussion of the recent history of facial recognition technologies in Leslie, 2020b.

[79] Morozov, 2013, p. 36

[80] Canguilhem, 1966:1989; Desrosières, 1993:1998; Foucault, 1966: 1989, 1972, 1978; Hacking, 1975, 1990, 2016; Porter, 1995; Weber, 1922: 1978.

[81] Rosenthal, 2018. For details on the scientific management of slavery, see ch. 3.

[82] Hanna et al., 2020; Zuberi, 2000, 2001

[83] Jacobs, 2016

[84] Haddon et al., 2015

[85] Hacking, 2015, p. 67

## Who

Besides the need to relocate the 'where' and the 'when' of data justice research and practice, these is one final site of broadening: the 'who'. Initially, the work to define data justice was undertaken by academic scholars working with Global North institutions. However, were this 'who' of data justice to remain by and large a product of the Northern Academy, a universe of insights and wisdom that come from the lived experiences of members of impacted communities and from data advocacy and policymaking knowledge would be largely excluded from the development of a globally and socially inclusive vision of data justice.

For this reason, this literature review relocates the 'who' of data justice in order to amplify perspectives that are rooted in the lived experience of individuals and communities who are impacted by datafication – especially members of those groups which have been historically disempowered and marginalised. It must give an appropriate parity of voice to the academic articles and books, policymaking outputs and activist papers, statements, and declarations – each of which contribute to the advancement of data justice research and practice. Consequently, in the present literature review we horizontally integrate this range of sources.

# What is Data Justice?

## Brief History of Data Justice Literature

In light of this triple relocation, the understanding of data justice adopted throughout this literature review will be informed both by a diverse range of actors and ideas from across the globe and by concerns that draw on longer histories of injustice, inequality, and discrimination. We emphasise that while the term 'data justice' may have recent origins within academic institutions in the Global North, the movement itself has a longer and more extensive history that pulls in the critical insights and energies of adjacent social justice movements from around the world.

The emergence and trajectory of recent academic literature on this topic has nevertheless been significant in shaping our understanding of data justice. Between 2014 and 2016 three distinct ideas of 'data justice' were proposed.[86] Each responded to the social, political, and practical implications of datafication but did so in different contexts. In 2017, these three perspectives were brought together by Linnet Taylor to create a data justice framework with three core pillars (Figure 2 below). It is in this work and through these three pillars that data justice came to be understood as a 'conceptual framework for organizing freedoms'.[87] It has been defined by Taylor as 'fairness in the way people are made visible, represented and treated as a result of their production of digital data'.[88]

---

| Taylor's three pillars of Data Justice[89] | | |
|---|---|---|
| Visibility | Engagement with technology | Non-discrimination |
| - Access to representation through data<br><br>- Informational privacy | - Share in data's benefits<br><br>- Autonomy in technology choices | - Ability to challenge bias<br><br>- Preventing discrimination |

*Figure 2: Taylor's three pillars of Data Justice.*

In addition to these three pillars of data justice, Taylor (2017), following Heeks and Renken (2016), calls for an 'ecosystemic approach on capabilities'.[90] This is a view of social justice, borrowed from the work of Amartya Sen and Martha Nussbaum, that centres human flourishing and that aims to create material conditions where people can realise their full potential and live freely.[91] Extending this rationale to the milieu of 'global data justice', Taylor (2019) places such a capabilities aspproach in the context of 'people's subjective needs with regard to data and representation', arguing that models of governance for data technologies should be aligned with these contextually specific demands.[92] This alignment can be achieved, on Taylor's account, through public dialogue and ethnographic research into how people experience global datafication.

Early accounts of data justice (like those of Johnson, Dencik et al., Heeks and Renken, and Taylor) share an underlying frustration with existing responses to datafication. Debates which focused on security, data protection, privacy, and individual rights were criticised as too narrow.[93] Fundamentalist rights framings were critiqued for requiring that harms be clearly visible and for assuming that these harms could be addressed on an individual level, even though data can impact groups as much as individuals.[94] Additionally, the predominant focus on datafication's impacts in the Global North was criticised and data justice was proposed as a global framework to combat this.[95] Finally, existing responses to datafication were critiqued for only focusing on negative and not positive rights.[96] Overall, the data justice framework was presented as an approach which addressed broader social justice concerns surrounding datafication.

---

[89] Taylor, 2017, p. 9

[90] Taylor, 2017, p. 10

[91] Nussbaum, 2006; Sen 1999; Taylor 2019

[92] Taylor, 2019, p. 203

[93] Dencik et al, 2016

[94] Dencik et al., 2016; Taylor, 2017

[95] Heeks & Renken, 2016

[96] Taylor, 2017

Since the publication of Taylor's 2017 data justice framework, the literature has expanded. Dedicated institutions including the Data Justice Lab at Cardiff University and the Global Data Justice Project at the Tilburg Institute for Law, Technology, and Society have been established. The concept has been interrogated in a range of specific global contexts such as policing in Iran, activism in South Africa, Indigenous agriculture in Africa, humanitarian work in post-earthquake Nepal, and more.[97] These academic understandings of data justice will continue to inform this work while additional perspectives, collected through our Policy Pilot Partners, our data justice survey, and the present literature review will be used to broaden this definition even further.

---

Johnson identifies power asymmetries in the governance and administrative functions of data which can lead to normatively coercive data structures and forms of extraction. He argues in favour of 'information justice' in the context of open data as a framework to address these power dynamics.

Heeks and Renken propose that a framework of data justice is needed to account for local and global variations in how datafication impacts individuals and communities. While data justice needs to be applied differently in different contexts, human rights and fundamental freedoms are important guideposts. Heeks and Renken argue such a global approach is lacking.

Linnet Taylor defines Data Justice as 'fairness in the way people are made visible, represented and treated as a result of their production of digital data'.

Global Data Justice Project launched at Tilburg Institute for Law, Technology, and Society.

2020 – Global Partnership on Artificial Intelligence (GPAI) is established. Its aim is 'to bridge the gap between theory and practice on AI by supporting cutting-edge research and applied activities on AI-related priorities'. GPAI's 15 founding members are Australia, Canada, France, Germany, India, Italy, Japan, Mexico, New Zealand, the Republic of Korea, Singapore, Slovenia, the United Kingdom, the United States, and the European Union. They were joined by Brazil, the Netherlands, Poland, and Spain in December 2020.

**2014   2015   2016   2017   2018   2020   2021**

World leaders adopt 17 Sustainable Development Goals (SDGs) at a UN Summit. These goals provide an important framing for the responsible adoption of AI.

Dencik et al. propose that a data justice framework is needed to broaden the conversation around datafication to account for concerns beyond security, privacy, and data protection. They argue that the pursuit of data justice must include the involvement of activists and advocates in civil society.

Data Justice Lab officially launched at Cardiff University's School for Journalism, Media, and Cultural Studies.

Data Justice literature takes on increasingly globally oriented and intercultural approaches as authors explore local and contextual understandings of how social justice intersects with datafication.

*Figure 3: Timeline of the data justice literature 2014-Present.*

## Decidim Analysis

As part of our research, we developed an online participatory engagement platform (through the *decidim* digital interface) for individuals and communities not only to provide feedback on the themes of this integrated literature review, but also to complete an extensive survey on topics such as defining data justice. The survey additionally asked participants to consider and share examples of challenges to data justice. We received 28 responses in total. The link to the *decidim* platform was shared across social media, as well as through existing networks and partnerships. Most of the respondents indicated that they were replying as members of the public. There were responses from 13 countries with the majority coming from the United Kingdom, Chile, and Uganda. Seven respondents indicated that they had completed post-secondary school and 21 respondents stated that they held an advanced degree. When asked their familiarity with data- and algorithm-related technologies, the answers were concentrated in 'Moderately familiar', followed by 'Very familiar'. As part of our Pilot Partner Programme, we also asked the Policy Pilot Partners to complete this survey, and their responses are included within the total count. The responses to prompts about defining and situating data justice help to answer the question of 'What is data justice?' and will be explored here.

When participants were initially asked about their familiarity with the term data justice, most respondents stated 'Somewhat' on a scale of 'Not at all' to 'To a great extent'. Participants were then asked to define the term data justice using three words of their choice. There was diversity in the responses, but after running a top keyword search, we determined the top 10 most common words used to define data justice were: equity, fairness, data, ethic* (ethics and ethicality), access, inclusion, right(s), representation, collection, and social. This keyword analysis can be found in Figure 4 below.

What are the first three words that come to mind when thinking of data justice?

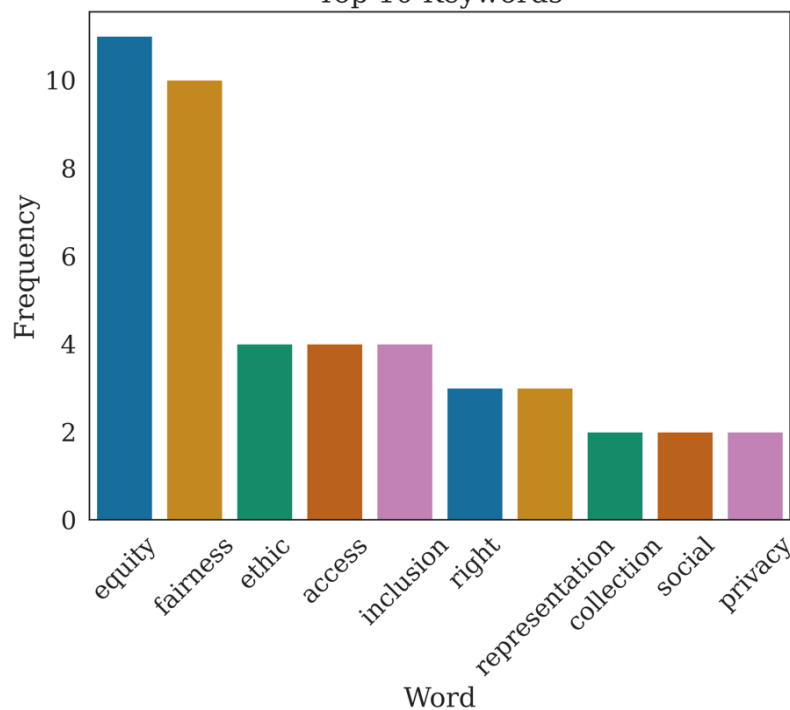

*Figure 4: Decidim results for the question: What are the first three words that come to mind when thinking of data justice?*

Next, we presented the frequently referenced 2017 definition of data justice offered by Linnet Taylor, 'fairness in the way people are made visible, represented and treated as a result of their production of digital data'.[98] Participants were then asked whether or not this definition captured what they think of when they think of data justice. 16 respondents answered 'Agree', 10 with 'Strongly Agree', and 2 were 'Undecided'. Afterwards, participants were asked what they believed was missing as well as the strengths and weaknesses of this definition, and these responses are summarised below.

| Strengths of the definition | Weaknesses of the definition |
|---|---|
| - Clear, short, and simple.<br><br>- Human-centric.<br><br>- Includes fairness as a key concept. | - Too broad.<br><br>- Does not expand on the terms utilised within the definition (fairness, representation, visibility, and treatment).<br><br>- Key concepts, aspects, and implications related to data justice are not fully highlighted or addressed (see below).<br><br>- Its anthropocentrism disregards other entities that are subjects of data justice.<br><br>- It links digital rights to just freedom. |

*Figure 5: Decidim participants identify the strengths and weaknesses of Taylor's 2017 definition of data justice.*

Whilst for some participants the human-centric aspect of the definition was a strength, for others it disregarded organisations, communities, and systems, which can also be subjects of data justice. As such, tensions between individuals and the collective as well as power imbalances were pointed out as necessary considerations. Participants also highlighted the need to include the role of colonialism in entrenching historical inequalities between and within countries and entities. For many, the definition did not adequately address historical, cultural, and economic aspects that reflect biases and discrimination in data processing nor how inequality and the exclusion of individuals and groups may be replicated, automatised, or created through data-driven processes and tools. Several participants mentioned that a comprehensive data justice definition should, therefore, acknowledge other rights and values, such as equity and non-discrimination, and how data justice encompasses inclusion and diversity.

Participants mentioned two other missing points. On the one hand, the definition focuses on people as passive objects of data collection, processing, and use, and does not capture their agency as a key factor in data justice. On the other hand, the definition disregards responsibility and ethics as they relate to the implications and use of data. As a result of these gaps, participants stressed that the definition should include concepts of access, understanding, and consent to data collection processes.

---

[98] Taylor, 2017

Also worth noting, later in the survey, participants were shown several examples of data justice activism occurring across the globe. After being shown these examples, participants were again asked to define data justice using three words of their choice. The top 10 keywords from this question were compiled (Figure 6), and while there are some similarities between the first and second instances of the question with words such as fairness, equity, right, and social present, there are several key differences. Following the presentation of several examples, keywords such as transparency, sovereignty, power, respect, and agency replaced previous top 10 keywords of access, ethic* (ethics/ethicality), inclusion, representation, collection, and privacy.

If I were to use three words to describe data justice now, what words would come to mind?

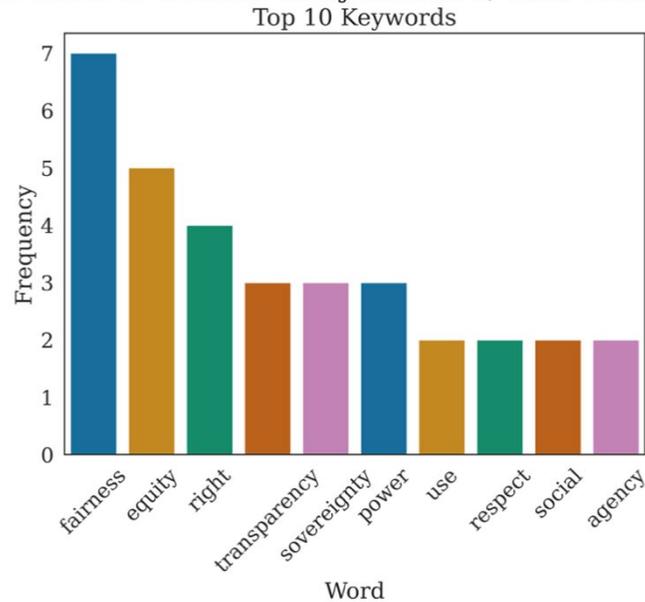

*Figure 6: Decidim results for the question: If I were to use three words to describe data justice now, what words would come to mind?*

# Six Pillars of Data Justice Research and Practice

Taken together, our analysis of the *decidim* survey results, our critical exploration of the important conceptual work carried out in the first years of the academic data justice literature, our interactions with our Policy Pilot Partners, and our other desk-based research have led us to propose six pillars of data justice research and practice. These are the guiding priorities of *power, equity, access, identity, participation, and knowledge*.

While such pillars build on and expand previous attempts to delineate the term "data justice," they are not offered here as part of a definition *per se*. Key to the re-orientation of data justice prioritised throughout this literature review is the idea that it is *contextually determined*. It should be seen as a set of critical practices and procedures that respond to—and enables the transformation of—existing power asymmetries and inequitable or discriminatory social orders rather than as a collection of abstract formulae, principles, or prescriptions. Consequently, instead of answering the question "what is data justice" directly, the pillars are meant to be tools for orienting critical reflection and for prompting the generation of constructive insights into how to transform data justice research and practice to redress the data inequities of the past and present in the ends of building more just societal and biospheric futures. Ultimately, the pillars are merely intended to sharpen the analytical and practical lenses both of those researching issues of justice and equity surrounding the collection and use of data and of those working to advance data justice globally—whether as advocates, policymakers, developers, or members of impacted communities.

It is useful to note, as well, that the six pillars are offered as a re-orientation and re-envisioning of current data justice perspectives. As our survey results show, a framing of data justice that focuses specifically on 'fairness in the way people are made visible, represented, and treated as a result of their production of digital data' (i.e. on 'visibility', 'engagement with technology', and 'non-discrimination') is valuable but may not be sufficiently nuanced to address the myriad power imbalances and the complex social, political, cultural, institutional, economic, and ideological manifestations of historically entrenched inequity that need to be spotlighted in data justice considerations. For instance, though a focus on visibility—which stresses the importance of 'privacy and representation'[99] and highlights the challenges of exposure and opportunities to be counted that are occasioned by datafication—should remain an important part of data justice considerations, this view narrows the critical lens in ways that may limit access to the multi-faceted and chameleonic operation of power. Power indeed crops up in the dynamics of inclusion and exclusion that explicitly shape the politics of personal visibility. However, as the power pillar elaborates, it also arises at many other levels that condition access to representation and exposure to coercion. A suitably refined data justice perspective should be able to pin down and distinguish actionable entry points for critical engagement of the existing power dynamics and legacies of structural inequity, systemic racism, coloniality, and inequality, in all their elusive forms.

Likewise, while clearly essential, an understanding of data justice that conceives of the issue of 'engagement with technology' as having to do with the 'freedom to adopt' or opt out of data systems (as well as the ability to access their benefits) may not sufficiently address transformative possibilities for empowerment and participation. Beyond the more passive aspects of benefitting from, adopting, or opting out of data systems, the constructive potential of individual agency and social action also opens empowering possibilities for the co-creation and democratic steering of technology agendas, standards development, and the fashioning of governance protocols and policy regimes. It creates possibilities, moreover, for the participatory co-design of data systems deemed worthy of pursuit. A data justice perspective that adequately discerns the *spectrum of*

---

[99] Taylor, 2017, p. 9

*engagement* as a *spectrum of empowerment* should mind these kinds of transformative opportunities for the exercise of autonomy, the actualisation of freedom, and the expansion of social license, democratic governance, and public consent.

The six pillars introduce another significant site of re-orientation and re-envisioning that is worth underscoring. Taylor's three pillars—and other similar framings that initiate data justice considerations by focusing predominantly on the political economy of datafication and on the development of data-intensive technologies in the era of big data—run the risk of falling into the kind of information-centrism, economism, and tech-focused short-sightedness that we have highlighted so far. Such standpoints are liable to having a cramped view of the moral horizons and emancipatory potentials of data justice because they implicitly accept that social justice considerations should respond to problems raised both by the seemingly independent march of technology and by the existing political economic incentives and structures that appear to unavoidably steer it. In our formulation of the six pillars, we deliberately endeavour to step out of this tech-centred and economistic circle and, instead, to take up a more historically conscious, whole-of-society, pluriversal, and agency-centred perspective. The pillars broach how wider social and ethical dimensions of power, access, equity, identity, participation, and knowledge are drawn into (and can be brought to bear on) data innovation ecosystems not *vice versa*. In doing so, we propose a multi-dimensional idea of just data practices that includes aspects of redistribution, recognition, restoration, and human and biocentric flourishing.

## Six Pillars of Data Justice

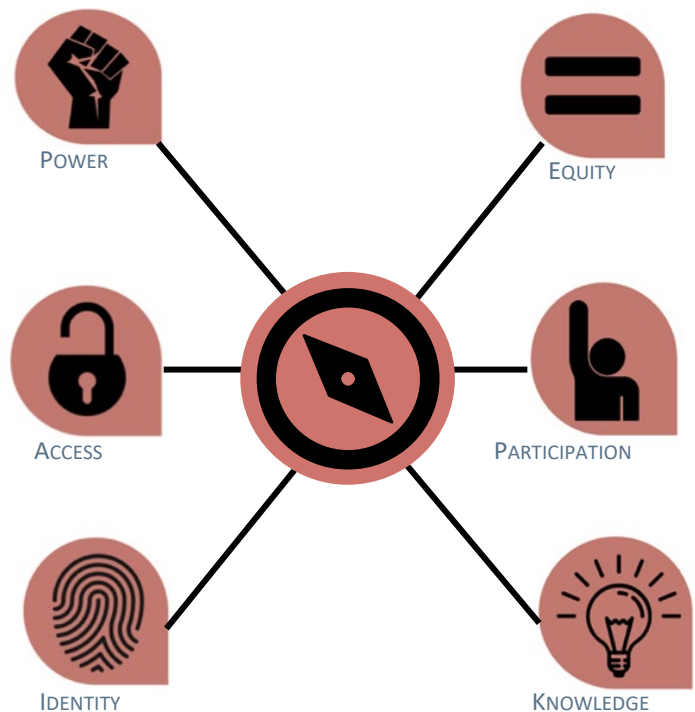

*Figure 7.*

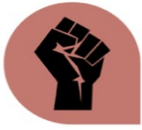

# Power

**1. Interrogate and critique power:** Power dynamics can be present in many different places and in several different ways. It is therefore important to *understand the levels at which power operates in data innovation ecosystems; to understand how power manifests and materialises in the collection and use of data in the world; and, to use this understanding to question power at its sources and to raise critical awareness of its presence and influence:*

*Understand the levels at which power operates in data innovation ecosystems*. For instance, at the *geopolitical*[100] *level*, high-income nation-states and transnational corporate actors—which control access to high entry cost technological capabilities and pursue their own interests on the global stage—can exercise significant influence on which countries or regions are able to develop digital and data processing capacities. At the *infrastructural*[101] *and socioeconomic*[102] *levels*, large tech companies can decide which impacted communities, domestically and globally, are able to access the benefits of connectivity and data innovation. They can control the provision of essential digital goods and services that directly affect the public interest without, at the same time, being subject to corresponding constraints on their pursuit of private benefits and their optimisation of shareholder value and market share. At the *policy, legal,*[103] *and regulatory*[104] *levels*, large international standards bodies, transnational corporations, trade associations, and nation states, can exercise disproportionate amounts of influence and network power[105] in setting international policies, standards, and regulation related to the governance of digital goods and services and data innovation. At the *organisational and political*[106] *levels*, governments and companies can control data collection and use in intrusive and involuntary ways—especially where the public have no choice but to utilise the services they provide or must work in the environments they manage and administer. At the *cultural level*, power can operate through the way that large tech companies use relevance-ranking, popularity-sorting, and trend-predicting algorithms to sort users into different, and potentially polarising, digital publics or groups. At the *psychological*[107] *level*, tech companies can use algorithmically personalised services to reach into the private, inner lives of targeted data subjects with the aim of curating their desires and controlling their consumer behaviour but thereby also play an active and sometimes damaging role in their identity formation, mental wellbeing, and personal development.

---

[100] Ciuriak, 2021; Crampton, 2018; Deibert and Pauly, 2019; Gray, 2021; Lobato, 2019; Miailhe, 2018; Parks, 2009; Pauwels, 2019; O'Hara and Hall, 2021; Rosenbach and Mansted, 2019

[101] Abdalla & Abdalla, 2020; Amodei and Hernandez, 2018; Birch, 2020; Birch et al., 2020; Frank et al., 2019; Gupta et al., 2015; Lohr, 2019; Riedl et al., 2020; Roberge et al., 2019

[102] Zuboff, 2015, 2019; van Dijck et al., 2018; Jin Yong, 2015; Srnicek, 2017; Sadowski, 2019, 2020

[103] Cohen, 2019

[104] Chomanski, 2021; Baik, 2020

[105] See fn. 18.

[106] Ciuriak & Ptashkina, 2020; Eubanks, 2018; Fourcade & Gordon, 2020; Tréguer, 2019

[107] Lupton, 2016; Bucher, 2017; Agger, 2012; Crawford, 2014

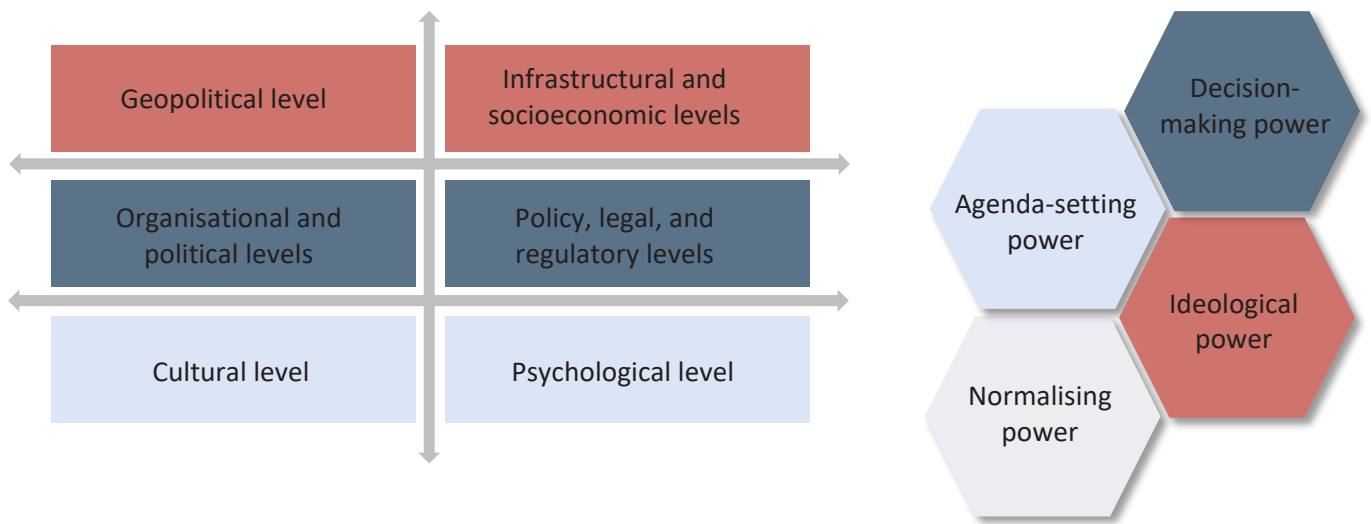

***Understand how power manifests and materialises in the collection and use of data in the world.***
Power can surface in several different ways. Social scientists have been helpfully exploring and taxonomising how power manifests and materialises in everyday life for over half a century. For instance, it can manifest as ***decision-making power***.[108] Here, an individual or organisational actor A has power over B to the extent that A can get B to do something that they would not otherwise do. Decision-making power is therefore identified in terms of the individual or organisational agency that wields it and is explicit, visible, and readily observable. It is easily seen, for instance, in the way that government agencies collect and use data to build predictive risk models about citizens and data subjects or to allocate the provision of social services (and then act on the corresponding algorithmic outputs).

Power can also manifest as ***agenda-setting power***.[109] Here, an individual or organisational actor A has power over B to the extent that A sets the agenda that B then must fall in line with by virtue of A's control over the terms of engagement that delimit practical options within A's sphere of influence and interest. Agenda-setting power means that A is able to shoehorn the behaviour of B into a range of possibilities that is to A acceptable, tolerable, or desired. This kind of power is explicit, for example, in practices of regulatory capture, where large tech corporations secure light touch regulation through robust lobbying and legal intervention. It is also apparent in the ways digitally mature and well-resourced corporate or geopolitical actors can secure extractive markets and rents for their provision of data and compute infrastructure in digitally developing nations or regions by commanding terms of engagement and regulatory structures that overwhelmingly serve the interests of the provider.

---

[108] Dahl, 1961, 1968, 2007
[109] Bachrach & Baratz, 1962

Power, however, can also operate less visibly than these decision-making and agenda-setting types. *Ideological power*[110] is exercised where people's perceptions, understandings, and preferences are shaped by a system of ideas or set of beliefs in a way which leads them—frequently against their own interests—to accept or even welcome their place in the existing social order and power hierarchy. This can happen 'either because [impacted people] can see or imagine no alternative to [the existing social order] or because they see it as natural and unchangeable, or because they value it as divinely ordained or beneficial'.[111] Because this kind of power often operates unconsciously or invisibly through the influence of ideas and ideological framings of society and culture, it is seen as subversive and difficult to identify and overcome. As Amartya Sen puts it, 'the underdog learns to bear the burden so well that he or she overlooks the burden itself. Discontent is replaced by acceptance, hopeless rebellion by conformist quiet, and suffering and anger by cheerful endurance'.[112] An example of the presence of ideological power can be seen in the way that priorities of attention capture and screen-time maximisation (pursued by certain social media and internet platforms) have groomed users within the growing ecosystem of habit- and compulsion-forming reputational platforms to embrace the algorithmically manufactured comforts of life-logging, status-updating, and influencer-watching all while avoiding confrontation with realities of expanding inequality and social stagnation.

The subtlest and least detectable form of power is, however, *normalising power*.[113] Normalising power manifests in the way that the ensemble of dominant knowledge structures, scientifically authoritative institutions, administrative techniques, and regulatory decisions work in tandem to maintain and 'make normal' the status quo of power relations. Thinkers like Michel Foucault have referred to this kind of power as disciplinary power—a form of social order in which dispersed mechanisms of discipline, supervision, and behavioural regulation function to form and norm the identities of passive subjects who embody and enact the prevailing state of normalcy. Where tools of data science and AI come to be used as techniques of knowledge production (within a wider field of statistical expertise) that yield a scientific grasp on the inner states or properties of observed individuals, forms of normalising or disciplinary power can arise. Individual human subjects who are treated merely as objects of prediction or classification and are thereby subjugated as objects of authoritative knowledge become sitting targets of disciplinary control and scientific management.

*Use this understanding to question power at its sources and to raise critical awareness of its presence and influence.* Interrogations of where and how power operates are first steps in a longer journey of questioning and critical analysis. An active awareness of power dynamics in data innovation ecosystems should also lead to examinations of what the interests of those who wield power or benefit from existing social hierarchy are and how these interests differ from other stakeholders who are impacted by or impact data practices and their governance. This line of inquiry involves scrutinising how power imbalances shape the differential distribution of benefits and risks as well as result in potentially unjust outcomes for marginalised, vulnerable, or historically discriminated against groups.

---

[110] Sen, 1984; Lukes, 1974, 2015

[111] Lukes, 1974

[112] Sen, 1984

[113] Foucault, 1990/1976, 2003/1976

**2. Challenge Power:** *Mobilise to push back against societally and historically entrenched power structures and to work toward more just and equitable futures.*[114] While the questioning and critiquing of power are essential dimensions of data justice, its purpose of achieving a more just society demands that unequal power dynamics that harm or marginalise impacted individuals and communities must be challenged and transformed.

**3. Empower People:** *People must be empowered to marshal democratic agency and collective will to pursue social solidarity, political equity, and liberation.* It is common to think about power primarily in a negative way, that is, in terms of the unjust exercise of control, coercion, or influence. But this tells only half the story. Power can also be understood, more positively, in terms of *empowerment*. It can be understood in terms of *collective empowerment through democratic action*. When people and communities come together in the shared pursuit of social justice through mutually respectful practices of deliberation, collaboration, dialogue, and resistance, power becomes constructive and opens transformative possibilities for the advancement of data justice, social solidarity, and political equity.

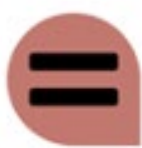 **Equity**

**1. Consideration of equity issues should begin before any data are collected or used. Issues of equity should be confronted by developers and organisations at the earliest stage of project planning and should inform whether data innovation practices are engaged in at all:** Data equity is only partially served by seeking to improve data and data practices, such as by pursuing data quality, or increasing its representativeness and accuracy. While errors and incompleteness are obstacles to data equity, the choice to acquire and use data can itself be a question of justice, particularly where the goal or purpose of a data practice is to target and intervene in the lives of historically marginalised or vulnerable populations. Here, the question may not be 'how can we repair an imperfect system or make it more effective', but rather 'does a particular use or appropriation of data enable or disable oppression?'; and 'does it preserve or combat harmful relations of power?' A perfectly engineered system employed by an oppressive regime (either governmental or commercial) can facilitate and potentially amplify data injustice.

**2. Focus on the transformative potential of data equity:** Data equity demands the transformation of historically rooted patterns of domination and entrenched power differentials. Concerns raised surrounding elements of data innovation practices like data security, data protection, algorithmic bias, and privacy are an important subset of data equity considerations. However, the transformative potential of data equity to advance social justice comes in a step earlier and digs a layer deeper. It starts with questions of how longer-term patterns of inequality, coloniality, and discrimination seep into and penetrate data innovation practices and their governance. Data equity, in this deeper context, is about overhauling power imbalances and forms of oppression that manifest in harmful, unjust, or discriminatory data practices. To realise this sort of equity, those with power and privilege must be compelled to respond to and accommodate the claims of people and groups who have been marginalised by existing political and socioeconomic structures.[115]

---

[114] These three dimensions of power draw heavily on D'Ignazio & Klein, 2020

[115] D'Ignazio & Klein, 2020; Kapoor & Whitt, 2021; Jagadish et al., 2021a, 2021b

**3. Combat any discriminatory, racialised, or 'single axis' forms of data collection and use that centre on disadvantage and negative characterisation:** Following work in critical Indigenous studies, we need to confront and combat statistical representations of marginalised, vulnerable, and historically discriminated against social groups that focus mainly or entirely on measurements of 'disparity, deprivation, disadvantage, dysfunction, and difference', the '5 D's'.[116] Approaches to statistical measurement and analysis that centre on disadvantage and negative characterisation produce feedforward effects which further entrench and amplify existing structures of inequity, discrimination, and domination.

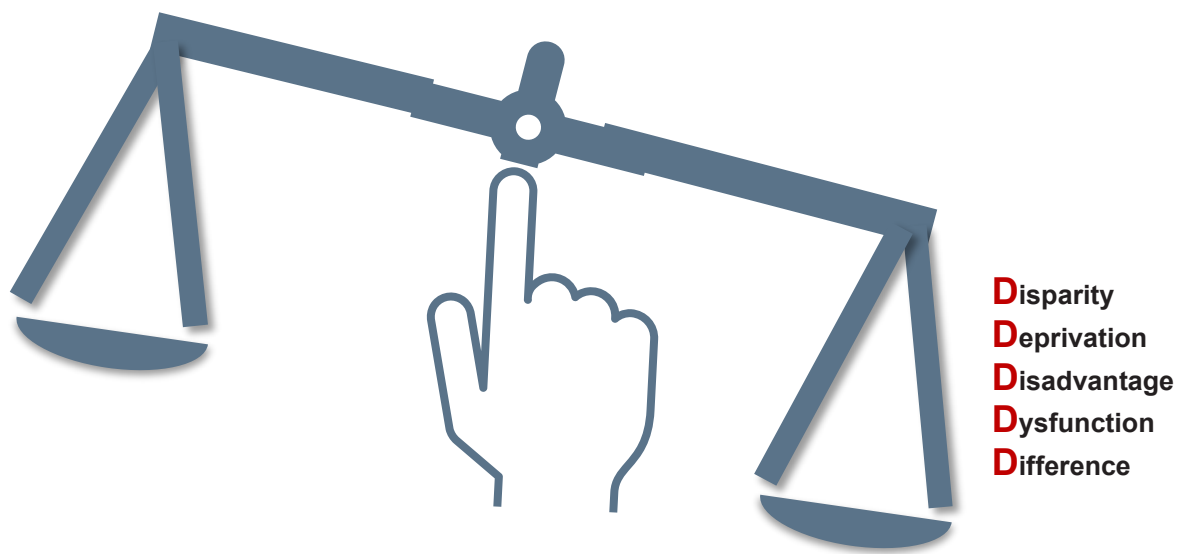

**D**isparity
**D**eprivation
**D**isadvantage
**D**ysfunction
**D**ifference

*Figure 9: Single axis modes of statistical representation; adopted from the 5 D's presented by Kukutai and Taylor (2016).*

**4. Pursue measurement justice and statistical equity:** Measurement justice and statistical equity involve focusing on collecting and using data about marginalised, vulnerable, and historically discriminated against communities in a way that advances social justice. It requires using data in ways which draw on their strengths rather than primarily on perceived weaknesses, and approaches analytics constructively with community-defined goals that are positive and progressive rather than negative, regressive, and punitive. This constructive approach necessitates a focus on socially licenced data collection and statistical analysis, on individual- and community-advancing outcomes, strengths-based approaches, and on community-guided prospect modelling.[117]

---

[116] Kukutai & Taylor, 2016
[117] Leslie et al., 2020

More proactive research is needed to explore how positive (individual- and community-advancing) outcomes can be integrated into data analytics that involve marginalised, vulnerable, and historically discriminated against communities. Part of developing such prospect assessment models would involve inclusive, community-integrating processes of objective setting, problem formulation, and outcome definition as well as multi-stakeholder and interdisciplinary approaches to model planning and implementation. Through these processes of co-creation, the analytics would come to better reflect the best interests of the communities to which they apply. Exploring the possibilities of strengths-based, prospective approaches would also involve creating a better data landscape capable of capturing the lived experience of impacted communities, as well as patterns indicative of positive outcomes that foster their wellbeing and flourishing. At the same time, those working toward cultivating this data landscape would have to safeguard the interests of affected data-subjects—in particular, those most vulnerable to over-collection and the potential harms of data misuse—by working through privacy-preserving and consent-based programming.

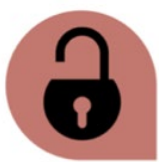

## Access

**1. Confronting questions of equitable access involves starting from real-world problems of material inequality and structural injustice. Access is about providing people tangible paths to data justice by addressing the root causes of social, political, and economic injustice:** Applied concepts of data equity should not be treated as abstractions that can be engineered into data-intensive technologies through technical retooling or mathematisation. This approach will produce a limited range of vision whereby only the patterns of bias and discrimination in underlying data distributions that can be measured, formalised, and statistically digested are treated as actionable indicators of inequity. This leads to the exclusion of the under-the-surface dynamics of sociocultural domination.[118]

Rather, the existing sociohistorical, economic, and political patterns and qualities of disadvantage that create material conditions of injustice and a lack of access to the benefits of data processing must be taken as the starting point for reflection on the impacts and prospects of technological interventions. The beginning of any and all attempts to protect the interests of the vulnerable through the mobilisation of data innovation should be anchored in reflection on the concrete, bottom-up circumstances of justice, in its *historical and material preconditions*. From this more pragmatic point of view,[119] there must be a prioritisation of the real-world problems at the roots of lived injustice—problems that can then be treated as challenges 'remediable'[120] by concerted social efforts and struggles for rectification, redistribution, and recognition.[121] Only then will true-to-life demands for data equity and social justice become properly visible as struggles against the moral injuries inflicted by unjust social arrangements that obstruct the participatory parity of all indviduals in pursuing their unique paths to flourishing and in fully contributing to the moral and political life of the community.[122]

---

**2. Start from questions of access and capabilities:** Beyond the critical demand to advance 'access to representation', data justice thinking must focus on *equitably opening access to data through responsible data sharing; equitably advancing access to research and innovation capacity; equitably advancing access to the benefits of data work; and equitably advancing access to capabilities to flourish*.

*Equitably opening access to data through responsible data sharing* involves moving beyond calls for 'open data' that sometimes run the risk of oversimplification and appropriation by market forces which could end up curtailing equitable access. In general, the opening of access to data is crucial for the reproducibility and reusability of research, the iterative improvement of datasets, and the amplification of research impacts for the public benefit.[123] However, the concept of 'open data' itself must be bounded and qualified.[124] Data sharing does not occur in a sociocultural, economic, or political vacuum but is situated amid an interconnected web of complex social practices, interests, norms, and obligations. This means that, at all times, those who share data ought to remain critically aware of the moral claims and rights of the individuals and communities whence the data came, of the real-world impacts of data sharing on those individuals and communities, and of the practical and sociotechnical barriers and enablers of equitable and inclusive research.

There is also a need to consider the right of communities to access and benefit from the use of their data. Such proposals for community-rights based approaches to data access and data sharing have been proposed from an economic standpoint, such that communities can become monetary beneficiaries of their aggregate data while leveraging their collective rights to access over multinational corporations.[125] Beyond this, community-rights based approaches to data access and data sharing have a strong participatory component whereby by equitably opening access to community data entails the democratic governance of data collection and use as well as robust regimes of social license and public consent.

*Equitably advancing access to research and innovation capacity* involves rectifying long-standing dynamics of global inequality that may undermine reciprocal sharing between research collaborators from high-income countries (HICs) and those from low-/middle-income countries (LMICs).[126] Given asymmetries in resources, infrastructure, and research capabilities, data sharing between LMICs and HICs, and the transnational opening of data, can lead to inequity and exploitation.[127] For example, data originators from LMICs may put immense amounts of effort and time into developing useful datasets (and openly share them) only to have their countries excluded from the benefits of the costly treatments and vaccines produced by the researchers from HICs who have capitalised on such data.[128] Moreover, data originators from LMICs may generate valuable datasets that they are then unable to independently and expeditiously utilise for needed research, because they lack the aptitudes possessed by scientists from HICs who are the beneficiaries of arbitrary asymmetries in education, training, and research capacitation.[129] This creates a two-fold architecture of inequity wherein the benefits of data production and sharing do not accrue to originating researchers and research subjects, and the scientists from LMICs are put in a position of relative disadvantage vis-à-vis those from HICs whose research efficacy and ability to more rapidly convert data into insights function, in fact, to undermine the efforts of their disadvantaged research partners.[130] In redressing these access barriers, emphasis must be placed on 'the social and material conditions under which data can be made useable, and the multiplicity of conversion factors required for researchers to engage with data'.[131] Equalising know-how and capability is a vital counterpart to equalising access to resources, and both together are necessary preconditions of just data sharing. Data scientists and developers engaging in international research collaborations should focus on forming substantively reciprocal partnerships where capacity-building and asymmetry-aware practices of cooperative innovation enable participatory parity and thus greater research access and equity.

---

[123] Borgman, 2015; Burgelman et al., 2019; Leslie, 2020a; Molloy, 2011; Piwowar et al., 2011; Tenopir et al., 2011; Whitlock, 2011

[124] Dove, 2015; Jasanoff, 2006; Leonelli, 2019

[125] Singh & Gurumurthy, 2021

[126] Drawn directly from Leslie, 2020a

[127] Bezuidenhout et al., 2017; Leonelli, 2013; Shrum, 2005

[128] Goldacre et al., 2015

[129] Bull et al., 2015; Merson et al., 2015

*Equitably advancing access to capabilities to flourish* involves a prioritisation of well-being at individual, community, and biospheric levels. It also involves the stewardship of the human capabilities and functionings that are needed for all to freely realise a life well-lived.[132] A capabilities- and flourishing-centred approach to just access demands that data collection and use be considered in terms of the affordances they provide for the ascertainment of wellbeing, flourishing, and the actualisation of individual and communal potential for these.[133] It demands a starting point in ensuring that 'practices of living' enable the shared pursuit of the fullness, creativity, harmony, and flourishing of human and biospheric life (what Abya Yala Indigenous traditions of Bolivia and Ecuador have called 'living well' or *sumak kawsay* in Quechua, *suma qamaña* in Aymara, or *buen vivir* in Spanish).[134]

**3. Confronting questions of equitable access involves *four dimensions of data justice*:** Concerns with equitable access should:

1. Concentrate on the equitable distribution of the harms and benefits of data use.
   This is the dimension of *distributive justice*.

2. Examine the material preconditions necessary for the universal realisation of justice.
   This is the dimension of *capabilities-centred social justice*.

---

[130] Bezuidenhout et al., 2017; Crane, 2011. Notably, such gaps in research resources and capabilities also exist within HICs where large research universities and technology corporations (as opposed to less well-resourced universities and companies) are well positioned to advance data research given their access to data and compute infrastructures.

[131] Bezuidenhout et al., 2017, p. 473

[132] From Robeyns & Byskov, 2021: 'Capabilities are the doings and beings that people can achieve if they so choose, such as being well-nourished, getting married, being educated, and travelling; functionings are capabilities that have been realized. Whether someone can convert a set of means - resources and public goods - into a functioning (i.e., whether she has a particular capability) crucially depends on certain personal, sociopolitical, and environmental conditions, which, in the capability literature, are called 'conversion factors.' Capabilities have also been referred to as real or substantive freedoms as they denote the freedoms that have been cleared of any potential obstacles, in contrast to mere formal rights and freedoms'.

[133] Nussbaum 1988, 1992, 2020; Sen 1992, 1993, 1999, 2009; Walsh 2000

[134] The idea of *sumac kawsay/buen vivir* anchors the 2008 Ecuadorian Constitution and the 2009 Bolivian Constitution. As Walsh (2018) explains: 'In its most general sense, buen vivir denotes, organizes, and constructs a system of knowledge and living based on the communion of humans and nature and on the spatial-temporal-harmonious totality of existence—that is, on the necessary interrelation of beings, knowledges, logics, and rationalities of thought, action, existence, and living. This notion is part and parcel of the cosmovision, cosmology, or philosophy of the Indigenous peoples of Abya Yala but also, and in a somewhat different way, of the descendants of the African diaspora'. Walsh 2018, p. 188. For helpful expansions on the concept of *sumac kawsay/buen vivir*, see: Huanacuni 2010; Walsh 2009, 2018.

3. Rectify the identity claims of those who have faced representational injury.
   This is the dimension of *representational and recognitional justice*.

4. Right the wrongs of the past so that justice can operate as a corrective dynamic in the present.
   This is the dimension of *restorative and reparational justice*.

This *four-dimensional approach to data justice* should use the ethical tools provided by the principles of social justice to assess the equity of existing social institutions, while also interrogating the real-world contextual factors that need to change for the universal realisation of the potential for human flourishing and reciprocal moral regard to become possible. It should likewise enable the reparation of historical injustices by instituting processes and mechanisms for reconciliation and restitution. While the first three of these facets remain integral to the advancement of access as it relates to data justice research and practice, they tend to focus primarily on addressing present harms and making course corrections oriented to a more just future. Restorative justice reorients this vision of the time horizons of justice. It takes aim at righting the wrongs of the past as a redeeming force in the present.

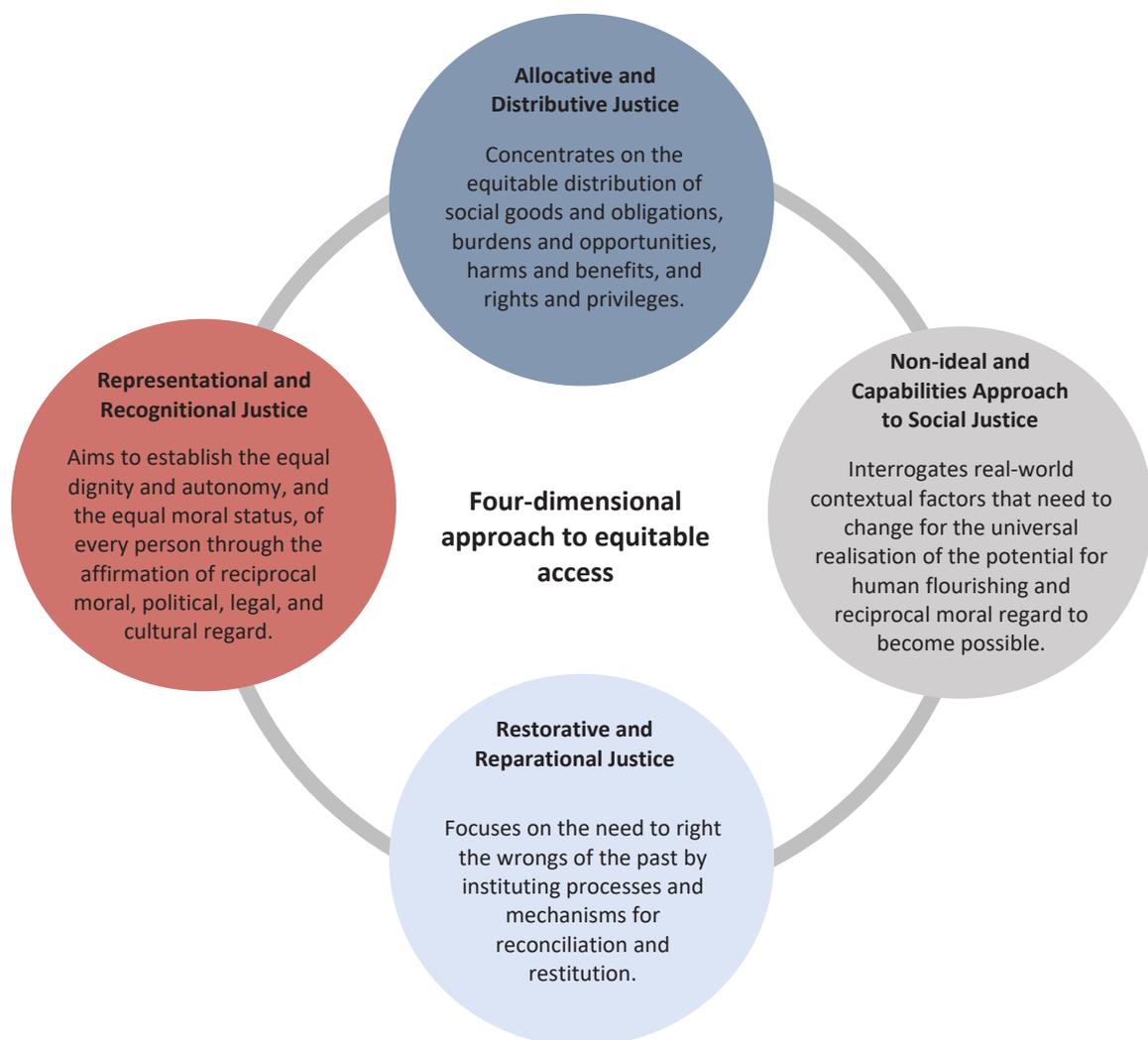

*Figure 10: Four-dimensional approach to equitable access.*

**4. Promote the airing and sharing of data injustices across communities through transparency and data witnessing:**[135] Datafication makes possible the greater visibility of everyday social experience. It has been argued that this expanded visibility is a 'double-edged sword' that calls for a balancing of 'the need to be counted, and thus potentially served and represented, against the potential for the abuse of power over those who are identified and monitored'.[136] However, increasing visibility should also be harnessed in positive ways to promote emancipatory transformation by exposing lived injustices, historical abuses, and moral harms.[137] The growth of a networked and connected global society multiplies the transformative power of observation and communication, enabling the far-reaching airing and sharing of previously hidden inequities and mistreatment. The witnessing of injustice both through proximate data work and through the employment of digital media at-a-distance should be marshalled as a force for change and as an opportunity to expand justice by means of transparency and voice.

The role of transparency in the airing and sharing of potentially unjust data practices must also be centred. Transparency extends both to outcomes of the use of data systems and to the processes behind their design, development, and implementation.[138] The latter component, *process transparency,* requires that the design, development, and implementation processes underlying the decisions or behaviours of data systems are accessible for oversight and review so that justified public trust and public consent can be ascertained. Process transparency also requires *professional and institutional transparency*. At every stage of the design and implementation of a project, responsible team members should be identified and held to rigorous standards of conduct that secure and maintain professionalism and institutional transparency. These standards should include the core, justice-promoting values of integrity, honesty, and sincerity as well as reflexively-fortified and positionality-aware modes of neutrality, objectivity, and impartiality. All professionals involved in the research, development, production, and implementation of data-intensive technologies are, first and foremost, acting as fiduciaries of the public interest and must, in keeping with these core justice-promoting values, put the obligations to serve that interest above any other concerns.

The second element of transparency, *outcome transparency*, demands that stakeholders are informed of where data systems are being used and how and why such systems performed the way they did in specific contexts. Outcome transparency therefore requires that impacted individuals are able to understand the rationale behind the decisions or behaviours of these systems, so that they can contest objectionable results and seek effective remedy. Such information should be provided in a plain, understandable, non-specialist language and in a manner relevant and meaningful to those affected.

---

[135] Gray, 2019
[136] Martin & Taylor, 2021
[137] Gray, 2019
[138] Leslie, 2019

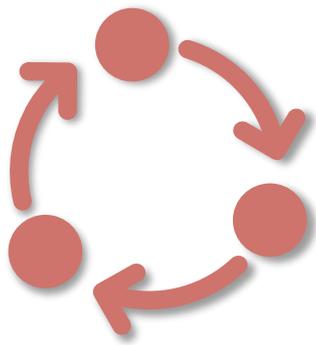

Process Transparency

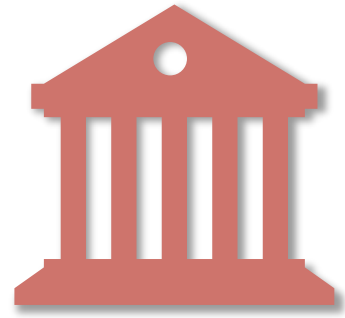

Professional and Institutional Transparency

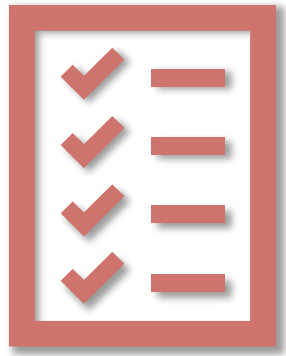

Outcome Transparency

*Figure 11: Various types of transparency.*

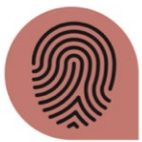

## Identity

**1. Interrogate, understand, and critique harmful categorisations and modes of othering:** The construction and categorisation of data, particularly when it is about people, is a fundamentally social activity that is undertaken by humans whose views of the world are, in part, the product of cultural contexts and historical contingencies. As such, the construction and categorisation of data is shaped by the sociocultural conditions and historical contexts from which it is derived. The social character of data coupled with the sorting and clustering that results from its cleaning and pre-processing can lead to categorisations that are racialised, misgendered, or otherwise discriminatory. This can involve the employment of binary categorisations and constructions—for example, gender binaries (male/female) or racial binaries (white/non-white)—that are oriented to dominant groups and that ought to be critically scrutinised and questioned. Data justice calls for examining, exposing, and critiquing histories of racialisation and discriminatory systems of categorisation reflected in the way data is classified and the social contexts underlying the production of these classifications.

**2. Challenge reification and erasure**: *Resist the reification of identities as a convenience of computational sorting and optimisation,*[139] *and contest the erasure of identities and the risk of intersectional harm from incomplete and mistargeted data innovation practices.* In the construction and categorisation of data, system designers and developers can mistakenly treat socially constructed, contested, and negotiated categories of identity as fixed and natural classes. When this happens, the way that these designers and developers categorise identities can become naturalised and reified, thereby running the risk that categories or classes that they may have excluded, missed, or grouped together are erased or rendered invisible. For instance, the designers of a data system may group together a variety of non-majority racial identities under the category of "non-white", or they may record gender only in terms of binary classification and erase the identity claims of non-binary and trans people.

In a similar way, designers and developers can produce and use data systems that disparately injure people who possess unacknowledged intersectional characteristics of identity which render them vulnerable to harm, but which are not recognised in the bias mitigation and performance testing measures taken by development teams. For instance, a facial recognition system could be trained on a dataset that is primarily populated by images of white men, thereby causing the trained system to systematically perform poorly for women with darker skin. If the designers of this system have not taken into account the vulnerable intersectional identity (in this case, women with darker skin) in their bias mitigation and performance testing activities, this identity group becomes invisible, and so too do injuries done to its members.[140]

Implicit practice of erasure      Corrective practices of inclusion

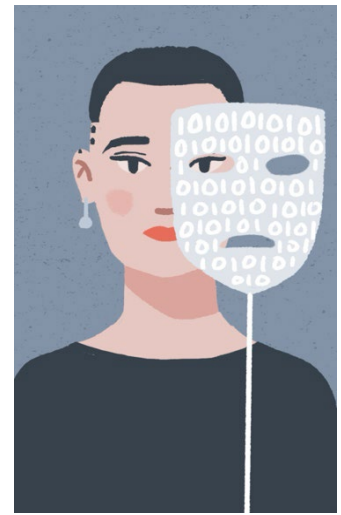

*Figure 12: Practices of erasure that take place during project lifecycle.*

---

**3. Focus on how struggles for recognition can combat harms of representation:** Struggles for the rectification of moral injuries to identity claims that are suffered at the hands of discriminatory data practices should be understood as struggles for recognitional justice—struggles to establish the equal dignity and autonomy, and the equal moral status, of every person through the affirmation of reciprocal moral, political, legal, and cultural regard.

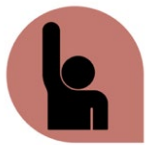 **Participation**

**1. Democratise data and data work:** *Prioritise meaningful and representative stakeholder participation, engagement, and involvement from the earliest stages of the data innovation lifecycle to ensure social licence, public consent, and justified public trust.* The democratisation of data scientific research and data innovation practices involves bringing members of impacted communities, policymakers, practitioners, and developers together to collaboratively articulate shared visions for the direction that data innovation agendas should take. This entails the collective and democratically based determination of what acceptable and unacceptable uses of data research and innovation are, how data research and innovation should be governed, and how to integrate the priorities of social justice, non-discrimination, and equality into practices of data collection, processing, and use.

**2. Understand data and data subjects relationally:**[141] Data collection and use should not be pursued in a way that reifies, objectifies, or commodifies data or data subjects. Where data innovation practices focus only on an individual's relationship to data (as a possession or form of property) or on an individual's privacy or data protection rights, these practices lose sight of the wider contexts of their social effects, their population-level impacts, and the interconnectedness of the people and communities who are affected by data innovation ecosystems. A relational view of data practices, which starts from this broader vantage point, recasts them as involving horizontal and interwoven social relationships in addition to vertical relationships between the individual data subject and the data collector, processor, or user. Understanding data and data subjects *relationally* entails recognising that data practices need to be situated in their social environments and governed democratically through horizontal, participation-based forms of collective action that provide coverage of a complex and multi-stakeholder ecology of interests, rights, obligations, and responsibilities.

**3. Challenge existing, domination-preserving modes of participation:** Where current justifications and dynamics of data practices reinforce or institutionalise prevailing power structures and hierarchies, the choice to participate in such practices can be counterproductive or even harmful. When options for a community's participation in data innovation ecosystems and their governance operate to normalise or support existing power imbalances and the unjust data practices that could follow from them, these options for involvement should be approached critically. A critical refusal to participate is a form of critical participation[142] and should remain a practical alternative where extant modes of participation normalise harmful data practices and the exploitation of vulnerability.

---

[141] Viljoen, 2020
[142] Ahmed, 2012, 2018; Benjamin, 2016; Hoffman, 2021; Cifor et al., 2019

**4. Ensure transformational inclusiveness rather than power-preserving inclusion:**[143] Incorporating the priority of inclusion into sociotechnical processes of data innovation can be detrimental where existing power hierarchies are sustained or left unaddressed. Where mechanisms of inclusion normalise or support existing power imbalances in ways that could perpetuate data injustices and fortify unequal relationships, these should be critically avoided. *Transformational inclusiveness demands participatory parity* so that the terms of engagement, modes of involvement, and communicative relationships between the includers and the included are equitable, symmetrical, egalitarian, and reciprocal.

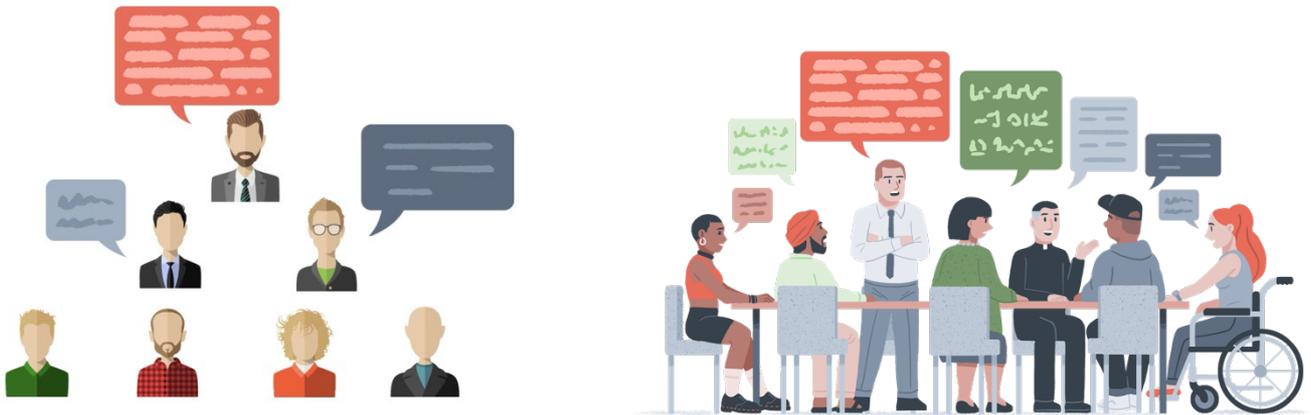

*Figure 13: Moving towards transformational inclusiveness.*

---

[143] Hoffman, 2020

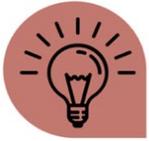 **Knowledge**

**1. Embrace the pluralism of knowledges:** Different communities and sociocultural groups possess unique ways of seeing, understanding, and being in the world, and this plurality of knowledges and of lived experience should inform and be respected in practices of data collection, processing, and use as well as in the policymaking practices surrounding the governance of data technologies. Embracing the pluralism of knowledges involves recognising that diverse forms of knowledge, and ways of knowing and understanding, can add valuable insights to the aspirations, purposes, and justifications of data use—including on the local or context-specific impacts of data-intensive innovation. Moreover, inclusion of diverse knowledges and ways of being can open unforeseen paths to societal and biospheric benefits and maximise the value and utility of data use across society in ways which take account of the needs, interests, and concerns of all affected communities.

**2. Challenge the assumed or unquestioned authority of technical, professional, or "expert" knowledge across scientific and political structures:** Processes of knowledge creation in data science and innovation are social processes which require scrutiny and wider public engagement to hold "expertise" to account and to ensure that data science and innovation progress in ways which align with wider societal values. This means that data technology producers and users have a responsibility to communicate plainly, equitably, and to as wide an audience as possible. Clear and accessible public communication of research and innovation purposes/goals and data analytic and scientific results, should enable the public to interrogate the claims and arguments being put forward to justify data-driven decision-making and data innovation agendas. This also means that members of the public have a corollary responsibility to listen to—i.e. to pay attention to, engage with, and critically assess – the scientifically authoritative knowledge claims and technological systems that impact them.

**3. Prioritise interdisciplinarity:** Approach the pursuit of understanding of data innovation environments—and the sociotechnical processes and practices behind them—through a holistically informed plurality of methods. This involves placing a wide range of academic disciplines and specialised knowledges conceptually on par, enabling an appreciation and integration of a wide range of insights, framings, and understandings. Ways of knowing that cannot (or are not willing to) accommodate a disciplinary plurality of knowledgeable voices that may contribute to richer comprehensions of any given problem cease to be knowledgeable *per se*.

**4. Pursue 'strong objectivity':**[144] A robust approach to objectivity demands that knowers have positional self-awareness, which acknowledges the limits of each individual's personal, historical, and cultural standpoint. It also demands that knowers carry out critical and systematic self-interrogation to better understand these limitations. This launching point in *strong objectivity* can end up leading to *more objective and more universalistic understandings* than modes of scientific or technical objectivity which stake a claim to unobstructed neutrality and value-free knowledge that evades self-interrogation about the limits of standpoint and positionality. One reason for this has to do with power dynamics. Strong objectivity starts from a reflective recognition of how differential relations of power and social domination can skew the objectivity of deliberations by biasing the balance of voices that are represented in those deliberations. It then actively tries to include and amplify marginalised voices in the community of inquiry to transform situations of social disadvantage where important perspectives and insights are muted, silenced, and excluded into situations that are scientifically richer and *more advantaged*. Such richer and more inclusive ecologies of understanding end up producing more comprehensive knowledge and more just and coherent practical and societal outcomes. Strong objectivity amplifies the voices of the marginalised, vulnerable, and oppressed as a way to overcome claims of objectivity, impartiality, and neutrality that mask unquestioned privileges.[145]

---

[144] Harding, 1992, 1995, 2008, 2015

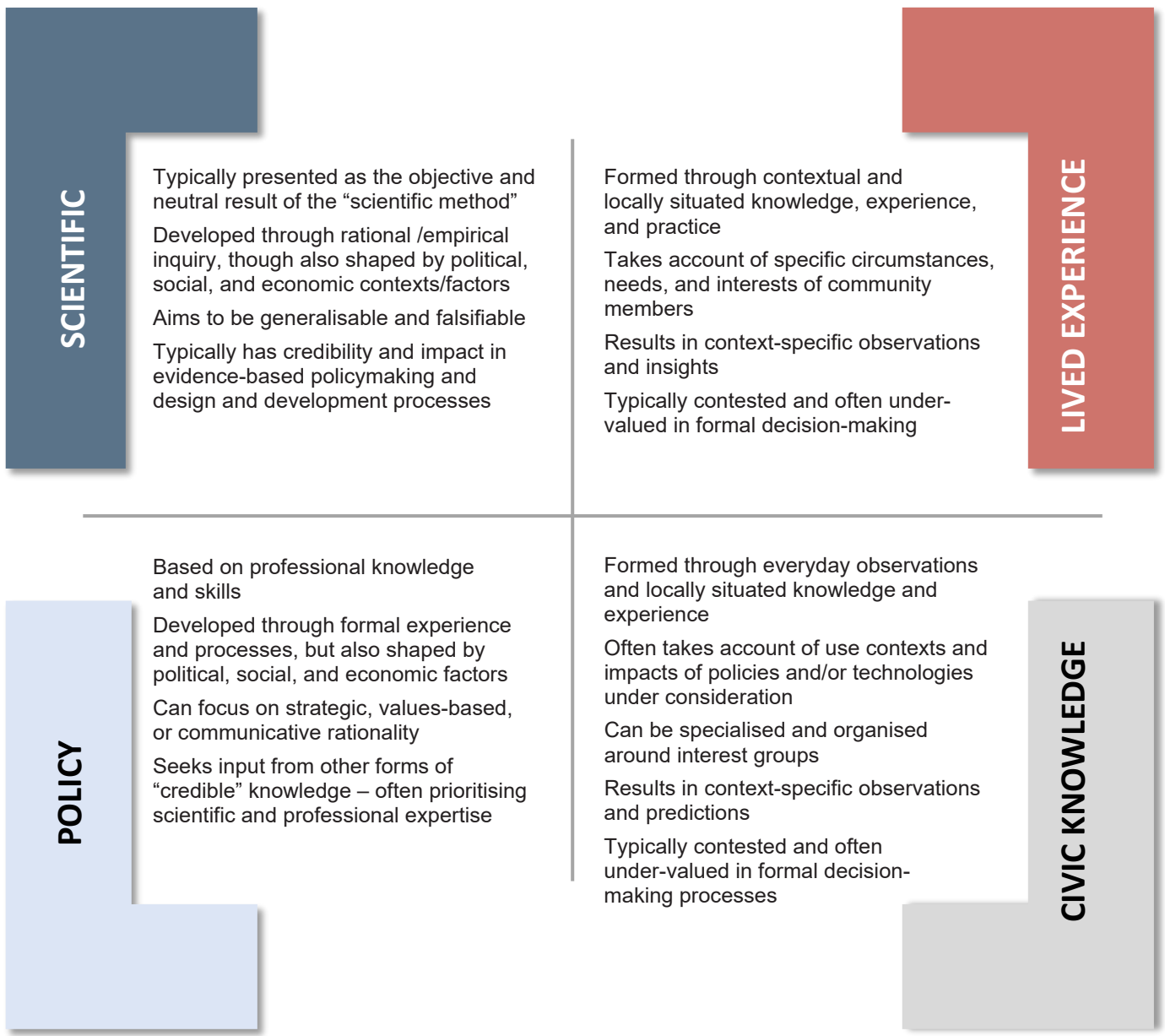

*Figure 14: Pluralism of knowledges.*

---

**5. Cultivate intercultural sharing, learning, and wisdom:** A plurality of insights, learning, and wisdom from a diverse range of communities and sociocultural groups should inform the values, beliefs, and purposes behind data research and innovation agendas and practices. Cultivating intercultural sharing, learning, and wisdom serves this end by bringing a multitude of ideas and perspectives into conversation. This involves setting up and sustaining networks of communication and collaboration between communities and sociocultural groups, so that they can come together to cultivate shared understandings and constructively explore differences. At the same time, cultivating intercultural sharing, learning, and wisdom *as a way to advance data justice* involves drawing on the principles and priorities of social justice to find commonality and to build solidarity among communities and sociocultural groups.

# Thematic Review of Literature

This thematic review of literature draws on a broad base of contributions across academic, policy, and activist literatures. To incorporate this range of contributions, seven themes will be addressed in turn in order to lay out the multiple perspectives from which transformational research and work to advance data justice is taking place.

First, literature on 'the geopolitics of data power, essential digital infrastructures, and data flows' is addressed to introduce the ways in which powerful networks on a national, international, and infrastructural level are determining datafication trajectories while bottom-up coalitions begin to find ways of advancing their own transformational priorities.

Second, 'data colonialism, data activism, and de-colonial AI' are addressed through literatures which expose data extraction and other colonial practices. Literatures which propose tackling this through various modes of activism are also analysed.

Third, 'economic and distributive justice' literatures are reviewed, spanning both accounts of existing economic policies to tackle inequality resulting from datafication and novel proposals to resolve these injustices through collective approaches to the economic governance of data.

Fourth, literatures covering topics of 'identity, democratic agency, and data injustice' are addressed as harms of representation inflicted by datafication are laid out and contextualised alongside movements such as data feminism, design justice, and intersectionality which tackle these harms.

Fifth, this review incorporates 'adjacent justice literatures' and the lessons they offer to the advancement of data justice. Here literatures across the themes of environmental justice, culture-centred communication, restorative justice, global public health justice and participatory learning and action are linked directly to the themes and pillars of data justice.

Sixth, 'knowledge, plurality, and power' is addressed through an exploration of pluriverse and post-development theory and of science and technology studies.

Finally, 'non-Western and intercultural approaches to data justice and injustice' are incorporated to include advancements to data justice provided by intercultural communication, the Indigenous data sovereignty movement, and more.

# The Geopolitics of Data Power, Essential Digital Infrastructures, and Data Flows

## Geopolitics of Data Power

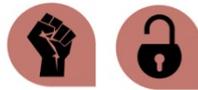

| Key Points | Gaps Identified |
|---|---|
| - Data and data-driven technologies can offer nation states strategic advantages by increasing economic affluence, national security capability, and social, cultural, and political influence.<br><br>- The rise of datafication has shifted the global balance of power, allowing China to leverage new technologies while Europe, and sometimes the United States, are predicted to lag.<br><br>- More complex international dynamics—for example regarding the role played by low-and-middle-income countries in the development of international technical standards—are often obscured by the narrative of conflict between the duopoly of the United States and China or the triumvirate of the US, the European Union, and China.<br><br>- Literature suggests significant action from international bodies, in particular the UN, is required to de-escalate tensions and the potential for runaway technological competition. | - Literature which adopts the geopolitics framing tends to focus on the United States and China, with Europe, the Americas, Oceania, and Africa sometimes appearing as background players or not mentioned at all. More work is needed to explore the relative positionality of other nations and to consider geopolitical issue in a more globally inclusive way.<br><br>- There is a significant gap regarding future recommendations. While there has been a focus on the possible role the UN could take, little attention has been paid to the collective power of national governments, regional bodies, and potential alliances outside of the United States and China.<br><br>- More work is needed to integrate recommendations for the private sector, national governments, and multilateral organisations. |

The role played by 'information power' in international affairs is longstanding as those who hold large quantities of information are considered to have the ability to make 'more effective decisions'.[146] However, data-driven technologies and largescale data collection have transformed this power.[147] Data power now operates in numerous spheres to further nations' economic, defence, social, cultural, and political influence.[148] This power shapes international relations as national governments are incentivised to increase investments in data-driven technologies, to collect a growing quantity of data from citizens and data subjects, and to place strict limitations on the collection of personal data by foreign actors. Governments have responded to such incentives in contrasting ways, resulting in significant shifts in power.

---

[146] Rosenbach & Mansted, 2019

[147] Ibid.

[148] Rosenbach & Mansted, 2019; Pauwels, 2019

Literature on the global state of play regarding the geopolitics of data power frequently centres China. Within this narrow focus, most attention is given to two key elements of China's rise. First, authors emphasise the success of China's strategic investments in tech giants.[149] Second, authors underline information authoritarianism or autocracy as a dominant factor contributing both to domestic control and international influence.[150] Others argue the Chinese approach cannot be captured by accounts of top-down, monolithic policies which neglect ethics entirely.[151] The importance of understanding China's AI strategy from a Chinese rather than a Western perspective has also been emphasised.[152] Finally, other authors have exposed less conspicuous contexts through which China's influence is growing, in particular through their agenda setting power on technical standards, an arena which is 'too often considered as benign'.[153]

In contrast to narratives on China's rise, literature on the United States is divided regarding who holds data power. A common approach has been to debate whether the US or China will win what is seen as a 'zero-sum game' towards AI leadership. However, this has been critiqued by those who do not see the success of China and the US as mutually exclusive.[154] Another approach is to focus not on competition between the US and China but the gulf that lies between the 'American and Chinese digital empires' and the rest of the world.[155] Within literature which addresses the US' data power, there is debate surrounding whether Silicon Valley or the US government is the key power broker. The 'Four Internets' model goes so far as to distinguish the government's model of the internet from that of Silicon Valley.[156] Others have emphasised that, to be competitive, the US government has no choice but to coordinate with the private sector.[157]

Receiving relatively less attention within literature on geopolitics and AI, Europe is often cast as 'the champion of ethics' and human rights.[158] Some have suggested that the European Union could learn from the innovation centred approach taken by China, while others have instead called upon China to learn from the EU's AI ethics strategies.[159] Additionally, while the EU has at times been praised for a relatively ethics-focused and rights-based approach, critics have called for the competitive edge of EU companies to be advanced.[160] Be that as it may, some authors point out, by contrast, that the geopolitics of data 'is made to order for large firms and large economies that generate data at scale and that invest at scale', stressing that 'the main contest is thus between the three major economies – the United States, the EU and China – which have the capacity to structure data realms'.[161] Other nations which frequently get cast as secondary players include the United Kingdom, Israel, Japan, South Korea, Australia, and Russia.[162]

---

[149] Lee, 2018

[150] Pauwels, 2019

[151] Ding, 2018

[152] Roberts et al., 2020

[153] Seaman, 2020

[154] Villasenor, 2018

[155] Miailhe, 2018

[156] O'Hara & Hall, 2018

[157] Rosenbach & Mansted, 2019

[158] Miailhe, 2018

[159] Roberts et al., 2021a

[160] Roberts et al., 2021b

[161] Ciuriak, 2021, p.11; See also Ciuriak & Ptashkina, 2018b; Aaronson & Leblond, 2018; McDonald & An, 2018

[162] Villasenor, 2018; Williams, 2019

The inclusion of other nations in the debate surrounding geopolitics and AI is rare. Africa has been referenced as a 'battleground' for the colonising behaviour of China and the United States—the renewed site for 21st century version of the 'Scramble for Africa'[163]—but the interests of individual nations or regional alliances on the geopolitical stage are rarely addressed.[164] More broadly, developing nations have been incorporated into the geopolitical framing as 'vulnerable links', states whose underdeveloped cyber-security strategies render them vulnerable to attacks.[165]

To the extent that literature addresses future directions, authors often focus on recommendations for one nation. This has been framed within the context of consolidating or gaining power through policies such as 'avoiding overregulation', or prioritising 'coordination with the private sector'.[166] Internationalist framings of solutions to geopolitical tensions focus instead on the role multilateral organisations can play in responsible innovation, de-escalation, and power distribution. Pauwels argues the United Nations can play a unique role in 'ensuring that wider participation, from more vulnerable States or underserved groups, is enabled and supported within strategic foresight'.[167] The UN has also begun to facilitate global cooperation on AI governance through the High-Level Panel on Digital Cooperation. This led to the adoption in 2019 of five core recommendations: 'build an inclusive digital economy and society', 'develop human and institutional capacity', 'protect human rights and human agency', 'promote digital trust, security and stability', and 'foster digital cooperation'.[168]

---

[163] Ciuriak, 2021

[164] Villasenor, 2018

[165] Pauwels, 2019

[166] Villasenor, 2018; Rosenbach & Mansted, 2019

[167] Pauwels, 2019

[168] United Nations, 2020

## Digital Infrastructures

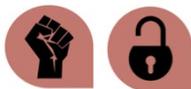

| Key Points | Gaps Identified |
|---|---|
| - Digital infrastructures provide the underpinnings for data-driven communication, applications, and services and therefore have significant societal impact.<br><br>- There is much consolidation in the digital infrastructures space, with the majority being offered by a few dominant US-based tech firms, with some in China gaining presence globally.<br><br>- Datafication at scale is a common feature of digital infrastructures, capturing and extracting data on individuals for organisational-gain.<br><br>- The opacity surrounding data infrastructures—largely shaped by claims for trade secrecy, intellectual property, and security—hinders the ability for oversight, contestation, and challenge.<br><br>- The power of infrastructure providers raises issues in terms of gatekeeping, value exportation, lack of representativeness, among others. | - Though there is significant literature questioning the power of these firms, there are real opportunities for work offering practical and pragmatic alternatives.<br><br>- A means for tackling issues of socio-technical opacity throughout data innovation supply chains is necessary moving forward. This should consider more than just the organisation/individual relationship.<br><br>- A need for interventions that recognise the fundamental role digital infrastructures play in society, including those that prevent extractive business models, and treating the infrastructures as a necessary public good. |

Data-driven systems have the potential to infringe on a wide range of fundamental rights and freedoms, regardless of the field of application. Though many appreciate that such systems can raise issues around privacy, data protection, and non-discrimination, less discussed are other rights, such as human dignity, access to justice, social and economic rights, and consumer protection, among others. While much of the rights discussion tends to focus on 'AI', ultimately AI is driven by the processing of data, where any practical application of AI operates not on its own, but as part of a wider socio-technical system.[169]

Importantly, as many scholars have emphasised, technological systems are not neutral; not only does any AI or data-driven system inherently encapsulate the values of its designers, but it also springs from a field designed to serve powerful decision-makers demanding tools that are predictive and efficient. Comparatively little research, however, focuses on the data subjects on the other side of the power structure. For this reason, researchers like Kalluri argue that what is needed are ways for these individuals to investigate AI systems, to contest them, to influence them, or to even dismantle them.[170]

---

[169] Cobbe et al., 2021

[170] Kalluri, 2020

Data processing, and the systems that enable it, are largely systemically opaque because users of digital services often lack (or are prevented from gaining) knowledge or understanding of aspects including: (1) the practices of organisations capturing and processing their data, including the details, reasons for, and implications of holding particular data or performing particular computation; (2) the data sharing practices of those organisations with third parties and beyond; (3) the technical details and complexity of the systems involved; (4) the data-driven, and indeed, often surveillance-driven business models; (5) the insights and power that organisations can gain through having access to data, particularly where data is aggregated or computation occurs at scale (collective computation); and (6) the legal issues that can extend from copyright, trade secrecy, and so on that restrict the information available.[171]

From a transnational-perspective, data infrastructures are dominated by the major US tech-firms. Arguably, the pre-eminence of these firms suggests that we are still living in the imperialist era. The United States, which has utilised its imperial power, not only with military force, capital, and cultural products, but also with technologies, continues to dominate most of the world by way exercising control over the provision of digital technologies and infrastructures.[172] China is also home to several dominant firms providing substantial digital infrastructures – though these primarily focus their operations within China – they are increasingly expanding their offerings to other markets.

One primary effect of this dominance is the way business practices, whose values are those of a handful of enormous and hugely influential firms, become the values that drive the priorities and experience of millions. In the 'era of corporate gigantism',[173] as Frank Pasquale calls it, tendencies to the centralisation and consolidation of power over data and compute resources translates into the amplification of the societal influence and voice of a precious few.[174] Some scholars see an unavoidable conflict here between dominating market forces and the public good.[175] This is exemplified by the distribution of cultural products by social media platforms. The power of platforms creates optimisation mandates for news organisations and other cultural producers. News sites produce content based on perceived and measurable social media uptake. Rather than seeing cultural products merely reflected and reproduced in platform content, production is targeted to follow the flow of social media design and usage.[176] Related are issues of representativeness, where we have already seen many services built for (and thus favour) particular demographics, cultures, and so on, reflecting aims, values, backgrounds, and geographies of the dominant firms. This is seen, for instance, in a computer vision algorithm that labels a photograph of a traditional US bride dressed in white as 'bride', 'dress', 'woman', 'wedding', but a photograph of an Indian bride as 'performance art' and 'costume', likely a result of having models built using insufficiently diverse and unrepresentative datasets.[177]

---

Importantly, many of the business models of these dominant firms entail the 'datafication' of people. In terms of regulation, most consider a vertical relationship, whereby organisations and individuals interact directly to exercise obligations and rights. Yet, the more common relation is the horizontal, in which people are treated as members of groups rather than individuals. Datafication along the horizontal relation, where group identities and collectivities are captured in classification, prediction, and inference, is little addressed in regulation but warrants attention as it works not only to erode the capacity for subject self-formation, but also materialise unjust social conditions: data relations that enact or amplify social inequality and injustice, inscribing forms of oppression and domination over marginalised or discriminated against groups.[178]

At a lower infrastructural level, the control of internet service providers (ISPs) and telecom companies over broadband infrastructure is one in which private actors possess the means to undermine the public value of essential goods and services upon which many businesses, communities, and individuals depend. Dangers arise where private actors accumulate outsized control over those goods and services that form the vital foundation or backbone of any given political economy and social infrastructure.[179] Consequently, there are arguments that the major information platforms should be viewed as essential infrastructure and could potentially be regulated as public utilities.[180]

## Jurisdictions and Data Flows 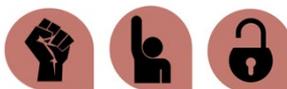

| Key Points | Gaps Identified |
|---|---|
| - Data law/regulation and associated legal rights and obligations are tied to some 'presence' within a geographic boundary.<br><br>- Data readily flows across jurisdictional boundaries.<br><br>- Some jurisdictions assert control over the data and infrastructures in their territories, including the data flows in/out.<br><br>-This means that an understanding of the political and legal contexts of data localisation is an important part of discerning potential data injustices and harms. | - More work is needed on developing global data governance norms that address the issues surrounding data localisation and the equitable transfer, sharing, and management of data flows.<br><br>- Discussions are often driven by 'Western' perspectives, much more diversity of input is required. |

Digital infrastructures both enable and are enabled by the flow of data. Applications, tools, services, and other functionality delivered by technology are inherently data-driven, which entails the movement of data within and across digital infrastructure. Given the Internet's global nature, this means that from a technical perspective, data can readily travel across regional borders, and therefore across jurisdictional boundaries.

---

[178] Viljoen, 2021
[179] Rahman, 2018
[180] Teachout & Rahman, 2020

Though information is often spoken about as virtual, intangible, and 'know[ing] no bounds',[181] digital infrastructures do, however, have some physical presence. This can include the geographic location in which the particular technical infrastructure (such as a server or network) resides.[182] This may also concern the regional presence of the organisation (and the people involved) that is deploying and operating the infrastructure, e.g., where they are incorporated, where they 'do business', as well as the location of users (or individuals to which data otherwise relates). This regional presence or connection provides a means through which states can assert their jurisdictional authority over data, be it to enforce regulation or to engage in certain behaviours, such as surveillance, access to data, content suppression, etc.

This is relevant to data justice for a variety of reasons. Specifically, from a legal (and geo-political) standpoint, several issues arise surrounding data moving across borders. One high-profile Western example of this are the questions of legality regarding the flow of data from the European Union to the United States, where a key concern has been that US law could allow personal data to be accessed on national security grounds, in a manner that misaligns with EU principles of data protection.[183] There are also issues concerning the exporting, or indeed imposition, of particular legal, governance (and other) norms, which can arise given that the data infrastructures that many are forced to rely on are consolidated around a few entities from particular jurisdictions.[184]

Relevant here is the concept of *data localisation*. In essence, this is where a state requires that certain data—and therefore data infrastructure—should or must reside within a particular region,[185] working to encumber the transfer of data across national borders.[186] By keeping (some) data and its infrastructure—including the people and organisations involved—within its jurisdictional reach, this gives the state a lever by which to exert some control over the local data landscape and its interactions with those in other regions. However, the implementation, enforcement, and other practicalities of a data localisation regime is challenging given the global nature of technology and data infrastructures.[187]

The goals of a data localisation regime naturally vary by the aims and circumstances of the particular state. Localisation-measures may be imposed to achieve different goals, e.g., concerning control over aspects of privacy, trade, law enforcement, national security, and so on.[188] It follows that some interventions might work to support the rights and interests of citizens, and assist data justice, while others less so. Moreover, a localisation regime can also serve to 'export' particular law and norms, where, for example, the restrictions on data transfers outside of a state places certain obligations on those on the receiving end of such data. It follows that any localisation measures must be considered in their political context.

---

[181] Ness, 1996
[182] Lakhina et al., 2004
[183] Tracol, 2020
[184] Internet Society, 2019
[185] Cory, 2017
[186] Chander & Le, 2014
[187] Hon et al., 2016
[188] Chander & Le, 2014; Cory, 2017

# Data-Driven State and Surveillance

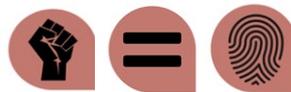

| Key Points | Gaps Identified |
|---|---|
| - Surveillance has evolved from a largely visual practice to more subtle data-driven strategies through the expansion of digital capacities. It is undertaken both for profit and corporate advantage and for the state's exercise of administrative control.<br><br>- Development of surveillance tools are proliferating at a faster rate than the reaction necessary for adequate regulation of practices.<br><br>- Potential for the data subject to reflect on and renegotiate terms of agreement is limited.<br><br>- Gender, race, class, caste, and so forth are embedded in the power relations of surveillance capitalism and state surveillance machinery.<br><br>- Based on a security and privacy trade-off, national and international strategies of data surveillance have resulted in the othering of individuals and groups while algorithmic processes developed within limited characteristic parameters can reify extant marginalisation.<br><br>- Surveillance has been found to overlap with challenges such as privacy, data colonialism, critical theory, and power dynamics. | - Literature is predominantly focused on Transatlantic and Western-centric challenges of data, privacy, and surveillance through generalised perspectives that insufficiently cover the plurality of potentially harmful or inequitable experiences, particularly from the 'Global South'.<br><br>- Prospective solutions to mitigate harms identified in surveillance modalities, while dependent on eliminating historical marginalisation, is narrowly focused on regulatory mechanisms for data processors; overlooking the diverse ways surveillance and datafication influence data subjects across the world and limiting the exploration of transformative possibilities for public participation and democratic governance. |

Surveillance capitalism as conceptualised by Shoshana Zuboff is a 'new economic order that claims human experience as free raw material for hidden commercial practices of extraction, prediction, and sales'.[189] Economic value is generated through the monetisation of extracted data. This is supported by an array of systems and technologies feeding on the nudging, curating, and shaping of human behaviour and social experience. Data on behaviour and preferences are gathered and consolidated in databases. These are later transformed into algorithmic systems that aim to alter or modify behavioural patterns for commercial profitability. To advance data justice, the three pillars of Power, Equity, and Identity can be employed to critique and confront the corresponding dangers associated with surveillance capitalism. The pillars can also be applied to discern and address similar digital surveillance and control tactics of data-driven states.

---

[189] Zuboff, 2015, p. ix

Some scholars have observed that, in both corporate and state digital surveillance, data are collected almost entirely from interactions and self-disclosures through the 'tyranny of convenience', wherein individuals are compelled to divulge personal information and data because of the increasing digitisation of essential everyday activities that are captured by sensors and measurement mechanisms controlled by corporate-owned services or political authorities.[190] Data extraction is so deeply rooted within social processes that disconnecting oneself is nearly impossible. In addition to this, the data subject often has too limited a knowledge of data processes to effectively engage with or (re)negotiate terms of agreement, where this may be possible.[191]

Much of contemporary scholarship on power and surveillance capitalism has tended to situate the operation of the global political economy in the context of the relationship of datafication and biopower (i.e. the mechanisms or techniques of power that are directed at the regulation and control of living bodies and human populations). This perspective has been largely derived from the foundational work of Michel Foucault, who argued that a full understanding of power needs to move beyond the centralised and territorialised exercise of political sovereignty or legal coercion to include the dispersed forms of scientific authority, technological tactics, and knowledge regimes that have a regulative influence on human body, psyches, and groups.[192] In the data justice context, power in this Foucauldian sense is increasingly de-territorialised, extending beyond the state to 'anywhere by any organisation through information gathering and data-management processes and tools'. A use case frequently cited as an instance of biopower in data innovation ecosystems is the deployment of live facial recognition technologies by states or profit-oriented companies to identify, securitise, and control individual human bodies.[193]

The rise of digital surveillance has introduced a host of other problems. The accelerating rate of release of new surveillance capabilities has resulted in a regulatory lag; technological developments outpace the social processes of reflection, criticism, and revision that are needed for the development and formulation of effective regulatory policy. Additionally, disadvantaged or marginalised groups within society – those differentiated by race, gender, class, region, etc. – are disproportionately affected by dataveillance and information gathering at scale. From the use of zip-codes as predictive variables or proxies in modelling and government census data, surveillance capitalism commodifies information to alter consumer demand trends, restrict access to goods and services, or even target individuals with predatory services.[194]

---

[190] Lester, 2001; Andrejevic, 2007

[191] van Dijck, 2014

[192] Ceyhan, 2012

[193] Leslie 2020b; Selinger & Hartzog, 2020; Starks et al., 2019

[194] Browne, 2012

The evolution of data-intensive surveillance technologies has also occurred in the context of coloniality. Such technologies have been developed in 'Western', gendered, and racialised frameworks where individuals are *othered* by these systems and by the institutions designing and implementing them. These data processing tools facilitate the gathering of mass data within and across national boundaries. Notably, such mechanisms reveal the historical issues of Western colonialism and globalisation associated with the movement of refugees, asylum seekers, and immigrants. Not limited to the Transatlantic or 'Global North', surveillance capitalism is also expanding to the markets of formerly colonised economies. Simultaneously, states themselves, including those in the 'Global South', employ surveillance tools to monitor, discipline, or deter. For example, India's biometric identity card and China's social-credit system which academics have found to be fraught with issues of privacy, exclusion, and power asymmetries.[195] Thus, resulting in the creation of new forms of exclusion described as 'digital discrimination'.[196] Moreover, the "social sorting" reveals how 'the social practices of surveillance and control sort out, filter and serialise who needs to be controlled and who is free of that control'.[197]

Edward Snowden's revelations on the US National Security Agency's (NSA) as well as the British Government Communications Headquarters' (GCHQ) access to the content and traffic data of Internet users has been found to result in what Dencik and Cable have identified as 'surveillance realism'.[198] Not only are certain consumers and activists affected with a sense of resignation in the face of limited transparency and knowledge, but mass surveillance has also restricted avenues of resistance. Instead, individuals increasingly self-regulate patterns of behaviour and 'outsource' concern to experts.[199]

---

[195] Banga, 2019; Campbell, 2019
[196] Lyon, 2003
[197] Bigo & Guild, 2005, p. 5
[198] Dencik & Cable, 2017
[199] Ibid.

# Human Rights in Data Infrastructures and AI Systems

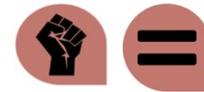

| Key Points | Gaps Identified |
|---|---|
| - From the perspective of international human rights doctrine, several harms to fundamental rights and freedoms are potentially implicated by data and AI, including violations of labour rights as well as of liberty, security, privacy, and freedom of expression and belief.<br><br>- The energy requirements to train machine learning models (e.g. language models) influence climate change and thereby present a hazard to the sustainability of the biosphere.<br><br>- Supply chain demands for data-intensive technologies contribute to forced labour and other labour abuses.<br><br>- Cross-border data flows to facilitate commerce, security, and public health create risk due to incompatible national regulatory models.<br><br>- While individuals are increasingly required to provide data about themselves and their communities in order to participate fully in society, trust in corporations and governments to use data responsibly is a major hurdle.<br><br>- Value-sensitive and participatory-design frameworks may translate fundamental human rights into context-dependent design through structured, inclusive, and transparent participatory processes. | - The general application of human rights doctrine to data infrastructures and AI systems is nascent and requires further development.<br><br>-International bodies who have historically borne responsibility for human rights doctrine (e.g. Council of Europe, United Nations) are only just beginning to produce guidance on the human rights implications of AI and data infrastructures.<br><br>- The collateral effects of data infrastructures on the natural environment, the allocation of raw materials, and upon labour conditions are only just beginning to be recognised and documented.<br><br>- While trust in data infrastructures and AI systems has been documented in the Global North, additional work is needed to understand this trust relation in less-commonly studied regions, economic levels, and other overlooked categories. |

Prominent critiques of AI research and development focus on the effects of data-driven technologies on human well-being with implications for social issues including discrimination, equality, economic justice, and environmental harm. A range of dignitary rights and interests articulated by human rights doctrine, as interpreted by international bodies, are potentially implicated by data infrastructures and AI, including violations of labour rights as well as of liberty, security, privacy, and freedom of expression and belief.[200] The pillar of power is centrally involved in human rights considerations insofar as these are protected, and sometimes prejudiced, through governmental intervention in human lives. The Equity pillar is also involved, because the distribution of the benefits and harms of AI and data-intensive technologies fall unevenly across social, economic, and political differences between individuals and groups.

Data-driven technologies create demand for natural resources and energy production with profound impacts. As Bender et al. state regarding popular natural language processing models, 'training a single BERT base model…on GPUs was estimated to require as much energy as a trans-American flight'.[201] Many machine learning (ML) models are similarly energy-intensive. Whittaker et al. are among scholars who have demonstrated the link between this kind of energy consumption and climate change and human rights – with long term environmental effects posing serious threats to life, biodiversity, planetary health, etc.[202] Consequently, the energy expenditure of ML models must also be considered as an area of grave human rights concern.

The supply chains of the hardware and software used in development of these technologies also generates risks to human rights, for example, where unjust labour practices contribute to conditions of modern slavery. For instance, raw materials found in mobile phones and other devices are mined by child labour in the Democratic Republic of the Congo and may be funding armed militias there.[203] International bodies have prescribed due diligence guidance in an attempt to stem the potentially adverse human rights impacts of using raw materials of this kind in the production of data-driven technologies.[204]

Public bodies face challenges when they exploit data in efforts to improve living conditions. The 2018 report 'A Human-Rights Based Approach to Data' by the UN Human Rights Office of the High Commission[205] reminds us that the 2030 Agenda for Sustainable Development is to be implemented in a manner that is consistent with the rights and obligations of States under international law. As the UN Report, *A World that Counts*, states: 'Any legal or regulatory mechanisms, or networks or partnerships, set up to mobilise the data revolution for sustainable development should have the protection of human rights as a core part of their activities, specify who is responsible for upholding those rights, and should support the protection, respect and fulfilment of human rights'.[206]

---

Cross-border sharing of data has long been a feature of trade, global and national security, and the management of public health. The Open Data Institute suggests that data sharing to promote collective well-being, such as pandemic management, entails the need for a 'data diplomacy' in which efforts to use data to better coordinate in international crisis management are built from foundations of open and rights-respecting data practices.[207] The World Economic Forum provides steps to hardwire accountability by establishing cooperation mechanisms that hold governments responsible for the security and confidentiality of the data they collect and share.[208]

As patterns of data collection and use evolve – with interactions in the workplace, at home or with public services increasingly being shaped by digital technologies, there is pressure on individuals to 'opt in' and provide detailed data to a range of data controllers if they are to be able to participate in society. The Ada Lovelace Institute states: 'Today's data environment is characterised by structural power imbalances…patterns of data use can create new forms of vulnerability for individuals or groups'.[209] Examples through this report demonstrate how data can be used to target individuals in ways that may lead to discrimination or social division, further exacerbating inequalities and risks to human rights and fundamental freedoms. Recent work by the Council of Europe, exploring how AI and data-intensive innovation can be aligned with human rights, democratic principles, and the rule of law, paints a sobering picture of the full spectrum of potential adverse impacts that datafication and digital transformation can have on human dignity, equal respect before the law and protection from discrimination, access effective judicial remedies, full and equitable participation in community life, and freedoms of thought, association, assembly, and expression.[210]

It has been suggested that international human rights law may provide strategic guidance in the AI context, such as definitional clarity for assessing the violation of key rights and specifying clear obligations for states as well as commercial actors regulated by state power.[211] Human rights doctrine also includes a system of monitoring and oversight that could be brought to bear on AI, such as requirements for supply chain transparency, reporting, and other mechanisms of human rights impact assessment and due diligence.

As a further strategy for increasing trust and reducing the potential for human rights abuses, scholars have advocated for methodologies that promote value sensitive and participatory design.[212] These approaches potentially support the translation of fundamental human rights into context-dependent design requirements through a structured, inclusive, and transparent process.[213] However, as reported by Nesta, 'participatory design alone will not be enough to address all of the critiques of AI in humanitarian settings when developed alongside other complimentary measures it can help to strengthen the ecosystem for responsible AI'.[214]

## Reflection Questions

| Academic Researchers | Policymakers | Developers | Impacted Communities |
|---|---|---|---|
| - To what extent is my research aware of and responsive to the geopolitical and infrastructural power dynamics that have been identified by these literatures?

-How does current research about the geopolitics of data, data infrastructures, and artificial intelligence lifecycles overlook or underplay human rights implications?

- How can current research about the geopolitics of data, data infrastructures, and artificial intelligence lifecycles better illuminate or draw attention to the human rights dimensions of my research?

- To what extent has my research been aware of and responsive to the shifts in international relations which have been influenced by and continue to shape the development of data-driven technologies?

- To what extent has my research on the global political economy focused on a Western/Global North centric understanding of global data flows? Is the exclusion /inclusion of select narratives influencing the research outcome? | - What are the human rights implications of policymaking regarding data infrastructures, international standards mechanisms, and AI regulation?

- How can policymaking better account for the human rights dimension of the geopolitics of data, data infrastructures, and AI regulation?

- To what extent are the policies I draft and pursue reflective of my responsibility to protect and promote human rights—especially as these relate to global power asymmetries at geopolitical, infrastructural, and socio-economic levels?

- How can I introduce policy safeguards to limit against unjust data extraction, monetisation, exchange or appropriation by corporate entities and foreign states? | - To what extent are my data innovation practices aware of and responsive to the geopolitical and infrastructural power dynamics that have been identified by these literatures?

- How do the systems I design and develop affect the human rights of system users and other affected people?

- How can design and development practices better reflect awareness of and promote human rights—especially as these relate to global power asymmetries at geopolitical, infrastructural, and socio-economic levels?

- What are the current gaps in my understanding of how data collection and use could be aligned with human rights? Am I sufficiently aware of wider power dynamics in the economy, the market, and in infrastructural means of technology production and consumption? Have I considered the possible implications and adverse impacts of dual use technologies? | - To what extent are members of my community and I aware of and responsive to the geopolitical and infrastructural power dynamics that have been identified by these literatures?

-What are the current gaps in my and my community's understanding of how political or corporate actors (both domestic and international) might influence the way data is collected and used? How could these actors influence the future trajectory of data-driven technologies in my community, nation, or region?

- How can human rights doctrine support efforts to advocate for the well-being of my community in relation to data-driven technologies?

- Has the use of digital surveillance methods been identified within my community? How can I challenge or address the negative outcomes from such surveillance? Does my community have access to avenues to challenge invasive surveillance tools? |

# Data Colonialism, Data Activism, and De-Colonial AI

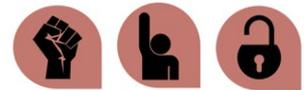

| Key Points | Gaps Identified |
|---|---|
| - Extractive data collection processes are common, and data are often collected and used without consent and in ways that are considered by many scholars and activists as exploitative.<br><br>- Data subjects are often not the beneficiaries of the algorithmic systems that are trained on their data and employed.<br><br>- Colonial practices of 'philanthro-capitalism' are widespread and dominated by Big Tech.<br><br>-De-colonial AI scholars focus on how algorithmic oppression, exploitation, and dispossession cause the unjust subordination of one social group while privileging another and allow institutional actors to take advantage of already marginalised people. | -There is a lack of substantial discussions/progress surrounding data being collected without consent of individuals.<br><br>- Data colonialism must be confronted in other ways besides narrowing digital divides as the existing economic, social, and political structures behind these divides often have power and control implications that exacerbate already existing inequities.<br><br>- Researchers and practitioners must continue to work on new ways to use data collection and solving the missing data problem as a form of resistance against exploitative data practices. |

'Data colonialism' is a term that was coined by Couldry and Mejias to mean a new form of colonial practices distinctive to 21st century contexts of data capture and capitalisation. These practices are extractive in nature and stem from the notion that data is 'free and available for appropriation'.[215] Their definition is, however, not meant to serve as an analogy to colonialist practices that took place in previous periods of imperialism, but rather they seek to 'explore the parallels with historic colonialism's function within the development of economies on a global scale, its normalization of resource appropriation, and its redefinition of social relations so that dispossession came to seem natural'.[216] Couldry and Mejias call attention to at least two poles of colonial power, China and the United States, that are leading the charge on finding ways to profit off global flows of data. Global data flows are the objects profiteering interests, because the rise in global digital platforms has made them easier to gather and more widespread.[217] Continuing along a similar narrative of data colonialism is Thatcher et al.'s characterisation of widespread asymmetries of data capture as the 'accumulation as dispossession that colonises and commodifies everyday life in ways previously impossible'.[218] The authors argue that asymmetries of power in data collection practices exacerbate not only data colonialism but increase the ways in which communities are dispossessed of their own data. These expositions of data colonialism have consequently laid the foundation for thinking about ways to challenge this narrative through advancing data justice research and practice.

---

[215] Couldry & Mejias, 2018

[216] Ibid.

[217] Ibid.

[218] Thatcher et al., 2016

Extractive data colonialist practices include but are not limited to the collection and use of individuals' data without their consent and the failure to consult them about how their data are being used. It also includes the obstruction of individuals from gaining knowledge about where and how their data are being used and who is profiting from this. Coleman approaches this topic from a data protection perspective calling these extractive practices a 'modern day "Scramble for Africa"' in which social media and other scaled digital platforms are built, 'churning a profit, and/or storing data as raw material for predictive analytics'.[219] Populations which have non-consensually had their data collected and used are often not beneficiaries of the data-driven systems that are trained on the resulting datasets, and many times data collected on these individuals can serve to 'exclude, miscount, or distort those individuals or groups'.[220]

Al Dahdah and Quet explicate the concept of technological positivism – the notion that technology is neutral, provides benefits to humanity, and is separated from politics.[221] They also explain how philanthro-capitalism expands commercial agendas into 'developmental' markets, and how movements of data and the control of physical infrastructures are managed by only a few actors from developed countries.[222]

Pieterse explains how these monopolies and concentrations of knowledge and norms are not only exclusionary but create 'paths of dependency of certain 'territories', social groups, and individuals on others'.[223] Marten and Witte point out that many Big Tech firms sit on governance bodies and are active contributors to 'digital development policies in the Global South'.[224] Thus, a cycle is created in which Big Tech plays the role of advancing technological capacities in countries where they simultaneously exercise control over the regulatory, policy, and governance environments thereby creating exploitative forms of dependency and reliance. Al Dahdah and Quet elaborate on this by explaining how metropole-periphery ways of looking at the world have led to colonial data innovation practices which purport to deliver 'benefits to humanity' – all while bolstering Big Tech profits under the guise of promoting the 'social good' and simultaneously gaining more control over data flows, infrastructure, and technologies.[225]

However, as other scholars have observed, data colonialist practices extend well past data collection and into data processing and modelling. Models that involve local data are often created externally, in a different country context, but are meant to be used at a local level. Organisations running the analyses are often far removed from the local context and needs of the individuals whom the model will impact. Andrejevic analyses these power asymmetries as well as gaps in data collection and refers to them as the 'big data divide'.[226] These gaps demonstrate the need for building up technical infrastructure and skillsets in countries and regions that have traditionally been on the wrong side of the digital divide, while simultaneously allowing such countries (and the individuals who live in them) to have control over their data and agency over how they are used.

---

[219] Coleman, 2019

[220] Cinnamon, 2019

[221] Al Dahdah & Quet, 2020; Bijker et al., 2012

[222] Al Dahdah & Quet, 2020

[223] Pieterse, 2010

[224] Marten & Witte, 2008 as cited in Al Dahdah & Quet, 2020

[225] Al Dahdah & Quet, 2020

[226] Andrejevic, 2014

*'The proposition that Big Tech, based in one part of the world and benefiting from a very particular concentration of resources, can judge how social problems should be interpreted and resolved across all the world's societies is, in the light of colonial history, an astonishing usurpation of power that claims the capacity to see all the world's social differences and similarities in terms of one single data-driven logic that justifies corporate intervention anywhere'.*[227]

While the current landscape demonstrates the harmful and extractive processes that take place, there has simultaneously been calls to move away from the discussion on datafication towards an increased focus on data justice and data activism, and through this to move the attention towards communities that are resisting datafication processes.[228] Milan and Treré call for a de-Westernisation of not only the field of critical data studies, but also a deeper understanding of the South as a 'composite and plural entity, beyond the geographical connotation ("Global South")'.[229] Mohammed et al. also argue against 'reducing the reality of lived experiences to oversimplified binaries' such as the 'North and the South' and 'the powerful and the oppressed'.[230] The theme of data colonialism, data activism, and decolonial AI relates to several pillars, but calls specific attention to power, participation, and access as decolonial AI practices consist of combatting existing data processes that are used to marginalise communities. Arora explicates how engaging with the decolonial approach means properly investigating how datafication impacts 'individuals and communities at the bottom of the data pyramid' and simultaneously moves away from treating this population as 'consumers' rather than 'beneficiaries'.[231] *Decolonial AI: Decolonial Theory as Sociotechnical Foresight in Artificial Intelligence* by Mohammed et al. is a critical piece about decolonial AI that provides several recommendations on forming a decolonial field of AI by extending the framing of data colonialism to 'algorithmic coloniality'. The authors call attention to algorithmic oppression, exploitation, and dispossession which relate to the unjust subordination of one social group while privileging another, institutional actors using algorithmic tools to 'take advantage of (often already marginalised) people' and the centralisation of power that thereby leads to the deprivation of power for another group.[232]

There are many examples of decolonial research being conducted in efforts to combat data colonialist practices across the globe. Several of these examples can be found in the figure below with the inclusion of specific cases cited within Milan and Treré's work.[233]

---

| Author | Combatting data colonialist practices |
|---|---|
| Ricaurte (2019) | • Discusses coloniality of power and data colonialism practices that are present through 'structural violence' that occurs and is reinforced by the lack of data on feminicides.[234] <br><br> • Highlights citizen resistance to data colonialism and gender violence in Mexico such as the creation of Disappeared People (Personas Desaparecidas)[235] that demands justice for missing people, as well as a database of feminicides that have occurred in Mexico since 2016.[236] |
| Chenou and Cepeda-Másmela (2019) | • Examines #NiUnaMenos feminist activism movement, as well as exploring the creation of a 'national index of sexist violence in Argentina in 2016'.[237] |
| Gutiérrez (2019) | • Advocates for 'proactive data activism' as a form of resistance in which technologies that are often used to exacerbate social inequality are used instead as a tool for creating positive social change.[238] |

*Figure 15. Select research on efforts to combat data colonialist practices as cited by Milan and Treré, 2019.*

---

[234] Ricaurte, 2019

[235] https://personasdesaparecidas.mx/db/db

[236] https://feminicidiosmx.crowdmap.com/

[237] Chenou & Cepeda-Másmela, 2019

[238] Gutiérrez, 2019

## Reflection Questions

| Academic Researchers | Policymakers | Developers | Impacted Communities |
|---|---|---|---|
| - To what extent is my research aware of and responsive to the structural injustices that have been identified by the data colonialism and decolonial AI literatures?<br><br>- What are the current gaps in my understanding of how my research could exacerbate some of the issues identified by the data colonialism and decolonial AI literatures such as data extraction, technological positivism, and polarising binaries such as 'the North and the South'?<br><br>- How can I challenge data colonialist practices through my scholarship and in my role within academia? | - To what extent do I see the themes of extractive exploitation, technological positivism, and data capture (which have been identified in this literature) present within my policymaking remit?<br><br>- What are the current gaps in my understanding of how the policies I shape can contribute to data colonialist practices (for instance by failing to address the exploitation of data subjects, the use of their data without their consent, the imposition of models onto communities without consultation, among others)?<br><br>- How can I ensure that the policies created within my remit incorporate local context and the needs and perspectives of impacted data subjects? | - To what extent do I see the themes of extractive exploitation, technological positivism, and data capture, (which have been identified in this literature) present in my work as it relates to the data innovation ecosystem?<br><br>- What are the current gaps in my understanding of how processes of data collection and use could contribute towards the harms resulting from data colonialist practices?<br><br>- Having read about the hazards raised by the data colonialism literature, such as objectifying and exploiting through data extraction and use, how can I ensure that my data innovation practices are equitable and responsible with regard to the identified risks of harm? | - To what extent has my community been impacted by data colonialist practices? To what extent am I impacted by exploitative practices of data extraction or the use of my data without my proper consent?<br><br>- What are the current gaps in my understanding of how my data could be used without my consent, how companies are inequitably profiting off my data, and how I could challenge these practices?<br><br>- What can I and my community do to challenge existing unjust data extraction and datafication practices and take back ownership over my/our data? |

# Economic and Distributive Justice

## Innovation, Diffusion, and Redistribution

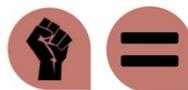

| Key Points | Gaps Identified |
|---|---|
| - Factors such as the increasing returns from scaling up data collection have resulted in a data-driven economy which favours Big Tech over small and medium-sized enterprises (SMEs).<br><br>- Competition law has been framed as one strategy to address this, but its impact is limited.<br><br>- A disproportionate focus on innovation over diffusion or redistribution has limited the impact of AI and data policies intended to benefit SMEs.<br><br>- Current legislation is outdated and unable to respond to the specific features of 'data' as an economic resource. | - Policy literature needs to focus on diffusion as well as innovation in order to facilitate the participation of SMEs in the data-driven economy.<br><br>- More research is needed on the integration of economic policy with other concerns such as identity, empowerment, and participation, with union perspectives requiring more attention in academic literature.<br><br>- Further work is needed to understand the levels at which power operates and the levels at which economic policy could contribute to global justice. The distinct roles played by the private sector, states, and multinational bodies in setting the economic agenda is rapidly evolving and can contribute to a vacuum of responsibility. |

The digital economy includes businesses that 'rely upon information technology, data, and the internet for their business models'.[239] Across the sectors in which these businesses operate, datafication has contributed to significant power imbalances between monopolists and individuals, platform dependent actors, and small and medium-sized enterprises (SMEs). Factors from the increased returns of scaling up to the greatly improved predictive power provided by big datasets, have been identified as contributing to the success of large businesses, often resulting in the exclusion of SMEs.[240] In addition, SMEs face numerous obstacles to market entry. These include a lack of required skills, the cost of transition to digitally enabled technologies, and the time taken for this transition to result in profit or increased productivity.[241] Analyses that address these challenges are therefore closely related to the pillars of power and equity.

In addressing this domination of big business over SMEs, policymakers have frequently focused on innovation and diffusion.[242] It has been argued that they can contribute to economic justice through tackling the dominance of 'big tech' and redistributing economic profit among smaller businesses.

---

'Innovation' is a central priority in 'Advanced Economies' national strategies on AI.[243] Particular attention has been paid to competition law, which is intended to foster competition by preventing one business from dominating a particular market. Khan and Vaheesan have argued that a 'counterrevolution' is needed in competition law for it to effectively contribute to redistribution.[244] Others argue that however much it is reformed, its impact on economic justice will be limited. Competition law only comes into play once significant harm has been done by the dominant player in the market. And because it only operates across a single market, small businesses dependent on platforms cannot appeal to competition law.[245]

Innovation for SMEs is also encouraged through other mechanisms, such as financial incentives.[246] However, due to identified limitations of the innovation-centred approach, diffusion of data-driven technologies through society and small enterprises have been proposed. OECD research suggests it is crucial to address the multiple barriers to SMEs.[247] In addition, economic policies which address the consequences of diffusion have also been identified as a priority. Diffusion policy must therefore be complemented by re-education policies and social safety nets, with suggestions often focusing on the potential of universal basic income.[248] Academic and activist literatures suggest that the adaptation and expansion of existing regulation on innovation and diffusion policy are inadequate methods for advancing economic justice. Alternative approaches have thus proposed a shift toward the economic governance of data itself.[249]

First, innovation and diffusion on a global scale can in fact contribute to economic injustice.[250] Mann argues that the deployment of new technologies for development is typically presented as a 'win-win': new technologies diffuse to developing nations while innovation is encouraged among multinational corporations who deploy their data-driven technologies in new contexts and benefit from such uses of data for development. Mann proposes that governance structures in both developing and advanced economies must be updated to prevent economic injustice. Second, innovation and diffusion policy do not address the long-term economic trajectories which have resulted in economic injustices.[251] Cath et al. argue that to differing extents, national AI strategies from the United States, United Kingdom, and European Union rely on 'liberal notions of the free market'.[252] Srnicek links the current chasm between Big Tech and SMEs to longer term capitalist trajectories.[253] He argues that transformative approaches would be required to alter the status quo. Third, activist literatures also make clear that innovation and diffusion policy must take a locally sensitive and participatory approach, combining economic priorities with sociological and political concerns. Work by the Digital Empowerment Foundation on 'empowering handloom clusters' demonstrates the need to consider the theme of identity. Rathi and Tandon suggest participation in unions can play an important role in achieving positive economic conditions for workers.[254]

---

[243] Cath et al., 2018
[244] Khan & Vaheesan, 2017
[245] Singh & Gurumurthy, 2021
[246] OECD, 2021
[247] Ibid.
[248] Agrawal et al., 2019
[249] Singh & Gurumurthy, 2021; Spiekermann et al., 2021; Arrieta-Ibarra et al., 2018; Singh, 2020
[250] Mann, 2018
[251] Srnicek, 2017
[252] Cath et al., 2018
[253] Srnicek, 2017
[254] Rathi & Tandon, 2021

## Economic Governance of Data

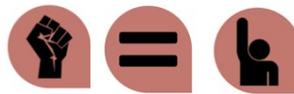

| Key Points | Gaps Identified |
|---|---|
| - The accumulation of data by big businesses leads to economic inequities and imbalances of power between consumers, employees, communities, and nation states versus the multinational corporations collecting their data.<br><br>- Numerous proposals for individualised economic governance of data have been proposed, whether these exert property rights over data, encourage data sharing by large holders of data, or propose data subjects receive economic recompense for their personal data.<br><br>- However, flaws in these individualised approaches to the economic governance of data have been identified and proposals for collective rights-based approaches have been proposed. | - There is an overreliance on existing market regulations over the economic governance of data itself.<br><br>- Some efforts have been made to incorporate union perspectives and the collective power of workers in these debates, but more work is needed to ensure knowledge sharing and effective collective action.<br><br>- Further work is needed to explore the role unilateral and international organisations, as well as national governments, must play in enacting an approach to the economic governance of data which respects collective rights. The leveraging power of workers through unions has been addressed, but more work could look to how international groups such as the International Telecommunication Union or other United Nations bodies can be used to leverage the collective power of governments to advance the interests of multiple developing nations. |

A reliance on adapting existing legislation to the data-driven economy has resulted in a significant gap regarding regulations governing data directly. Singh and Gurumurthy identify two significant reasons for this. First, the 'nebulous' nature of data presents a challenge to policymakers. Second, it suits global corporations and governments of advanced economies to rely on existing, inadequate legislation.[255]

Yet, it is argued that it is the collection of data itself which results in power imbalances within economic relationships. First, the economic discrimination experienced by consumers of digital platforms has been documented by Zuboff, who accounts for the origins of the business model dominant within 'surveillance capitalism', and O'Neil who outlines how those already experiencing economic hardship can be directed towards services which drain their economic resources further.[256] Second, employees are impacted by the volume of data collected and a lack of transparency on how data is used to assess performance.[257] Adler-Bell and Miller as well as Singh and Gurumurthy focus on 'platform-dependent' actors working in the gig economy.[258] The authors state that current legislation is frequently insufficient. Protections set out under GDPR cannot be applied to small businesses as with individuals. And because platforms create the market in which platform dependent actors operate, they cannot appeal to competition law.[259] Finally, according to scholars, the accrual of data for economic profit by multinational corporations can lead to a lack of access for domestic corporations and results in disadvantages for national economies, communities, and individual data subjects.[260] Even initiatives of 'data for development' have been problematised, on the basis that they frequently involve the extraction of data from communities which are then used as a source of economic profit for large multinational firms.[261]

---

[255] Singh & Gurumurthy, 2021

[256] Zuboff, 2019; O'Neil, 2016

[257] Adler-Bell & Miller, 2018

[258] Adler-Bell & Miller, 2018; Singh & Gurumurthy, 2021

While there is a significant policy gap regarding the direct regulation of data, possible resolutions have been articulated. These proposals often advocate for the redistribution of the economic benefits of data, hence contributing to both equity and redistributions of power as discussed in the data justice pillars. In addition, ideas for the direct economic governance of data often advocate for participatory approaches, involving unions and data subjects, consequently responding to concerns raised by the participation pillar.

First, proposals which advocate for individualised ownership explore how data subjects might take a share in economic profit.[262] Arrieta-Ibarra et al. argue in favour of treating data as labour rather than capital to acknowledge the contribution made by the original producer of the data.[263] However, Spiekermann et al. question the feasibility of the approach given the number of small transactions that would be required.[264] They also claim that it places profit in the hands of 'data producers', which may be a more diverse group than current data collectors. This could cause significant harm to those wishing to retain their right not to be represented in the data economy.

Singh and Gurumurthy argue instead that these proposals are focused on individual rather than community rights. They state that the economic value of data arises in its aggregate form and consequently a collective or commons-based approach is required.[265] Spiekermann et al. propose a progressive data tax to regulate the 'Global Information Commons' that would provide subsidies for non-profit uses of data and targeting those who profit most.[266] Finally, it has been argued that community-rights approaches should involve meaningful participation of impacted communities.[267] Unions are important sites where this can take place. Public Services International argue that trade unions could 'create 'worker data collectives' by pooling their members' data into data trust structures' which can be used as economic leverage.[268] Others propose a plurality of data trust structures, varying in the level of participation offered, each accounting for the needs and aspirations of different data subjects.[269]

---

## Reflection Questions

| Academic Researchers | Policymakers | Developers | Impacted Communities |
|---|---|---|---|
| - To what extent has my research been aware of and responsive to distributional injustices which result from datafication? Does my research understand these harms of allocation as entangled with harms of representation which may arise due to data-driven technologies?<br><br>- What are the current gaps in my understanding of how my research could contribute not only to economic theory but to practice through practicable proposals for policymakers?<br><br>- Does my research pay sufficient attention to community rights and the 'commons' understanding of data innovation ecosystems? | - How can I ensure that, where policies created within my remit cover the innovation, diffusion, or economic governance of data, these policies tackle the concentration of economic profit by Big Tech companies and related imbalances of power ?<br><br>- What are the current gaps in my understanding of how data-driven economies could contribute to economic inequities and imbalances of power between economic actors?<br><br>- How can I ensure that the policies created within my remit incorporate a locally sensitive and meaningful participatory approach to the economic governance of data? How could this approach address both individual and collective rights? | - To what extent does my work as a developer distribute economic benefits to users or to data subjects?<br><br>- What are the current gaps in my understanding of how technology platforms can result in unfair working conditions for those in the gig economy who become reliant on these technological platforms?<br><br>- Having read proposals for the economic governance of data, are there ways in which these proposals could be integrated into my work developing new systems and technologies? | - To what extent has my community been impacted by current data-driven business models? To what extent do existing market regulations address these impacts?<br><br>- What are the current gaps in my understanding of how data created by and about me and my community are governed?<br><br>- How can I and members of my community draw on the community rights and 'commons' understanding of data innovation ecosystems to advance more equitable data futures? |

# Identity, Democratic Agency, and Data Injustice

## Data Feminism

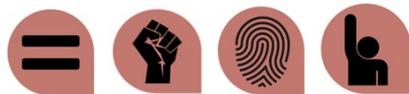

| Key Points | Gaps Identified |
| --- | --- |
| - Data feminism provides a very strong foundation for thinking about ways that historical and systemic biases and patterns of discrimination are drawn into practices of data collection and use.<br><br>- Data feminism literature refuses exclusionary and dehumanising data practices. It calls for a new future in which Latinx, Black, queer, trans, and Ingenious feminists are both celebrated and listened to.<br><br>- Data feminism calls for equitable representation in datasets and data practices especially as this relates to intersectional characteristics that can amplify discriminatory effects. | - The missing data problem is identified by data feminists as exacerbating inequality and signalling that some social groups are less important. This insight needs to be generalised across the study of data justice, so that it can become a more widely used critical tool.<br><br>- Training datasets are not representative which lead to biased models and adverse impacts for certain communities who are not represented.<br><br>- Feminist approaches to data science are often not considered in wider conversations about algorithmic bias or fairness and are excluded from mainstream dialogues about data science research. |

Data feminist literature and activism relate closely to issues of quantified identities and discriminatory systems of categorisations, especially when considering the intersectional approach provided by this field. A range of activist groups and academics related to data feminism have paved the way for envisaging new and more inclusive ways of advancing data justice research and practice including but not limited to the Feminist Internet, Mimi Onuoha's 'Our Missing Data Sets' project, the authors of *Data Feminism*, the Algorithmic Justice League, and the Feminist Data Manifest-No project, each of which will be introduced below.

Feminist Internet is a collective aiming to advance internet equalities, defined as 'equal rights to freedom of expression, privacy, data protection, and internet access regardless of race, class, gender, gender identity, age, belief, or ability', for women and other marginalised groups. They also facilitate educational engagements and develop frameworks such as a Feminist Design Tool[270] and a Trans-Competent Design Tool[271] that support developers in addressing feminist issues in technology.[272]

Mimi Onuoha's project 'On Missing Data Sets', calls attention to missing datasets that result from the entrenched social biases and indifferences of those who control factors of data collection. She gives four reasons for why a dataset seems like it 'should exist…but might not'.[273] This includes a lack of incentive to collect data by those who have the resources, the fact that the data that would be collected 'resist simple quantification', the perception that the work involved in collecting the data is not worth the benefit it will give, and the 'advantages to nonexistence' that some derive from not airing information that may adversely affect their interests.[274] The missing data problem is also explored in the book, *Data Feminism.* The authors, D'Ignazio and Klein, offer the example of the lack of data on maternal health outcomes, explaining how there is no financial gain in collecting more data on women who are dying, but there is when it comes to women being pregnant.[275] They conclude: 'Things we do not or cannot collect data about are very often perceived to be things that do not exist at all'.[276] The literature on data feminism draws attention to other similar forms of data injustice that occur, including search engines' discriminatory results that are rooted in a ranking prioritisation in which women are ordered and positioned in 'ways that underscore [their] historical, contemporary lack of status in society.'[277]

---

[270] Feminist Internet & Young, 2019

[271] Feminist Internet & Rincon, 2020

[272] Feminist Internet, n.d.

[273] Onuoha, 2018

The Algorithmic Justice League (AJL) is an organisation whose mission is to 'raise awareness about the impacts of AI, equip advocates with empirical research to bolster campaigns, build the voice and choice of most impacted communities, and galvanise researchers, policymakers, and industry practitioners to mitigate AI bias and harms'.[278] As discussed by Buolamwini and Gebru, founder of and collaborator with the AJL respectively, in their paper *Gender Shades*, the widespread existence of bias in facial analysis technologies such as a facial recognition system illustrates the kind of systemic discrimination and harm that AJL combats. They show how certain facial recognition classifiers perform the worst on darker-skinned female faces due to the underrepresentation of darker-skinned females and darker-skinned individuals in general in the training sets. The classifier performed best on lighter-skinned males.[279] AJL's 'Facial Recognition Technologies in the Wild: A Call for A Federal Office' proposes a new federal office as a model for state regulation of facial recognition technologies that categorises FRTs by degrees of risk with corresponding guidelines and redlines for control.[280]

The Feminist Data Manifest-No provides a clear depiction of both gaps and harmful practices in the existing data landscape. The Manifest-No is defined as 'a declaration of refusal and commitment…it refuses harmful data regimes and commits to new data futures'.[281] The Manifest-No sets out thirty-two declarations of refusal such as the refusal 'to be disciplined by data, devices, and practices that seek to shape and normalize racialized, gendered, and differently-abled bodies in ways that make us available to be tracked, monitored, and surveilled'.[282] It also calls for a new future in which Latinx, Black, queer, trans, and Ingenious feminists are both celebrated and listened to. Within the document, there are also commitments to mobilise data by working 'with minoritised people in ways that are consensual, reciprocal, and that understand data as always co-constituted'.[283] These practices are closely intertwined with the pillars of power, identity, equality, and participation, as they are engaging in critical refusal as participation, combatting discriminatory and racialised politics of data collection and use, questioning binaries, and critiquing existing forms of power.

---

The notion of feminism(s) in this context is especially important because it allows for various worldviews to coexist. As Cifor et al. (2019) state, 'Feminism is plural; there are many feminisms and they may differ in their positive visions, methodologies, collective ends, and situated concerns. Yet, what allows them to "hang together" as different but still feminist is the negative construction – a refusal of an inheritance'.[284] This refusal takes place in a variety of ways, from refusing to understand data as dehumanised,[285] to acknowledging inequalities that exist in both the data itself and data practices, so that they can be critiqued and transformed.[286]

## Design Justice 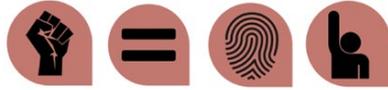

| Key Points | Gaps Identified |
|---|---|
| - Design justice works from the understanding that technology is not neutral; decisions about design and implementation flow from existing social structures and arrangements of economic, political, and social power.<br><br>- Design justice is a movement for opening up the terms of inclusion in technology design in ways that shift power to people and communities who are structurally marginalised and oppressed.<br><br>- Participatory design and co-design methods, repurposing and retargeting, and increasing the diversity of designers and artefacts are some of the strategies of this movement. | - Design justice is a relatively new area of research, and the literature is only just emerging. While it has roots in participatory design, action research, feminist Human Computer Interaction, and speculative design, the goals of design justice are expressly focused on broad goals of social liberation. More work and literature are needed in this space to develop and refine this body of knowledge and insight.<br><br>- Design justice emphasises participation as a path towards justice, but concerns have been raised that participation can itself create injustice where it enables extractive labour practices by researchers and tech companies, including those more interested in 'participation washing' problematic technologies and practices.[287] More work is needed to clarify the constraints for design justice to be a reliable enabler of data justice. |

---

Design justice is a relatively recent research topic but one with deep historical roots in various design and social science disciplines, including the Participatory Design movement of the 1970s, Science and Technology Studies (STS) since at least the 1980s, and more recent work in Human Computer Interaction (HCI), Data Ethics, Global Data Justice, and Feminist Technology Theory. Design justice scholars generally endorse the claim that technology is not neutral, or as the STS theorist Langdon Winner states, 'artifacts have politics'.[288] It is argued that choices made throughout the lifecycle of technological design, production, and implementation are made from within the constrained worldviews of designers, from the initial concept through to the moment of encounter with a system or artefact. These constraints have consequences for anyone or anything that is not considered within that frame. One example is a military aircraft cockpit design that did not account for smaller statures with the result that generations of female pilots were less likely to be able to fly them.[289] Another example is the cisgender-normative design of airport body scanners. Trans people whose anatomy do not match their assumed binary gender are likely to be flagged as 'anomalous', resulting in humiliating questions and additional screening.[290] When a piece of technology is being designed, deciding whose lives should be considered or ignored is not simply a functional choice; it has political implications and is a site of procedural and social justice.

Design justice describes efforts to broaden participatory inclusion in decision-making to those affected by data and technological design. Key literature invokes the pillars of power, equity, identity, and participation, emphasising that the inclusion of diverse identities in decision-making shifts power from the structures of social domination in which technological design frequently occurs to more marginalised groups. Design justice is a move to enhance social equity in decisions about the construction and use of data and systems.

Design justice is also a framework for analysis that recognises how larger systems – including norms, values, and assumptions – are encoded in and reproduced through the design of sociotechnical systems and how design distributes benefits and burdens and accounts for the reproduction and/or challenges the matrix of domination (white supremacy, heteropatriarchy, capitalism, ableism, settler colonialism, and other forms of structural inequality).[291] Machine learning (ML) technologies are sociotechnical systems that are special relevance in the data justice context. From a design justice perspective, ML technologies become sites of 'critical pedagogy where people and communities are involved in both setting the questions and determining the meaning of what is found'.[292] HCI scholars have provided actionable reconfigurations to design processes which engage and centre diverse perspectives and democratise decision-making through participation. In this connection, design justice uses participatory design methods, such as 'participatory action research', to centre the knowledge and experience of communities historically harmed by technological innovations in the creation of new interventions, and specifically those intended to shift socio-technical power downward. A goal is to elevate the 'situated knowledge' of communities to a status of influence in technology design and decision-making from which they are typically excluded.[293]

---

Real-world examples of design justice in practice include the Design Justice Network and Our Data Bodies. Machine learning technology is a design justice site of 'a critical pedagogy where people and communities are involved in both setting the questions and determining the meaning of what is found'.[294] HCI scholars have provided actionable reconfigurations to design processes which engage and centre diverse perspectives and democratise decision-making through participation. The Design Justice Network has developed shared principles aimed to ensure that those which may be overlooked within and marginalised by innovation processes are co-creators in collaborative and creative processes to sustain, heal, and empower communities, and to seek liberation from exploitative and oppressive systems.[295] The Our Data Bodies project combines community organising and academic research to develop participatory tools addressing surveillance and data-based discrimination in the United States, such as an examination of the impact of data collection and data-driven systems on the ability of marginalised people to meet their human needs and address surveillance and data-based discrimination.[296]

## Structural Racism, Intersectionality, and Data Injustice 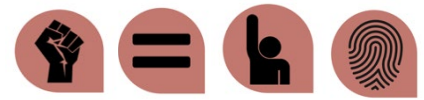

| Key Points | Gaps Identified |
| --- | --- |
| - Literature pertaining to structural racism, intersectionality, and data injustice provides a historical context for current forms of racialised data injustices situated within legacies of institutional racism dating back to European colonialism.<br><br>- The literature problematises what is termed 'data capitalism'. This is described, in part, as commercial activity between the public sector and white-dominated big business which generates profit via the disempowerment of racialised communities.<br><br>- Race is described as a political system of categorisation with colonial origins which bounds physical properties with cultural notions to create harmful distinctions between humans.<br><br>- According to scholars who explore this theme, race has historically justified the subjugation of people of colour to advance Euro-American interests. These forms of oppression have been maintained, on this view, through social theories and statistical methods throughout history.<br><br>- Forms of coded inequality are described as starting points informing demands to redress racialised data injustice. | - Data practices that may support structural racism are often erroneously perceived as acceptable or even benevolent given surrounding narratives that they are objective, impartial, and facilitate development.<br><br>- Critical perspectives on the way structural racism operates in data innovation ecosystems have not yet been integrated into mainstream technical literatures and communities of practice. This has led to the endurance of discriminatory attitudes and complacency in some corners of data science research and data work.<br><br>- Practices of data annotation and labeling have not yet widely integrated bias-aware protocols that stem racialised and intersectional harms. This includes critical reflection on how racial or ethnic attributes are identified, grouped, or disaggregated and how intersectional qualities may be a factor in the differential performance of machine learning systems. |

---

[294] McQuillan, 2018

[295] https://designjustice.org/

[296] Petty et al., 2018

Literature and activism pertaining to structural racism, intersectionality, and data injustice historically contextualise data practices and highlight how adapted mechanisms of racialisation and racial discrimination are used to maintain institutionalised white supremacy. Analyses of racialised harms of representation discussed by scholars and activists in this area are closely related to the pillar of identity as these highlight the reification of identity via the outputs of racially coded datasets. Their accounts of racialised distributive injustices within data practices maintain that the implementation of data systems today contributes to and amplifies hierarchically distributed rights and inequitable economic and labour conditions, both of which were first established under colonial rule. This examination of racial oppression and legacies of coloniality closely connects this theme to the pillar of power. Considering both the historic positioning of data practices against social movements advocating for racial liberation and the leveraging of the critical lens of representational harms to inform the redressing of structural racism, literature pertaining to structural racism, intersectionality, and data injustice is closely intertwined with the pillars of participation and equity.

Alexander argues that periods advancing the rights of Black people in the United States have been responded to through calculated mechanisms by white conservative elites who re-institutionalise racism within contemporary constraints, achieving a continual racial caste system.[297] She analyses the disproportionate number of Black people in the United States that are incarcerated and categorised as felons, which enables the legal restrictions of their rights and freedoms and creates similar conditions to those within the age of legally enforced racial segregation. Benjamin builds on Alexander's analysis by focusing on the use of racial categories in data.[298] She defines The New Jim Code as 'the employment of new technologies that reflect and reproduce existing inequities but that are promoted and perceived as more objective or progressive than the discriminatory systems of a previous era'.[299] Benjamin explains how racial hierarchies are perpetuated through coded representations in data systems and naturalised through cultural rhetoric presenting these as benevolent, desirable, and rising above human bias.[300]

Explanations for racialised harms of representation within technologies have been traced back to the construction of race itself, defined by Atanasoski and Vora as a political system of categorisation. They discuss how scientific notions about racial difference were instituted in 18th and 19th century European colonialism, creating a 'global sliding scale of humanity'.[301] This notion is supported by Braun, who defines 'racializing surveillance' as a technology of social control.[302] Congruently, Benjamin conceptualises race *as* a technology[303] and examines how race has been transposed on the plane of current data practices, drawing on Muhammad's work discussing 19th century 'racial data revolution'. Race-based notions of criminality, disease, and intelligence would emerge from practices that transformed pseudo-biological constructions of race into constructions within the social sciences, as a measure of Black inferiority, in turn, justifying white superiority.[304] Echoing these notions, Atanasoski and Vora argue that the construction of race is built into the

---

imaginaries of those that produce technologies, and in turn, technological developments carry forward these systems of categorisation.[305]

Outputs of data systems which reaffirm the marginalisation of racialised groups, such as predictive tools used for credit scoring, job applications, health diagnostic systems, facial recognition, and policing, have been documented (and sometimes advocated against).[306] Richardson, Shultz, and Crawford illustrate patterns of illegal and biased police practices (i.e. racially motivated stop-and searches) that distort the data used to build these models and produce discriminatory outcomes which are then used to justify the increased policing and surveillance of historically overpoliced communities.[307]

Racial discrimination within data systems is described as profitable for a variety of social actors who either ignore or leverage coded inequality to achieve their interests.[308] Indeed, activist organisations discuss systemic racism in data practices as a foundational component of an economic model pinned as 'data capitalism'—the extraction and commodification of data that exacerbates racial, class, gender, and disability-based inequality. On this view, data capitalism leverages data to concentrate and consolidate power within white-dominated big businesses.[309] Distributive injustices deriving from these data practices are discussed as 'a tool, not a bug'.[310] Evidence of the lucrative business relations between technology companies and public sector organisations that operate within this context of racialised power is provided by the Latinx activist organisations Mijente, the National Immigration project, and the Immigrant Defence Project.[311]

Activists and scholars working within structural racism, intersectionality, and data injustice discuss and respond to racialised representational and distributive injustices operationalised through data. Browne illustrates the possibility of mobilising the critique of surveillance through counter surveillance practices in order to appropriate, co-opt, and challenge surveillance technologies.[312] This claim is echoed by Benjamin, who highlights the need to envision new systems that cultivate safe and thriving communities by considering the need for education, employment, mental health, and broader support systems.[313] Struggles for liberation are often embedded within claims for recognition and redistribution that can inform more equitable cultural and legal structures and data practices. Gebru, in this respect, argues that a holistic approach must be taken, which includes the regulation of data-driven systems, but also increasing the diversity of who creates these tools and generating greater understandings of the historical factors that disadvantage individuals who are subject to them.[314] Benjamin stresses the need to rewrite cultural codes and prioritise 'equity over efficiency, and social good over market imperatives'.[315] Data for Black Lives outlines policy shifts pertaining to data and algorithmic transparency, regulation, data governance, and economic policy to help stakeholders address the challenge of deconstructing data capitalism to redress colonial legacies of structural racism and benefit Black futures. [316] They have also consolidated information on the disparate effects of COVID-19 on Black communities. [317]

## Reflection Questions

| Academic Researchers | Policymakers | Developers | Impacted Communities |
|---|---|---|---|
| - To what extent has my research been aware of and responsive to identity-based injustices that have been identified by these literatures pertaining to identity, democratic agency, and data injustice?<br><br>- What are the current gaps in my understanding of how my research could exacerbate some of the issues identified by this literature, such as data misrepresentation and exclusions, lack of stakeholder involvement in design processes, and data capitalism among others?<br><br>- In what ways may I use principles and practices discussed in this literature, such as participatory design methods, the refusal of harmful data regimes, and responsiveness to identity claims, to promote democratic agency and redress identity-based harms through my role within academia? | - To what extent do I see the identity-based injustices that have been identified by these literatures pertaining to identity, democratic agency, and data injustice within my policymaking remit?<br><br>- What are the current gaps in my understanding of how policies can contribute to addressing the issues identified by the literature, such as data misrepresentation and exclusion, lack of stakeholder involvement in design processes and data capitalism among others?<br><br>- How can I ensure that the policies created within my remit incorporate principles and practices discussed in these literatures, such as participatory design methods, the refusal of harmful data regimes, and responsiveness to identity claims? | - To what extent do forms of identity-based injustices that have been identified by literature pertaining to identity, democratic agency, and data injustice appear in my data innovation practices?<br><br>- What are the current gaps in my understanding of how data collection and use could contribute to the misrepresentation and exclusion of identity groups, a lack of stakeholder involvement, and data capitalism?<br><br>- Having read the hazards raised by the literature pertaining to identity, democratic agency, and data injustice, how can I incorporate the principles and practices discussed in these literatures, such as participatory design methods, the refusal of harmful data regimes, and responsiveness to identity claims? | - To what extent has my community been impacted by identity-based harms? To what extent am I impacted by misrepresentation or exclusion, lack of involvement in design processes, and data capitalism?<br><br>- What are the current gaps in my understanding of how I may be impacted by identity-based injustices that have been identified by literature pertaining to identity, democratic agency, and data injustice, and how could I challenge these practices?<br><br>- What can I do to combat identity-based harms through principles and practices discussed in these literatures, such as advocating for participatory design, refusing harmful data regimes, and mobilising in responds to identity claims? |

# Adjacent Justice Literatures and Social Mobilisation

## Environmental and Climate Justice

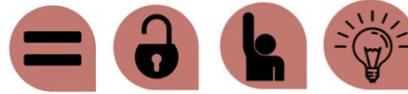

| Key Points | Gaps Identified |
|---|---|
| - Environmental and climate justice movements have been driven by coalitions of activist groups from varied backgrounds (including civil rights activists, Indigenous peoples, environmentalists, global justice campaigners, trade unionists, and scientists). <br><br> - These movements stress the importance of pursuing distributive, procedural and recognitional justice simultaneously. <br><br> - As the climate justice movement has received increasingly mainstream attention, tensions have emerged between more transformative and more pragmatic approaches. <br><br> - Other visions, which apply a restorative justice perspective on climate equity have started to emerge. | - There is currently a gap in recognising the consistencies and overlaps between environmental/climate justice and data justice. <br><br> - There is emerging interest in Environmental Data Justice; however, more work is needed to develop this new field. <br><br> - There need to be more concerted efforts to bring together the different climate justice viewpoints. |

The environmental justice and climate justice movements provide examples of activist-led movements which have challenged dominant power structures to achieve impacts on policy and practice. These movements have similarities with data justice in calling for interrogation and challenge of dominant societal structures, which has led to interest in the emerging field of Environmental Data Justice.[318]

The environmental and climate justice movements challenge existing power structures, noting that inequitable distribution of power (both within and between countries) has led to individuals and communities with the least power in decision-making processes being most severely affected by negative environmental impacts. The 'slow violence'[319] of environmental harm is committed against people who typically have the least power to control the conditions in which they live and work, whereas the perpetrators are powerful actors including private companies and governments typically from the Global North. The environmental and climate justice movements have sought to redress power imbalances in amplifying the voices and experiences of impacted communities (particularly Indigenous communities and people living in poverty).

---

[318] Longdon, 2020
[319] Nixon, 2011

Environmental justice (EJ) has its roots in the United States, examining the unfair distribution of environmental 'bads' (e.g. waste facilities and incinerators) in areas largely populated by low-income groups and Black and ethnic minority communities.[320] In the US, environmental justice grew out of the 'environmental racism' movement and the civil rights struggle of the mid to late 20th century.[321] The environmental justice frame has since been adopted by a range of civil society actors across the globe.[322] International successes include community groups successfully challenging plans for new developments (e.g. incinerators, coal-fired power stations or fracking);[323] raising awareness of environmental impacts within local communities,[324] and; advancing community-led environmental projects (e.g. community forestry and land management).[325]

Activist climate justice movements comprise a range of backgrounds and views, including Indigenous peoples, environmentalists, global justice campaigners, trade unionists, and scientists.[326] Climate justice addresses injustices and inequities in relation to responsibilities for causing climate change and simultaneously the likelihood of being impacted by it. Climate justice advocates recognise that 'the bitter effects of climate change will hit first and most powerfully the countries and people who did least to cause it'.[327] Additionally, climate justice calls for equity in relation to the costs and benefits of mitigation and adaptation strategies.[328]

Literatures relating to climate justice variously demonstrate conceptual, pragmatic, transformative, or restorative conceptualisations of climate justice leading to varying approaches and goals.[329] Conceptual approaches to climate justice (dominant in academic literature) tend to focus on ideal notions of justice, adapting theories of social justice to fit the features of climate change. Pragmatic approaches (which have received the greatest policy attention) seek to pursue climate justice through existing societal structures (e.g. using market mechanisms to tackle emissions). Transformative approaches to climate justice contend that climate injustices are a product of current capitalist systems, and that climate justice requires large, structural social and economic changes which go beyond pragmatic responses. Restorative approaches focus on righting climate harms done in the past through reparative dialogue, reconciliation, and restitution.

Recognising the complementarity of data justice and environmental and climate justice movements in calling for participatory and recognitional justice along with distributive and restorative justice, there is an emerging body of work pulling these areas together and proposing a new field of Environmental Data Justice.[330] Researchers in the emerging field of Environmental Data Justice have called attention to the importance of including diverse perspectives and interests within data collection or data science approaches in order to ensure data are used and interpreted in ways which reflect the interests and experiences of impacted communities. This requires combining community engagement with data collection to mobilise communities and incorporate local, situated, and contextual knowledge into environmental data science.[331]

---

[320] Bullard, 1993

[321] Benford, 2005

[322] Schlosberg, 2013; Walker, 2009

[323] e.g. https://communityactionworks.org/communities-in-action_page/

[324] e.g. https://ceh.org/about/protecting_communities/community-environmental-action-justice-fund/justice-fund-success-stories/

[325] e.g. http://www.onthecommons.org/success-stories-environmental-justice#sthash.Re21ZreL.JvgtbSj0.dpbs

[326] Bullard & Muller, 2012

[327] Sachs & Santarius, 2007, p. 53

[328] Aitken et al., 2016

[329] Ibid.

[330] Longdon, 2020

**Examples of Environmental Data Justice projects include:**

- The Environmental Data and Governance Initiative (EDGI) which has been developing approaches to establish community stewardship of data to 'make data more accessible and environmental decision-making more accountable through new social and technical infrastructure'.[332] They contend that: 'to challenge (environmental) injustice today, the question of data must be addressed, both to recognise how data enable that injustice and how data could be used by communities to name and contest it'.[333]

- Open Water Data which explores new ways to engage communities with environmental data to increase collective understanding and engagement. Open Water Data have projects to make open-source governmental data visible, accessible, and useful to community members, advocacy groups, and local governments and 'turns a critical eye on how open datasets about the environment are shared with the public and asks: Who do these datasets serve and who could they serve?'.[334]

- Public Lab works internationally to enable people to investigate their environment, finding and sharing knowledge across the wider community. Public Lab works to 'raise awareness about health impacts, improve scientific agency, build new scientific and technological skills, and mitigate certain exposures'. Public Lab develops and uses community-created and open-source tools to enable people to 'collaborate on and build upon community knowledge, and to share data about community environmental health'.[335]

---

[331] Ibid.

[332] Walker et al., 2018

[333] Ibid.

[334] e.g. http://datalanterns.com/

[335] e.g. https://publiclab.org/about

# Global Public Health Justice 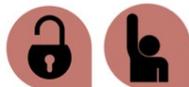

| Key Points | Gaps Identified |
| --- | --- |
| - Efforts to advance global public health justice do not define health as simply the absence of disease, nor as a purely biomedical phenomenon. Instead, literature on global public health justice considers the interdependence of health, economic development, stable government, education, climate, and more.<br><br>- To advance global health justice, it is necessary to consider how underlying assumptions and long-term trajectories have contributed to inequity. For example, the role played by neoliberalism in exacerbating health inequalities has been explored.<br><br>- Approaches to global public health justice have evolved as power has shifted from nation states to a multiplicity of actors in the private sector.<br><br>- Global public health justice cannot be understood by only looking at a global scale, as context and local understandings must be considered.<br><br>- The centring of participation, social mobilisation, and community empowerment in civic practices of public health is a crucial learning that aligns well with the data justice pillars. | - Further work is needed to map the ways in which states, international organisations, and private sector players operate within global public health. There is also a lack of consensus on the division of responsibility between different actors claiming to work towards global health justice.<br><br>- More work is needed to translate between the different kinds of evidence available on global health inequity, including economic, biomedical, and ethnographic evidence.[336]<br><br>- Additional consolidation is required to combine and compare the findings of different stakeholders, including those working in academia, policy, and activism.<br><br>- There is a gap in research which combines social approaches with technological measures rather than presenting these as contrasting approaches. |

Literatures exploring global public health justice focus on the origins of systemic differences in health outcomes, both on a global scale and within developed countries. Authors propose a variety of approaches for advancing equity from institutional proposals to transformative calls to action. It has been argued that inequality is rising, often because of the very structures set up to improve health.[337] Additional injustices in global public health have been both exposed and exacerbated because of the COVID-19 pandemic.[338]

---

[336] Biehl & Petryna, 2013

[337] MacDonald & Tamnhe, 2009

[338] Hotez et al., 2021

There is an increasing recognition in this literature that health is more than the absence of disease and social determinants of health must be addressed through a multidisciplinary approach. However, policy reports by the World Health Organisation (WHO) acknowledge a failure to address 'social determinants' and 'underlying systemic causes of inequality'.[339] Work in anthropology makes clear the need for social sciences to come together with biomedical approaches to address health as a social as well as biological phenomenon.[340] In the field of policymaking, more structures for collaboration between different agencies and departments is needed, while in academia more theoretical work is needed to understand how social and biomedical evidence can be combined. In each case, authors emphasise that improving healthcare access requires looking to material conditions beyond formal healthcare services.

More transformative approaches to global public health justice have also been proposed, drawing on longer term causes of inequality. MacDonald argues, drawing on the history of neoliberalism, that the economic and political system of free-market capitalism is responsible for significant health injustice, and that it is possible for it to be transformed.[341] A shift in the balance of power has also been detected as agency and resources move away from nation states toward private institutions and public-private partnerships and this is argued to present specific challenges.[342] COVID-19 has brought some of these complex relationships out into the open. There are differing approaches to how local and global approaches to global health justice should be operationalised. MacDonald, for instance, argues that there is a need to preserve the United Nations but empower people.[343]

Each of these approaches draw on the importance of participation to advancing justice, but there is variation in regard to who can and should participate. Some public health policymakers and scholars have stressed the importance of bottom-up social mobilisation as an essential means of empowering local communities to exercise agency in pursuing context-sensitive and culturally responsive health promotion.[344] Others have framed community participation in public health processes as an important element of the 'civic practice' of public health. That is, they have characterised this sort of community involvement as a collective activity that is oriented to the public good and that thus builds solidarity by creating a community of common interests and reciprocal duties and obligations among community members.[345] This centring of participation and community empowerment is a crucial learning that aligns well with the data justice pillars.

---

[339] World Health Organisation, 2021a

[340] Biehl & Petryna, 2013

[341] MacDonald & Tamnhe, 2009

[342] Biehl & Petryna, 2013

[343] MacDonald & Tamnhe, 2009; Biehl & Petryna, 2013; World Health Organisation, 2021a

[344] Airhihenbuwa & Dutta, 2012

[345] Jennings & Arras, 2016

Literature which draws together themes of global public health with data justice is also expanding as authors explore both the potential for data-driven methodologies to improve global health outcomes and to exacerbate existing inequities in health. Proposals frequently focus on data sharing and its potential to advance health equity with available data frames as a 'public health resource'.[346] Respectively, national governments and the WHO have been called upon to commit to data sharing and play a coordinating role.[347] Others have emphasised the potential for data sharing to improve health research in regions where there are severe limitations on resources.[348] The WHO's 'Global Strategy on Digital Health 2020-2025' calls for alignment with the UN's 2030 Sustainable Development goals.[349] Wide discrepancies in pandemic preparedness have also contributed to renewed calls for justice informed approaches to the deployment of digital technologies in the healthcare sector. In regard to the potential risks of such technologies, authors draw particular attention to the gaps in 'access, skills and motivation'.[350]

## Culture-Centred Communication for Social Change

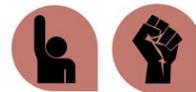

| Key Points | Gaps Identified |
|---|---|
| - Communication is key to the advancement of social change in several ways, from policy advocacy to community mobilisation and from raising awareness to deconstructing dominant systems of power.<br><br>- Approaches to communication for social change have typically adopted an individualistic approach, often focused on modernisation or on information dissemination.<br><br>- This approach has been challenged in favour of a 'culture-centred' approach that adopts radically 'bottom-up' strategies which challenge neoliberal assumptions surrounding international development.[351] | - Scholars have identified fundamental divisions between individualistic approaches to information dissemination and participatory approaches to community mobilisation. Some work has begun to explore the possible conversion of these theoretical understandings of communication[352] but more work is needed to understand the relationship between these parallel streams of work on communication for social change.<br><br>- Culture-centred communication envisions change emerging from partnerships between activists, academics, and civil society. Further research is required to understand how policymakers may fit within this framework. |

The importance of communication to development is rarely disputed. It is widely accepted that communication for social change must combine academic perspectives with on-the-ground insights and expertise.[353] Yet, the theoretical underpinnings of communication for social change, including the nature of the interdisciplinary, institutional, and cross-border collaborations required, remain disputed. Some favour an approach led by individual experts and multilateral institutions, while others propose community-led strategies which question

---

the basis of expertise to acknowledge multiple systems of knowledge.[354] Literature on culture-centred communication for social change relate to the pillars of power and participation, as they challenge neoliberal assumptions of communication and development that are insufficiently attentive to power structures and market-based inequities, and propose instead transformational inclusiveness.

Diffusion-based approaches have been influential and tend to focus on top-down information dissemination as the primary communicative means for affecting social change. For example, Rogers's 'theory of diffusion of innovations' takes an individualistic approach to influencing behaviours through 'information dissemination'.[355] Research has frequently focused on how the impact of communicative strategies can be maximised.[356]

This paradigm has been critiqued for making numerous assumptions about the nature of communication and development. First, critics argue that this definition does not account for how exchange and participation can contribute to the co-construction of knowledge. Second, they claim that it fails to consider non-Western practices where groups rather than individuals are the core agent of social change. Third, critics propose that 'participation and power' should be the focus, not 'behaviour'. Finally, critics question the hierarchies of knowledge which position some individuals as experts over others, with 'experts' often originating from positions of privilege.[357]

In accordance with several of these critiques, Dutta has also identified the underlying assumptions and institutional actors which, he holds, have contributed to the emergence of such a flawed system.[358] He argues that the top-down conception of communication for social change, largely predominant in US- and Eurocentric literature, is rooted in the 'war-military intelligence interests of the US empire'.[359] In post-World War II contexts, a managerial structure was promoted where individuals were understood to be the core agents of change on the path to development. This individualist framework has continued to dominate as traditions have shifted from imperialist framings of culture to the neoliberal transformation of social change.[360] In earlier frameworks, social change was presented as 'the solution to the problem of culture' through racist and imperialist 'modernization frameworks'.[361] Subsequent treatments reimagined culture as a 'key tool for the global implementation of neoliberal policies'.[362]

Dutta has proposed a different approach: culture-centred communication for social change.[363] He encourages researchers to use a 'deconstructive lens' in their work and to question the basis upon which experts' views should be prioritised over local understandings. Key to this approach is the necessity of 'unlearning privilege' and of facilitating collective action from below. He stresses the central role that marginalised groups need to play in bringing about meaningful social transformation:

---

[354] Obregon & Waisbord, 2012; Dutta, 2011
[355] Obregon & Waisbord, 2012
[356] Ibid.
[357] Ibid.
[358] Dutta, 2020
[359] Ibid.
[360] Ibid.
[361] Ibid.
[362] Dutta, 2020
[363] Dutta, 2011, 2012, 2020

*The culture-centered approach to social change envisions the capacity of communicative processes to transform social structures, and in doing so, it attends to the agency of the subaltern sectors in bringing about social change…The goal of the culture-centered approach is to create avenues and spaces of social change by listening to the voices of subaltern communities that have historically been marginalized. Participatory spaces are created so that these spaces offer co-constructive openings for listening to subaltern voices, foregrounding these voices in the discursive spaces of knowledge production. At the heart of the culture-centered approach is the theorizing of the intersections between culture, structure, and agency as the tripods that offer the base for meaning making and communicative enactment. Structure refers to the institutional roles, rules, practices, and ways of organizing that constrain and enable access to resources. Culture constitutes the local contexts where meanings are continuously negotiated. Agency is the capacity of individuals and collectives to enact their choices as they negotiate structures. The culture-centered approach builds upon subaltern studies and postcolonial theories to disrupt the hegemonic spaces of knowledge production with dialogues with the subaltern sectors that have historically been erased from the mainstream discourses of development and progress.*[364]

Research on culture-centred communication does not only critique multinational corporations and multilateral organisations for failures in achieving bottom-up communication for social change. Dutta also focuses on the role played by universities and academics themselves in reifying existing power structures. Consequently, he proposes 'academic-activist-community partnerships' as the way forward for the co-construction of effective advocacy.[365]

## Participatory Learning and Action Theory 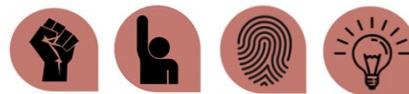

| Key Points | Gaps Identified |
|---|---|
| - Participatory Learning and Action (PLA) Theory is a participatory methodology traditionally employed within public health, where communities are engaged as equal partners in the solutioning of public health issues.<br><br>- PLA research illustrates an array of evaluations of PLA methods which demonstrate public health improvements and increased multi-stakeholder dialogue.<br><br>- Critical approaches to PLA motivate a politically informed, grassroots approach to PLA. | - Participatory Learning and Action Theory focuses on participatory methodologies within public health provision. There exist gaps in literature discussing the use or need of PLA in other non-health-related domains such as those service areas impacted by the design and deployment of data-intensive technologies.<br><br>- PLA projects have been critiqued for a lack of contextual awareness and consideration of power dynamics contextualising sites of research and intervention. |

---

[364] Dutta, 2011, p. 39-40

[365] Dutta, 2020

Participatory Learning and Action (PLA) Theory is a methodology employed to involve service users as equal partners and collaborators in research.[366] It has been utilised to improve health outcomes through community involvement in ongoing group discussion, action, trust-building, and collective problem solving in adjacent justice fields such as public health. PLA literature presents participation in health service provision as a human right.[367]

Multiple public health studies have involved a concrete mobilisation with local communities, assessing health outcomes. Costello discusses the adoption of 'Sympathy Groups' for addressing issues ranging from maternal and newborn death to pre-diabetes and diabetes in Bangladesh, India, Malawi, and Nepal. Group members have a common interest, agree on a focus, and meet regularly to work on strategies to solve a problem. Each trial demonstrated improved health outcomes, and from qualitative data, greater solidarity and spin-off activities beyond health concerns were observed.[368] De Brún et al. showed how the use of PLA methods in Austria, Greece, the Netherlands, and the United Kingdom enabled stakeholders with differential social status and power to offer perspectives to improve cross-cultural communication between primary healthcare workers and migrants[369] and foster shifts in understanding.[370]

PLA, however, has been critiqued for engaging poorly with underlying power and politics. Some critiques relate to placing the burden of solutioning on local communities and diverting attention from more powerful actors. Others discuss the role of Western NGOs in funding and operationalising their own notions of how lower- and middle-income countries should be managed.[371] These critiques call for an approach to PLA that is socially and politically informed, presented as a component of a greater program led by political actors from within countries engaged with broader aspects of change. Examples of such participatory approaches include participatory budgeting in Brazil, pregnancy groups campaigning for health plans, and the mobilisation of lower-caste groups in Tamil Nadu to gain political power and build schools.[372] PLA, when considering these factors, may be viewed as a method promoting the right to health.

---

[366] De Brún et al., 2017

[367] Costello, 2018

[368] Costello, 2018; Prost et al., 2013; O'Donnell et al. 2016; Roy et al., 2013

[369] De Brún et al., 2017

[370] Ibid.

[371] Costello, 2018

[372] Ibid.

# Restorative Justice 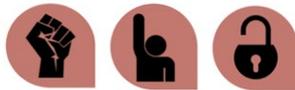

| Key Points | Gaps Identified |
|---|---|
| - Restorative justice has been defined in many ways, but at its core, it attempts to address the wrongdoings of the past.<br><br>- Restorative justice challenges traditionally Western views of criminal justice as punitive, and instead focuses on prioritising the needs of victims, offenders, and community members.<br><br>- While there are ongoing debates about what practices are encapsulated by the term restorative justice, some generally agreed upon examples include actions like offering a sincere apology – occurring through mediation such as conferencing or circles – offering restitution and engaging in community service.<br><br>- Restorative justice has been known by many other names and 'resonates with and draws from' various Indigenous and religious practices across the globe,[373] where justice, harmony, and balance are essential facets of the community. | - Restorative justice practices in instances of gender violence may imply that the relationship itself must be restored, when this, in fact, may neither be desired nor prioritise the security of the victim.<br><br>- There are many disagreements surrounding what practices should be considered under the umbrella of restorative justice practices.<br><br>- Instances of restorative justice in a data-driven context such as online content moderation require extensive training and resources which many Big Tech corporations are unwilling to provide. |

As outlined in the Access pillar above, promoting equitable access across data collection and use contexts involves a four-dimensional approach consisting of (1) a focus on harms of allocation and distributive justice, (2) the non-ideal/contextual and capabilities approaches to justice, (3) harms of representation and recognitional justice, and (4) restorative or reparational justice. While the first three of these facets remain integral to the advancement of access as it relates to data justice research and practice, they tend to focus primarily on addressing present harms and making course corrections oriented to a more just future. Restorative justice reorients this vision of the 'temporal horizons of justice'.[374] It takes aim at righting the wrongs of the past as a redeeming force in the present.[375]

---

[373] Van Ness, 2005

[374] Ackerman, 1997

[375] Restorative justice is not siloed in the domain of criminal justice but has also been extended to schools, institutions, and workplaces, demonstrating its positive impacts on conflict resolution.

While there are different interpretations of restorative justice,[376] Daly argues that the difficulty in defining the term stems from factors such as different views about what is (or is not) restorative justice, a geographically limited understanding of what it is, among other factors;[377] however, several foundational definitions have been offered. Howard Zehr signposts restorative justice as 'an alternative framework for thinking about wrongdoing'.[378] Daly defines restorative justice as 'a *contemporary justice mechanism* to address crime, disputes, and bounded community conflict. This mechanism is a *meeting* (or several meetings) of affected individuals, facilitated by one or more impartial people'.[379] Restorative outcomes according to Van Ness (2005) include things like offering a sincere apology – occurring through mediation such as conferencing or circles – restitution, and acts of community service, but they must occur after the admission of guilt through conviction or admission of responsibility. [380]

Departing from the punitive orientation of Western views of justice, restorative justice aims to prioritise the needs of the victims, offenders, and community members that should be addressed by the legal system, but many times are not.[381] Zehr and Gohar explore the notion that often victims involved feel 'neglected or abused by the justice process', which the authors argue stems from the legal definition of crime, which 'does not include victims [and is] defined as against the state, so the state takes the place of the victim'.[382] There has been a movement away from communities handling their own conflicts and instead transference of that role to the state, as seen in many Western countries.[383] Pratt explains that state punitive control was imposed on the 'indigenous peoples of colonised nations, suppressing their native restorative justice traditions'.[384]

---

[376] Fattah, 1998; O'Mahoney & Doak, 2009 as cited in Daly, 2016. See Menkel-Meadow (2007) for further elaboration on the different ways that restorative justice has been defined. See Batley (2005) for additional detail on how restorative justice aims to fill the gap in existing theories and approaches to justice including retributive theories of justice, prioritisation of the protection of society through punishment (the utilitarian deterrence approach), the rehabilitation approach, and the restitution approach. The Declaration of Basic Principles on the Use of Restorative Justice Programmes in Criminal Matters approach restorative justice by dividing it out into two sections: 'restorative process' – the collective participation in a facilitator-led resolution of matters – and 'restorative outcome' – 'an agreement reached as a result of the restorative process'. See Van Ness (2005) for more on this.

[377] Daly, 2016

[378] Zehr & Gohar, 2003

[379] Daly 2016, p. 14. See Archbishop Desmond Tutu's comment on restorative justice (Tutu, 1999, p. 51) as cited in Roche, 2002

*'I contend that there is another kind of justice, restorative justice, which was the characteristic of traditional African jurisprudence. Here the central concern is not retribution or punishment but, in the spirit of ubuntu, the healing of breaches, the redressing of imbalances, the restoration of broken relationships. This kind of justice seeks to rehabilitate both the victim and the perpetrator, who should be given the opportunity to be reintegrated into the community he or she has injured by his or her offence. This is a far more personal approach, which sees the offence as something that has happened to people and whose consequence is a rupture in relationships. Thus, we would claim that justice, restorative justice, is being served when efforts are being made to work for healing, for forgiveness, and for reconciliation'.*

[380] Van Ness, 2005. It is purported that around 80-100 countries use some form of restorative justice. There is some debate as to what practices can actually be considered restorative justice practices, especially as these practices vary across the globe. While this a very interesting aspect of the restorative justice field, this debate cannot be fully explored in this literature review.

[381] Zehr & Gohar, 2003; Johnstone, 2002; Zernova, 2007. For feminist critiques of restorative justice, see Cook, Daly, & Stubbs, 2006; Ptacek, 2005, 2010; Strang & Braithwaite, 2002 as cited in Daly, 2016. When considering the prioritisation of victims' needs, restorative justice processes have been met with feminist critiques for situations such as partner, sexual, and family violence, in which 'restorative' implies that the relationship itself must be restored, when this may not be desired nor place the security of the victim as a priority .

[382] Zehr & Gohar, 2003

[383] Roche, 2002; Johnstone, 2002, 1998

[384] Pratt, 1996

Zehr traces restorative justice back to the understanding that wrongdoings violate interpersonal relationships which is predicated on the assumption that we are all interconnected. This notion of the centrality of relationships stems from many cultures including 'Maōri's 'whakapapa', Navajo's 'hozho', and Bantu's 'ubuntu''.[385] Because of this violation of the community, there is an 'obligation to put right the wrongs'.[386] Restorative justice has been known by many other names and 'resonates with and draws from' various Indigenous and religious practices across the globe,[387] where justice, harmony, and balance are essential facets of the community. Scholars of restorative justice argue that Indigenous communities tend to approach justice from a lens of restoration which often contrasts the Western notion of retribution in the justice system.

Three examples of restorative justice are worth highlighting.[388] Various First Nations and Native American communities across the United States and Canada emphasise a form of restorative justice referred to as Circles, which were first introduced in the 1980s by the First Nations of Yukon. Circles are voluntary spaces for the victim and offender to have an encounter with the inclusion of community members and impacted families, in which all participants can voice their thoughts with the end goal to bring healing to the community.[389] In South Africa, the Community Peace Program based in Cape Town, according to Roche, builds upon some of the foundations created by the Truth and Reconciliation Commission and has established local peace committees, which are similar to Circles and provide a space for victims, community members, and offenders to determine how to move forward and find a consensual form of resolution for the injustice that took place.[390] New Zealand has paved the way for the restorative justice movement, as beginning in 1989, it situated restorative justice within its entire juvenile justice system through Family Group Conferences (FGC).[391] Zernova explicates how its proponents believe it is rooted in Maōri 'whanau conferences'. Additionally, it is believed that FGCs were also established due to the overrepresentation of Maōri youth in custodial penal institutions, causing concerns amongst the Maōri people.[392]

---

[385] Zehr & Gohar, 2003

[386] Ibid.

[387] Van Ness, 2005

[388] Some additional instances include: The Mohawk Nation of Akwesasne in Canada instituted an Indigenous people's court based on Mohawk values and principles. See Washington, 2018; Valiante, 2016; Mirksy, 2004a. The Mnjikaning peoples of Canada have taken steps to avoid the use of terminology such as "offender" and "victim", moving the focus towards how an individuals' actions negatively impacted on the community as well as creating a restorative justice programme entitled *Biidaaban* – 'new beginning' or 'new day'. *Biidaaban* takes the notion of restorative justice a step further and in addition to focusing on the present, it incorporates the future (*Bii*), the present (*daa*), and the past (*ban*). Please see Washington, 2018; Mirksy, 2004a for additional detail.

[389] Center for Justice and Reconciliation, n.d.; Mirsky, 2004b. Today, circles are used across the Yukon, Saskatchewan, and Manitoba peoples, as well as by Navajo peace-making courts.

[390] Roche, 2002. In South Africa, the most commonly cited instance of restorative justice is the Truth and Reconciliation Commission (TRC) which was instituted to deal with 'the nature, extent, and magnitude of the apartheid conflict between 1960 and 1994', but there is argument as to whether it can be considered a form of restorative justice. See Maepa, 2005; Roche, 2002; Llewellyn & Howse, 1999 for further elaborations.

[391] Zehr & Gohar, 2003. For more on Family Group Conferencing, see Zernova, 2007

[392] McElrea, 1994; Pratt 1996; Johnstone, 2002

Outside of the applications of restorative justice in the criminal justice system, Hasinoff et al. propose the principles of restorative justice as a mechanism for content moderation on social media platforms to challenge the traditional practice of removing offensive material as well as the users who post it.[393] These existing content moderation practices are viewed by many as punitive to ensure compliance, and they do not encourage the offender to take responsibility and acknowledge the harm they have done to the community. Salehi explains the notion of approaching online harms through restorative justice process by altering questions that are usually asked after a violation of platform policy has taken place – moving the focus to needs, obligations, and harms.[394] Hasinoff et al. claim that social media platforms 'provide even fewer opportunities than the criminal legal system for victims to participate in a process or access advocates, support, or reparations' and by implementing restorative justice practices they could make their spaces 'healthier and more resilient'.[395] The authors call for training and restorative justice facilitation within social media platforms as a means to address harms rather than the already existent flagging or reporting mechanisms.[396]

Another area where restorative justice has been applied is climate change. Motupalli, following Preston, has argued that a restorative justice framework can be used to redress environmental harm in a contextually specific and intergenerationally effective way that improves on existing environmental protection law.[397] Along similar lines, Robinson and Carlson have more recently maintained that, considering the failure of loss and damage cases in climate litigation, restorative justice mechanisms should be applied as 'an alternative, non-judicial approach to addressing loss and damage'.[398] On their account, these mechanisms should include instituting restorative dialogues that involve truth and reconciliation conferences and restitution. Likewise, they hold that restorative justice norms should be integrated into 'global climate governance as a pathway for progressing negotiations'. These potential applications of restorative justice processes to the technology policy and law may present attractive policy options for those who are endeavouring to build effective data governance frameworks that sufficiently cover the reparative, 'fourth dimension' of data justice.

---

[393] Hasinoff et al., 2020

[394] Salehi, 2020. The questions usually asked are, 'What content has been reported?, Is the content against the rules?,' or 'Should the content be removed, demoted, flagged, or ignored?' However, using a restorative justice approach, these questions would instead consist of 'Who has been hurt?','What are their needs?'," and 'Whose obligation is it to meet those needs?'

[395] Hasinoff et al., 2020

[396] Ibid.

[397] Motupalli, 2018; Preston, 2011

[398] Robinson & Carlson, 2021

# Reflection Questions

| Academic Researchers | Policymakers | Developers | Impacted Communities |
|---|---|---|---|
| - To what extent does my research demonstrate an understanding of the challenges surrounding social mobilisation and culture-centred participatory methodologies as these relate to data justice research and practice? To what extent does my research demonstrate an understanding of the challenges surrounding the restorative justice perspective and corollary governance possibilities as these relate to data justice research and practice? To what extent does my research embrace the kind of interdisciplinarity needed for learning to be transferred from data justice adjacent domains such as climate and global public health justice?

- What are the gaps in my current understanding of data justice adjacent domains of scholarship and activism such as climate and global public health justice? How responsive is my research to the need for bottom-up social mobilisation and culture-centred participatory efforts at social change? What are the gaps in my current understanding of the restorative justice perspective? | - To what extent do my policymaking practices demonstrate an understanding of the challenges surrounding social mobilisation and culture-centred participatory methodologies as these relate to data justice research and practice? To what extent do they demonstrate an understanding of the challenges surrounding the restorative justice perspective and corollary governance possibilities? To what extent do my policymaking practices embrace the kind of interdisciplinarity needed for learning to be transferred from data justice adjacent domains such as climate and global public health justice?

- What are the gaps in my current understanding of data justice adjacent domains of policymaking such as climate and global public health justice? How responsive is my policymaking to the need for bottom-up social mobilisation and culture-centred participatory efforts at social change? What are the gaps in my current policymaking understanding of the restorative justice perspective? How can impacted communities' knowledge and experience improve understandings and/or responses relating to policy areas (e.g. climate change, the environment, public health)? | - To what extent do my data innovation practices demonstrate an understanding of the potential role social mobilisation and culture-centred participatory methodologies in processes of data collection and use? To what extent are my data innovation practices responsive to the restorative justice perspective and corollary governance possibilities? To what extent do my data innovation practices embrace the kind of interdisciplinarity needed for learning to be transferred from data justice adjacent domains such as climate and global public health justice?

- How responsive are my data innovation practices to the need for bottom-up social mobilisation and culture-centred participatory efforts at social change? What are the gaps in my current understanding of the restorative justice perspective as this might apply to my data innovation practices? | - To what extent do I and members of my community possess an understanding of the potential role social mobilisation and culture-centred participatory methodologies could play in processes of data collection and use? To what extent do I and members of my community possess an understanding of the restorative justice perspective and corollary governance possibilities? To what extent do I and members of my community embrace transfer learning from data justice adjacent domains such as climate and global public health justice?

- How can I and members of my community work towards bottom-up social mobilisation and culture-centred participatory efforts at social change? What are the gaps in our current understanding of the restorative justice perspective as this might inform our demands for rectification of past wrongs and reconciliatory dialogue? |

| | | | |
|---|---|---|---|
| - What research methods could I use to enable community-centred approaches?<br><br>- To what extent can I use the concepts discussed in these data justice adjacent literatures to expand the scope and normative awareness of my research? | - What methods could be used to enable community-centred approaches to policymaking and implementation (e.g. incorporating PLA or restorative justice)? | - What methods can we use to enable community-centred approaches to development? How can we combine technical expertise with local or contextual knowledge and perspectives to ensure data practices and products are beneficial, appropriate, and take account of communities' interests? | - What methods can we use to enable community-centred approaches to data innovation? How can we combine technical expertise with local or contextual knowledge and perspectives to ensure data practices and products are beneficial, appropriate and take account of communities' interests? |

## Knowledge, Plurality, and Power

### Power and Science and Technology Studies Perspective

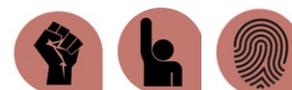

| Key Points | Gaps Identified |
|---|---|
| - Science, technology, and innovation are shaped by social, cultural and political factors.<br><br>- Public, or 'lay', knowledge has a valuable role to play in (re)framing and informing processes and practices relating to science and technology and the role these play in society.<br><br>- Public engagement with diverse communities is vital to ensure accountability and governance of science and technology. | - The Science and Technology Studies (STS) literature has largely focused on studying controversies around science and technology from neutral vantage points. This has led to calls for STS scholars to become more active in shaping scientific programmes of practices rather than critiquing from the sidelines.<br><br>- More work must be done to link the valuable critical analyses done by STS scholars to the constructive frameworks of practical ethics that are influential in the governance of responsible data innovation ecosystems. |

Science and Technology Studies (STS) is an interdisciplinary field focused on understanding the relationships between science, technology, and society. This includes examinations of the impacts of science and technology on society and also, importantly, of the ways in which social, cultural, and political factors shape science and technology.[399]

---

[399] e.g. Latour & Woolgar, 1979 [1986]; Mackenzie & Wajcman, 1985

Power is a central theme in much STS work. Research has focused on revealing the ways in which power operates in ordering societies and shaping which ideas, practices, and forms of knowledge have influence.

Within the varied field of STS, a significant focus has been on the role of social, cultural, or political factors within processes of knowledge creation and constructions of expertise. Bloor's (1976) seminal work Knowledge and Social Imagery which set out the 'Strong Programme' in Sociology of Scientific Knowledge (SSK), promoted the examination and consideration of interactions between the cultural setting in which claims to knowledge are made and those claims themselves.[400] STS research has identified the ways in which expertise is constructed and defended through established scientific and political systems, observing that scientific knowledge can act as an 'authoritarian force'.[401]

A number of prominent STS case studies have challenged modernist framings of technical expertise and highlighted the relevance and importance of diverse sets of knowledge within processes around science, technology, and innovation.[402] In challenging the hegemony of technical or professional knowledge, STS highlights the importance of engaging with diverse sources and forms of knowledge to inform justice-oriented discussions of new technologies. Jasanoff coined the term 'civic epistemology' recognising that members of the public can question and challenge experts' claims and present alternative knowledges based on their own experiences or expertise (which does not necessarily conform to the dominant notions of expertise or knowledge as they are conceptualised within scientific disciplines).[403] An individual or group's civic epistemology determines whether scientific claims are accepted, and which alternative ways of conceptualising issues or alternative knowledges are drawn on.

Recognising the value of public knowledge, STS scholars have pointed to the importance of public engagement as a valuable mechanism to inform the development and deployment of new technologies and as being crucial to ensure accountability and good governance of science and technology.[404] Public engagement serves important roles in 'test[ing] and contest[ing] the framing of issues that experts are asked to resolve',[405] facilitating scrutiny and accountability and ensuring professional "expertise" is not used to perpetuate unjust points of view or to bestow too much power on the organisations within which expertise is located.[406] Additionally, public engagement can ensure that science and technology conform to cultural standards and align with public values.[407]

---

[400] Bloor, 1976

[401] Hajer, 1995

[402] e.g. Wynne, 1992; Epstein, 1995; Kerr et al., 2007

[403] Jasanoff, 2005

[404] Irwin, 2006; Jasanoff, 2011

[405] Jasanoff, 2003, p. 397

[406] Ibid.

[407] Ibid.

Ribes has observed that data science is increasingly becoming an area of prominent interest to STS scholars.[408] Recent examples of STS research focused on topics relating to data justice include:

- Rahnama took an STS approach to explore uses of algorithms in court rooms in the United States. Rahnama proposes that due to uncertainty in the science used to develop and train algorithms and the potential for algorithms to reproduce or entrench pre-existing biases there is a need for greater inclusion of diverse perspectives and sources of knowledge in decision-making around the ways in which algorithms are used in courtrooms.[409]

- Egbert and Mann take an STS approach to examining discrimination in and through predictive policing algorithms. This STS approach highlights the importance of decentring the technologies to understand the broader socio-technical and historical contexts in which the technologies are developed and deployed. Engaging with this wider context is vital to move beyond techno-centric views of algorithmic systems and their impacts on society and to understand the ways in which discriminatory practices reflect, and have been enabled by, the social, cultural, or political contexts in which technologies are developed and deployed.[410]

STS scholars are increasingly working, or collaborating in the discipline of data science (e.g. in developing teaching or policy) and are well-placed to play an important role in shaping future data science practices and informing justice-oriented approaches.[411] STS scholars play roles as advisers or evaluators of scientific projects, and increasingly often STS scholars are recruited to facilitate public and/or stakeholder engagement within varied science and technology programmes.[412] While STS scholars have typically played the role of neutral observers in relation to science and technology projects, these more active roles have similarly at times being criticised as maintaining social scientists in a position of bystanders or facilitators in relation to data science programmes. This has led to calls for STS scholars to become more active in shaping scientific programmes of practices rather than critiquing from the sidelines.[413]

---

[408] Ribes, 2019

[409] Rahnama, 2019

[410] Egbert & Mann, 2021

[411] Ribes, 2019

[412] e.g. Marks & Russell, 2015; Aitken et al., 2016b

[413] Ribes, 2019

# Pluriverse and Post-Development Theory

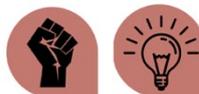

| Key Points | Gaps Identified |
|---|---|
| - From the perspective of many scholars who write about pluriversality and post-development theory, the struggle for 'under-developed' nations to emulate the Global North's economic template has come at an enormous ecological and social cost. The 'development as progress' paradigm is unsustainable and prioritises economic goods over planetary well-being. The cascading global crises we now face cannot be managed within existing intellectual and institutional structures.<br><br>- The problem lies, on this view, not in a lack of implementation, but in a combination of (1) Eurocentrism, in which financial wealth, geopolitical power, and epistemic superiority necessarily coincide, and (2) the conception of development as linear, unidirectional, teleological, and driven by economic modernisation.<br><br>- The 'pluriverse' is a world of worlds in which acceptable domains of thought and science are expanded from the narrow yet hegemonic perspective of European intellectual traditions to include ideas, conceptions, and traditions that span geographical, political, and epistemic boundaries. | - Post-development literature is primarily focused on the political and economic burdens placed on non-Western, financially impoverished nations by Western wealthy ones. The dominating influence of digital technologies and associated values is largely overlooked. While there is an emerging literature on post-colonial computing, there are tensions with post-development theory and its focus on centring non-European ideas. Where digital technology is discussed as a pluriversal or post-development issue, the primary focus on the environmental costs (e.g. resource consumption and electronic waste) and the role of technology in international development projects. A post-development critique of data extraction, exploitation, and capitalism would complement this literature and its central arguments. |

The term 'pluriverse' describes a 'world in which many worlds co-exist'.[414] In contrast with the idea of a 'universe' in which authoritative knowledge and expertise is defined narrowly and centred primarily in the Global North, the pluriverse conceptually highlights how such a Eurocentric, modern, and economically-driven worldview erases and eradicates other traditions and knowledge systems that range beyond itself. Scholars of the pluriverse reject the hegemony of prominent epistemologies of power in favour of broadly inclusive and integrative forms of knowledge. The pillars of power and knowledge are implicated in this work as interacting forces of both destruction and construction, oppression, and liberation.

The pluriverse is a 'broad transcultural compilation of concrete concepts, worldviews, and practices from around the world, challenging the modernist ontology of universalism in favour of a multiplicity of possible worlds'.[415] Pluriversality is an epistemic shift away from universal assumptions of Western, Eurocentric cosmology, consisting of ideas and traditions that span across geographical and political boundaries. It expands upon post-development theory as an alternate proposal to the extractive, homogenising, and unsustainable worldview promoted by the global north. Pluriversality suggests that by transcending the presumed universality of Western thought to embrace a 'mosaic epistemology',[416] solutions are more likely to be found to the world's overlapping crises.

---

[414] Reiter, 2018

[415] Kothari et al., 2019

Post-development theory follows a similar vein to confront the fallacies and assumptions behind the paradigm of wealthy nations altruistically 'developing' (and thereby 'civilising') the less well-off. The primary distinctions between 'developed' and 'under-developed' nations lean heavily on market-oriented indicators, such as gross domestic product without accounting for other values, such as cultural richness, quality of life, or other non-economic values. This paradigm of development follows easily from 'Eurocentrism', the belief that European culture and society is at the centre of civilisational maturity and modernity, itself an escape from socio-spatial provincialism to a state of rational arrival that 'immature' and 'peripheral' societies ought to aspire to.[417] And yet, despite decades of international development, the world's central problems have not been solved by the interventions of the presumptively "superior" nations. In many respects, the most critical problems facing the world's inhabitants have actually worsened.

The post-development view is that the overlapping crises of environmental degradation, political turmoil, discrimination, forced migration, and the extractive commodification of resources are enabled rather than resolved by market logics, the valuation of lives and resources through economic measurement, and individualist notions of political organisation, personhood, and responsibility. As with pluriversality, post-development theory suggests that the conceptually and materially violent ideologies of the former colonial powers that have created the development mindset represent not the best nor even the most ideal path for humanity. Instead, they are understood as the product of an 'entangled heterarchy' in which asymmetric power relations indicate what forms and sites of knowledge are authoritative and believed to affirm social progress.[418] Like pluriversality, post-development theory posits that answers to the world's problems may lie outside of the hegemony of Western thought and its narrow range of epistemic commitments (e.g. Western science as the only source of truth). Post-development theory and pluriversality therefore together argue for centring ideas, traditions, and forms of knowledge from putatively peripheral domains of discourse, bringing a broader set of practices and attitudes to a more central position in efforts to make life liveable and meaningful for all the world's inhabitants.

---

[416] Ibid.

[417] Dussel et al., 2000

[418] Ali, 2014

## Reflection Questions

| Academic Researchers | Policymakers | Developers | Impacted Communities |
|---|---|---|---|
| - To what extent does my research demonstrate an understanding of and account for questions of social, epistemic, and economic power? To what extent does it demonstrate an understanding of and account for systemic discrimination, and Eurocentrism, particularly in how they are reflected in data and data-driven technologies?<br><br>- What are the gaps in my current understanding of epistemologies and perspectives that differ from my own, particularly as they are expressed in data and data-driven technologies?<br><br>- To what extent can I use the concepts discussed in this literature to expand the scope of my research to include perspectives and demonstrate awareness of non-Western thought and experiences of marginalisation both from within and external to my own society? | - To what extent does my policymaking consider or include perspectives and experiences from the Global South and marginalised communities both locally and elsewhere?<br><br>- What are the current gaps in my understanding of how policymaking can contribute to issues identified in the literature, such as presumptions of where knowledge authority resides and historical failures to include or consider perspectives from the Global South and members of marginalised communities?<br><br>- To what extent can I ensure that my policymaking acknowledges and addresses histories of dominance and oppression by governments and institutions who hold the most economic, political, and epistemic power? | - To what extent do the technologies and datasets I use (or contribute to) reproduce the injustices and blind spots identified in the literature by either failing to include or account for the broadest possible range of perspectives and experiences or by channelling social, economic, or political power to those who already hold the most?<br><br>- What are the gaps in my understanding of how technological design and development can contribute to the issues identified in the literature, such as where technologies reflect the worldviews and structural power of their designers and deepen the marginalisation of the already marginalised?<br><br>- To what extent can my contribution to technology design and development and the construction of datasets promote greater inclusion of perspectives and experiences outside the Global North and identities/communities that already hold oppressive power? | - To what extent do I recognise how technological design and development and the collection and use of data affect the well-being, power, and authority of myself and my community?<br><br>- What are the gaps in my understanding of how data and the design and development of data-driven technologies can reproduce narrow worldviews and exclusionary conceptions of knowledge and expertise?<br><br>- How can I utilise an understanding of how data and the design and development of data-driven technologies reproduce structures of power to pursue strategies for increasing my community's inclusion in decisions about how data and technology are created and used? |

# Non-Western and Intercultural Approaches to Data Justice and Injustice

## Intercultural Communication and Contestation

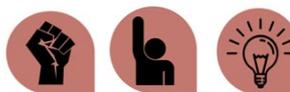

| Key Points | Gaps Identified |
|---|---|
| - Critical Intercultural Communication and Contestation literature presents frameworks for cultural analysis where culture is understood as the struggle between social entities for ideological control.<br><br>- Ideology is framed as the set of meanings composing social groups' worldview, which benefit their interests and priority.<br><br>- Differences of power in the articulation and circulation of ideology are situated within the context of globalisation, where the asymmetrical distribution and adoption of ideologies benefiting Western economic and cultural dominance is circulated via communication flows. These power asymmetries are extended to the globalising power of corporations owning digital communication platforms.<br><br>- The homogenising power of globalisation is contested, and instead the autonomy and unique circumstances of social entities are framed as adapting to globalisation through articulating hybrid cultures and plural ideologies.<br><br>- Articulation as an autonomous act and the potential for cultural hybridisation is exemplified by the varied 'cultures of contestation' that have emerged through social media use. These cultures adapted culturally resonant protest-based ideologies, creating versions of them which allowed the forging of links with and, in turn, the transformation of other movements.<br><br>- Limitations to the use of digital communication platforms as tools for contestation are attributed to existing power asymmetries between civil society, political, and private entities.<br><br>- Significantly, the power of US corporations owning and governing platform architectures challenges knowledge plurality, equal distribution of ideology, and ultimately, democratic processes. | - Critical Intercultural Communication and literature provides theoretical frameworks for analysis of the cultural influence. There are, however, gaps in literature focused specifically on digital communication platforms and their shaping of communication flows.<br><br>- Literature pertaining to Cultures of Contestation provides rich analyses of the distribution and influence of grassroots ideology via social media. There are gaps in literature analysing the role of platform architectures in depth. |

Critical intercultural communication studies move away from understanding culture as a set of characteristics held by a group of people in a geographic region. Instead, this field theorises culture as a struggle between differently positioned actors competing for ideological control.[419] Ideology is understood as the set of meanings that structure groups' worldviews (i.e. values, norms, assumptions).[420] The critical perspective adopted in this field presents communication as the process of articulating ideologies, which hold concrete effects.[421] Although all groups and individuals are understood to create and circulate ideologies that benefit their needs and priorities,[422] articulations are understood to contain varying degrees of power and influence. The power of specific articulations is determined by their relative historic and economic positioning.[423]

This body of literature discusses dynamics between larger structures of power and micro acts among cultural actors and groups.[424] Such analyses of macro-micro processes of communication connect with the pillars of power, knowledge, and participation. They provide a theoretical framework for understanding ideological circumstances facilitated by digital communication technologies. These include the cultural power of for-profit owners of digital communication technologies and the resistance of civil society members that repurpose these technologies for contestation.[425]

Critical Communication and Contestation studies explore globalisation as the economic structure shaping ideologies across cultures.[426] Characterised by the extractive financial involvement of Western nations in non-Western economies, globalisation is driven by nations expanding their market dominance and financial accumulation. Globalised economic structures privilege the circulation of ideologies by powerful nations that saturate international markets and media outlets with their commodities and cultural products. Branded with ideologies,[427] these products ease the cultural adoption of Western nations' economic and political systems and investments and strengthen their global power and presence.[428]

Some academics present globalisation as a structure that asymmetrically exposes populations to Western values, yet these scholars also challenge the deterministic power of exposure. They emphasise the autonomy and complexity of individual self-formation and social evolution. Through this lens, the formation of cultural identity is framed as vested with agency and the possibility of new ideology and hybridisation, rather than as the passive adoption of dominant ideology.[429] This perspective is supported, for these thinkers, by the swell of 'cultures of contestation' in the 2010s, when grassroots mobilisations across the globe used social media as a means of political contestation by circulating ideologies of protest.[430] The connective capacity of digital communication platforms created the conditions for widespread diffusion, but instead of yielding the replication of a single culture, multiple cultures of contestation were generated. Platforms made possible the adoption, translation, and 'domestication' of culturally resonant ideologies. Different cultural versions were articulated and passed on from one movement to the next, each of which connected to, learned from, and subsequentially influenced protests elsewhere.[431] Digital communication technologies have enabled movements countering dominant frames to articulate and circulate ideologies, gain visibility, garner support, forge strategic alliances, and ultimately mobilise against dominant actors.[432] Examples of contestation through digital platforms highlight micro-actors' agency in ideological formulation, the possibility for hybridisation, and ideological resistance to dominant ideologies and interests such as those driving globalisation.

---

[419] Nakayama & Halualani, 2010
[420] Ibid.
[421] Halualani, 2019
[422] Nakayama & Halualani, 2010
[423] Halualani, 2019
[424] Ibid.
[425] Nakayama & Halualani, 2010
[426] Peeren et al., 2018
[427] Halualani, 2019
[428] Ibid.
[429] Peeren et al., 2018
[430] Ibid.

Although the emergence of digital communications platforms may suggest a democratisation of cultural production and distribution, there are significant constraints on their use as tools for contesting power. Algorithm-driven platform architectures (such as recommender algorithms) have facilitated the emergence of new communication strategies adopted by for-profit and political actors alike.[433] The asymmetrical power of these actors informs their ideological tactics and their influence over mass digital communications. For example, the microtargeting of content towards groups challenges individual and group autonomy and has been adopted to influence democratic elections.[434] Politicians have exploited the spread of misinformation, governments have shut down the internet as a response to protests, censored online content, and used digital platforms for surveillance.[435]

Significantly, the monopolisation of essential digital infrastructures by US corporations has led the governance environments of platforms and the regulatory constraints on them (or lack thereof) to be driven by commercial objectives. Corporations managing digital communication architectures shape the mechanisms for modern-day cultural production and its material political and economic outcomes. The power of platform architectures is such that they have splintered regimes of truth—operationalised through traditional mass communication—into regimes of post-truths as groups engage with ideology via content exposure.[436] Digital communication platforms have become the context where culture is articulated, where differently positioned subjects and social entities compete[437] for communicative influence via channels shaped by corporate-governed algorithms. This centralisation of cultural governance threatens knowledge plurality, equal distribution of valid information, and ultimately, democratic processes.

---

[431] Ibid.
[432] Ibid.
[433] Ibid.
[434] Ibid.
[435] Ibid.
[436] Nakayama & Halualani, 2010
[437] Peeren et al., 2018

# Non-Western Values and the Transformation of Data Justice

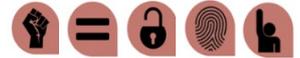

| Key Points | Gaps Identified |
|---|---|
| - Proponents of intercultural approaches to information ethics and data justice argue for the priority of centring non-Western values in the formulation of policies, ethics frameworks, and governance regimes as well as in the production of data-intensive technologies.<br><br>- Critically oriented and decolonial scholars and activists have emphasised the importance of amplifying non-Western and Global South(s) perspectives to combat the legacies of Western cultural hegemony that have, for them, created an unsustainable homogeneity of values in digital ethics and denigrated the worth of non-Western cultures and belief systems.<br><br>- Constructive forms of intercultural thinking about digital ethics have tried to elevate the worth and significance of non-Western belief systems, working from an understanding of the conceptual parity between Western philosophies and heretofore undervalued cultural frames of reference like Ubuntu and Indigenous thinking, Confucianism, Daoism, and Buddhism.<br><br>- Constructive intercultural approaches seek useable and accessible ways of connecting diverse value systems in the ends of human flourishing and aim to develop conceptually inclusive and pluralistic ethical frameworks for the governance of digital innovation that strive for an optimal degree of intercultural learning and generality across forms of life by drawing on societally beneficial and morally generative commonalities between manifold belief systems.<br><br>- At the same time, it is argued that this pursuit of ethical generality needs to start from a position of conceptual humility, charity, and openness that preserves and fosters the differences in value orientations that arise from unique sociocultural histories and ways of living. | - Though there is growing interest in the potential contribution of non-Western values to information ethics and data justice, intercultural approaches have not yet been incorporated into mainstream data justice thinking or prioritised in the shaping of policy, standards, and regulation.<br><br>- More global research needs to be funded and done in the area of comparative analyses between different value-, belief-, and knowledge-systems (non-Western and Western), so that intercultural learning and insight can make greater contributions to the policymaking and standards-setting environment.<br><br>- Existing research into possibilities for intercultural information ethics and data justice is still being undertaking within Western academic contexts and epistemic regimes. Data justice research and practice should therefore seek out novel, non-academic, and non-Western entry points into opening up intercultural dialogues. |

The priority of centring non-Western values in the transformation and advancement of data justice is a relatively novel and underdeveloped aspect of the current data justice literature. Be that as it may, intercultural approaches to the wider ethics of information technologies have been around for well over a decade. Building on Raphael Capurro's (2005) call for a new 'intercultural information ethics',[438] Hongladarom and Ess, in their 2007 book *Information Technology Ethics: Cultural Perspectives*, argued for the initiation of 'a new field of study' that introduced multiple 'dimensions of cultures into the deliberation on information and computer ethics' and 'into the discourse and discussions of not only academics but also policymakers and the various stakeholders in the area'.[439] More recently, scholars from Pak-Hang Wong,[440] Sabelo Mhlambi[441] and Abeba Birhane[442] to Peter Hershock,[443] Jason Edward Lewis,[444] and Shannon Vallor[445] have explored the prospects and pitfalls of drawing on non-Western values to frame the ethics and governance of information technologies.

---

[438] Capurro, 2005. See also Capurro, 2008 and Floridi and Savulescu, 2006 for early initiatives to introduce diverse cultural perspectives into information ethics.

From the start, the approaches taken to digital ethics by interculturally-oriented thinkers have ranged from critical and corrective to constructive and forward-thinking. On the critical and corrective side, some scholars and advocates have endeavoured to introduce and amplify non-Western perspectives to combat the legacies of Western cultural hegemony that have created an unsustainable homogeneity of values in digital ethics. On this view, monocultural anchorings of information and computer ethics (largely steeped in the predominant individualistic ethos of Anglo-European and Euro-American framings) have remained insufficiently responsive to the condition of cultural and ethical pluralism that typifies modern, interconnected global society.[446]

Critical and corrective approaches have been widely informed by an active awareness of post-colonial and decolonial contexts.[447] From this latter perspective, the exclusion of non-Western views from the dominant discourses that have shaped the ethics and governance of digital technologies up to the present reflects deeper legacies of coloniality and Western imperialism that have been increasingly typified by the assertion of extractive and hyper-individualistic ideologies, atomising free market logics, tendencies toward neoliberal "responsibilization,"[448] and the prioritisation of values of growth, efficiency, and profit optimisation.[449] The success of the globalising proliferation of this constellation of beliefs and values has hinged on the negative characterisation, demotion and proscription of non-Western values (such as Ubuntu or Indigenous beliefs) as the irrational, primitive, or mystical Other.[450] As Afolayan and Falola (2017) point out in the African context, the 'ideological cauldron of colonialism' produces erroneous demands to justify the very existence of non-Western philosophies and belief systems that have been widely represented over generations as the backwards, primordial and pre-modern inverse of civilised European and US cultures.[451] The crux of such an exclusion has been the brandishing of a weaponised notion of refined "Western reason" against an idea of non-Western values that are frequently caricatured as rooted in tribalistic collectivism and magical thinking—a power-consolidating dynamic whereby reason itself comes to be wielded by dominant Global North actors

---

as an ideological cudgel that is, in D.A. Masolo's (1994) words, 'believed to stand as the great divide between the civilized and the uncivilized, the logical and the mystical'.[452]

To recover the proper worth and relevance of non-Western values in the post-colonial/decolonial condition, critical thinkers such as Kwasi Wiredu,[453] Linda Tuhiwai Smith,[454] and Ngũgĩ wa Thiongo[455] have drawn on the earlier insights of Franz Fanon to argue for the decolonisation of the conceptual and psychic life of the colonised. 'Decolonizing the mind,' from this perspective, involves ending patterns of colonial domination that denigrate the value of non-Western cultures by creating mental dependencies on the "superior" products, personalities, and cultural forms that come from the United States or Europe while, at the same time, stifling and choking off the long-term development of Indigenous cultures and identities. Insofar as this conditioning of the colonised mind turned on the programmatic depreciation and repression of local histories, traditions, and forms of symbolic reproduction, the recovery of the worth and relevance of repressed non-Western values necessitates the re-vivification of cultural histories and identities that have become disregarded under colonial rule and its downstream effects.[456] As historic figures of decolonial liberation like Amilcar Cabral[457] and Kwame Nkruma[458] argued, formerly suppressed cultures can, in fact, be seen as living sources of social creativity and innovation, becoming thereby a fount for individual empowerment and civilisational advancement both locally and globally.

The launching point of the constructive and forward-looking side of intercultural thinking about digital ethics is, in effect, this elevation of the civilisational worth and significance of non-Western values and belief systems. Working from an understanding of the conceptual parity between Western philosophies and heretofore undervalued cultural frames of reference like Ubuntu and Indigenous thinking, Confucianism, Daoism, and Buddhism, those who have undertaken the programmatic cultivation of intercultural perspectives have constructively sought out commonalities, 'harmonies',[459] and 'resonances'[460] between such diverse belief systems that could support a more inclusive and pluralistic approach to the ethical governance of information technologies on the global plane. The idea here is that, while acknowledging and safeguarding value plurality both at the existential level (i.e. on the plane of each individual's lived experience and choices about a life well-spent) and at the cultural level (i.e. between different and irreducibly unique cultural histories and environments), intercultural approaches should seek serviceable ways of connecting diverse value systems in the ends of human flourishing. That is, they should aim to develop conceptually inclusive and pluralistic ethical frameworks for the governance of digital innovation that strive for an optimal degree of intercultural learning and generality across human forms of life by drawing on societally beneficial and morally generative commonalities between manifold belief systems.[461]

The challenge of bearing this torch of species-level ethical generality while simultaneously preserving the inextricable heterogeneity of diverse cultural frames of reference and values has, from the start, been acknowledged as one of the major hurdles faced by intercultural approaches. In this connection, Charles Ess (2008), has identified a two-pronged hazard that arises in the (even more basic) context of cross-cultural encounters: 'First of all, naïve ethnocentrisms too easily issue in imperialisms that remake "the Other" in one's own image—precisely by eliminating the irreducible differences in norms and practices that define distinctive cultures. Second, these imperialisms thereby inspire a relativistic turn to the sheerly local—precisely for the sake of preserving local identities and cultures'.[462] Ess' twin concerns with unreflective tendencies to reduce otherness to sameness or else to revert to divisive modes of cultural relativism signal the central importance of alterity as a departure point of intercultural ethics. Along these lines, aspirations to intercultural ethics should begin with a responsiveness to the moral demands of difference that arise amidst encounters with the other. The pursuit of species-level ethical generality needs to start from a position of conceptual humility,[463] charity,[464] and openness, namely, from the priority of 'preserving and fostering the irreducible differences that define our identities as distinct from one another, while simultaneously sustaining relations that, ideally, foster the flourishing of all'.[465]

---

[452] Masolo, 1994

[453] Wiredu, 1984, 1995, 1996

[454] Smith, 2021

[455] Wa Thiong'o, 1986, 1993

[456] Kiros, 2017

[457] Cabral, 1973, 1979, 2016

[458] Nkrumah, 1962, 1970

[459] Hongladarom & Ess, 2007

[460] Ess, 2008

[461] See the collected essays contained in Aggarwal, 2020

The starting point of intercultural ethics is underwritten by the relational orientation that is a common element of many non-Western value orientations like Confucian,[466] Buddhist,[467] Indigenous,[468] and Ubuntu[469] thinking (as well as more communitarian and dialogical schools of Western thought from those of Dewey,[470] Apel,[471] Habermas,[472] and Honneth[473] to those of Buber[474] and Levinas[475]). As Kitarō Nishida formulates it, the mutuality of otherness (be this the irreducible element of difference that constitutes the reciprocal individuation of interacting people or the cultural differences that frame such interactions) is an enabling condition of relationality as such. It is a generative alterity that makes connection and proximity possible *by simultaneously preserving distance and distinctness*.[476] If the dynamic of mutual otherness were dissolved into sameness or undifferentiated unity, the relationship between selves or cultures would cease to be, and possibilities for 'imperialisms that remake 'the Other' in one's own image' would arise in kind. However, if the interaction between selves and cultures is treated as resonant and complimentary—if difference is valued and preserved so that the horizon between selves and cultures can be sustained precisely by a caring labour that maintains otherness in the liminality of contact—then commonalities can become harmonies without dissolving difference and the common human predicament of coping with alterity can impel an unbounding of interpersonal and intercultural solidarity.

---

| Key Points | Gaps Identified |
|---|---|
| - Indigenous statistics primarily focus on the 5Ds – disparity, deprivation, disadvantage, dysfunction, and difference. <br><br> - Indigenous data sovereignty (IDS) calls for rights over ownership, collection, and application that extend to how decisions are made with their data. | - Indigenous communities are often excluded from conversations surrounding the use of their data <br><br> - Indigenous communities have not been offered the opportunity to voice their concerns/opinions about their data and the processes that involve their data |

Activists and scholars who write about and advocate for Indigenous data sovereignty aspire to enact changes in existing data practices that harm marginalised groups and thereby to advance data justice. Literature on Indigenous data sovereignty focuses on current disparities present in the way data is collected and used and involves communities making strides to demand control over these extractive processes and their own data.

Indigenous data sovereignty is situated in a landscape in which many disparities are present, specifically in how Indigenous communities are represented using data. Kukatai and Taylor explicate this through discussing the results of a Google search of 'Indigenous statistics' which resulted in a focus on 'statistical representations of the dire, and longstanding, socioeconomic and health inequities between Aboriginal and Torres Strait Islander peoples and non-Indigenous Australian people'.[477] They expand on this by defining 5 'Ds' of data on Ingenious people: 'disparity, deprivation, disadvantage, dysfunction, and difference'.[478] There are many sources that serve to inequitably portray Aboriginal and Torres Strait Islanders' marginal position.[479] This lens placed on Indigenous data creates a cycle in which the narrative perpetuated is one which claims that the only data available on these communities are those that reinforce their existing subaltern positionality with reference to the nation-state, thereby 'rationalising' any existing inequalities. As Kukatai and Taylor state, 'This racialised 'politics of the data', therefore, has powerful consequences in the determination, and practice, of the nation- state/Indigenous population relationship…5 D data provide an infinitely variable circular rationale for Aboriginal and Torres Strait Islander inequality, to the convenient exclusion of other less palatable explanations'.[480]

---

[477] Kukatai & Taylor, 2016

[478] Ibid.

[479] Ibid.

[480] Ibid.

Indigenous Data Sovereignty (IDS) has been defined by Rainie et al. as 'the right of Indigenous peoples to control data from and about their communities and lands, articulating both individual and collective rights to data access and to privacy'.[481] These rights to data access and privacy refer to not only Indigenous communities' direct rights in controlling their data, but also their involvement in decision making processes in which decisions about their communities using their data are often made without their involvement.[482] Another way of defining IDS comes from the Native Nations Institute at the University of Arizona. The definition states, 'Indigenous data sovereignty is the right of Native nations to govern the collection, ownership, and application of its own data'.[483] This definition is meant to relate to the inherent rights of Native nations to govern data in similar ways as they would govern their resources, land, and peoples.[484] It also touches on the application aspect which is a significant dimension of data governance processes in which Indigenous communities are often shut-out in addition to pre-existing exclusionary data collection and processing practices.

As a response to the problems made explicit through the IDS perspective, the Global Indigenous Data Alliance (GIDA) have formed out of a workshop with representatives from Maiam nayri Wingara Collective (Australia); Te Mana Raraunga Maori Data Sovereignty Network (Aotearoa New Zealand); and the United States Indigenous Data Sovereignty Network works on bridging gaps in existing UN Declaration on the Rights of Indigenous People (UNDRIP) regulation.[485] GIDA established four CARE principles for Indigenous data governance. These are: 'Collective benefit', 'Authority to control', 'Responsibility', and 'Ethics'.[486]

Another area of literature that demonstrates efforts towards achieving Indigenous Data Sovereignty is Canada's Open Government Plan which recently included more considerations of Indigenous data, moving from promoting data access for First Nations people to 'recognition of the developing nation-to-nation relationship between Indigenous nations in Canada, including over 600 First Nations, Metis Nations, Inuit, and the federal Crown'.[487]

Finally, as many themes within data justice are interrelated, it is important to note that work in data feminism highlights the importance of Indigenous Data Sovereignty. In the Feminist Data Manifest-No, the authors state, 'We refuse coercive settler colonial logics of knowledge and information organization; we commit to tribal nation sovereignties and Indigenous information management that values Indigenous relationality, the right to know, and data sovereignty'.[488] Thus, the overlap between Indigenous Data Sovereignty and Data Feminism reinforces the importance of the pillars of Power, Identity, and Participation.

---

[481] Rainie et al., 2019

[482] Kukutai & Walter, 2015, as cited in Rainie et al., 2019

[483] Rodriguez-Lonebear & Rainie, 2016; Rainie et al., 2017

[484] Rainie et al., 2017

[485] Kukutai et al., 2020

[486] Ibid.

[487] Canada Action Plan 2018-2020, as cited in Rainie et al., 2019

[488] Cifor et al., 2019

# Reflection Questions

| Academic Researchers | Policymakers | Developers | Impacted Communities |
|---|---|---|---|
| - To what extent does my research demonstrate an understanding of the challenges surrounding intercultural communication, the centring of non-Western values, and the equitable representation of Indigenous groups?<br><br>- What are the gaps in my current understanding of the power dimensions of intercultural communication? How responsive is my research to the set of problems raised by the colonial context, when considering the relationship of Western and non-Western value systems and beliefs?<br><br>- To what extent can I use the concepts discussed in this literature (in particular, notions of intercultural resonance and harmony as well as the CARE principles) to expand the scope and normative awareness of my research? | - To what extent do my policymaking practices demonstrate an understanding of the challenges surrounding intercultural communication, the centring of non-Western values, and the equitable representation of Indigenous groups?<br><br>- What are the gaps in my current understanding of the power dimensions of intercultural communication as these relate directly to my policymaking practices? How responsive are my policymaking practices to the set of problems raised by the colonial context, when considering the relationship of Western and non-Western value systems and beliefs?<br><br>- To what extent can I use the concepts discussed in this literature (in particular, notions of intercultural resonance and harmony as well as the CARE principles) to expand the scope and normative awareness of my policymaking practices? | - To what extent do my data innovation practices demonstrate an understanding of the challenges surrounding intercultural communication, the centring of non-Western values, and the equitable representation of Indigenous groups? How are the activities of data collection and use in which I am engaged implicated or affected by issues of intercultural communication and the 5 Ds of unjust data extraction?<br><br>- What are the gaps in my current understanding of the power dimensions of intercultural communication as these relate to the way I gather and/or use data? How responsive are my innovation practices to the set of problems raised by the colonial context, when considering the relationship of Western and non-Western value systems and beliefs? Do I address these problems as they might arise in my project planning, problem formulation, and impact assessment practices?<br><br>- To what extent can I use the concepts discussed in this literature (in particular, notions of intercultural resonance and harmony as well as the CARE principles) to expand the scope and normative awareness of my data collection and use? | - To what extent do I and members of my community possess an active understanding of the challenges surrounding intercultural communication, the centring of non-Western values, and the equitable representation of Indigenous groups? How can I and members of my community enhance my/our awareness of possibilities for transformational intercultural communication and contestation?<br><br>- What are the gaps in my and my community's current understanding of the power dimensions of intercultural communication? To what extent am I aware of the set of problems raised by the colonial context, when considering the relationship of Western and non-Western value systems and beliefs?<br><br>- To what extent can I and other members of my community use the concepts discussed in this literature (in particular, notions of 'cultures of contestation', intercultural resonance and harmony as well as the CARE principles) to expand the scope and awareness of possibilities for social change? |

# Appendix I: Policy Pilot Partners

| Policy Pilot Partner | Region | Mission |
|---|---|---|
| AfroLeadership | Africa | Founded in 2009, AfroLeadership is a civil society organisation based in Cameroon. Their aim is to 'strengthen human rights, government, and democracy by advocating for transparency, accountability, and citizen participation in public policies'. Their previous work surrounding data justice includes an ongoing partnership with Good of All. Through this partnership, they work to combat violence, hate speech and disinformation online through education. In their proposal, AfroLeadership emphasised the importance of participatory approaches to data justice which give visibility and representation to minorities. In particular, AfroLeadership drew attention to three factors which contribute to marginalisation, each of which will be reflected in their research as they explore the impact geographical situation in rural communities, gender and literacy can have in exacerbating data injustices. |
| CIPESA | Africa | The Collaboration on International ICT Policy in East and Southern Africa (CIPESA) was founded in 2004 with a mission 'to increase the capacity of East and Southern African stakeholders to participate in ICT policy-making'. They work to facilitate dialogue between stakeholder groups, to educate citizens on key issues and to collaborate with businesses, government officials and others with an interest in ICT policy. Prior work on data justice has seen CIPESA partner with the Internet Society to share knowledge and pool expertise on internet policy. They conducted stakeholder engagement throughout the region and aimed to 'work together for an open, secure and trustworthy internet for Africa.' In taking the Advancing Data Justice project forwards, CIPESA propose to draw on their experience of multi-country advocacy, network building and data governance in order to incorporate as many voices as possible into the data justice discourse. |
| CIPIT | Africa | The Centre for Intellectual Property and Information Technology Law (CIPIT) is a research institution based at Strathmore University, Kenya and was founded in 2004. Their mission is to 'study, create and share knowledge on the development of intellectual property and information technology, especially as they contribute to African Law and Human Rights'. CIPIT's previous work includes research focused on Kenya's Identity Ecosystem, specifically three identification systems that are critical to participation in both political and economic life. They have brought to life issues of accessibility, transparency, accountability, and inclusivity, as well as exclusionary practices that contribute to gender inequality. As CIPIT begins to conduct research as part of the Advancing Data Justice project, they plan to continue to explore how the African continent's unique social and cultural landscape can and must be foregrounded in global dialogues on AI. |

| WOUGNET | Africa | The Women of Uganda Network has worked since 2000 to 'promote and support the use of ICTs by women and women's organizations in Uganda in order to effectively address national and local problems for sustainable development'. WOUGNET has previously launched an initiative which focuses on increasing women's decision-making power and influence surrounding ICT policies. They engaged relevant stakeholders in conversations using the Feminist Principles on the Internet and the National Awareness Raising workshop on women's rights and technology. Now, as part of the Advancing Data Justice project, WOUGNET plan to ensure gender rights concerns are integrated with discourse on ICT policy and to empower communities both to use ICTs and demand their digital rights. |
|---|---|---|
| GobLab UAI | Americas | Founded in 2017 and based at the Universidad Adolfo Ibáñez in Chile, GobLab UAI works with 'government agencies, civil society organizations and businesses to ensure that data generates public value'. Their previous work has included a project titled "Market Opportunities for Technology Companies: Public Procurement of Accountable, Ethical and Transparent Algorithms". Through this work they have aimed to help build capacity among technology companies through training programmes aimed to incorporate ethical standards in automated decision-making services provided for the public sector. As part of the Advancing Data Justice project, GobLab has networks in place to engage an extensive range of both policymakers and developers in order to assess and advance Data Justice Guidelines for these groups. |
| Internet Bolivia | Americas | Internet Bolivia is a 'group of citizens committed to strengthening access to a safe, free and democracy-enhancing internet' who have been working to provide public resources since 2018. Their prior work on data justice includes a project undertaken in partnership with the Digital Defenders Partnership and Access Now. This project saw them establish a helpline, SOS Digital, which provided rapid responses to assist actors in situations of vulnerability to digital threats. As part of this Data Justice project, Internet Bolivia have set out extensive connections with each of the three stakeholder groups, including a wide range of impacted communities such as LGBTI people, feminist groups, indigenous peoples, parents' associations, small farmers and more. |
| ITS Rio | Americas | The Institute for Technology and Society of Rio de Janeiro was founded in 2013 to study 'the impact and future of technology in Brazil and worldwide'. One previous project saw them work to combat disinformation in Latin America through tutorials, blog posts, workshops and more aimed to support organizations and researchers tackling disinformation. Now, as part of the Advancing Data Justice Project, ITS Rio aim to address the lack of substantial participation of "intended Global South recipients" in international projects promoting data-based technologies as solutions for chronic global problems |

| Digital Empowerment Foundation | Asia | Based in India, the Digital Empowerment foundation have worked since 2002 'to empower marginalised communities in information dark regions to access, consume and produce information online using digital interventions and ICT tools'. They previously ran an initiative which helped introduce ICTs to India's traditional crafts sector where they trained over 10,000 people and introduced nine artisan clusters to digital interventions. In their proposal for the Advancing Data Justice project, the Digital Empowerment Foundation emphasised the networks it has established through its 1000 Community Information Resource Centres located across 24 states and 135 districts in "rural, tribal, marginalised, and unreached areas" of India. |
|---|---|---|
| Digital Rights Foundation | Asia | Digital Rights Foundation were founded in 2013. Their mission states that "DRF envisions a place where all people, and especially women, are able to exercise their right of expression without being threatened. We believe that free internet with access to information and impeccable privacy policies can encourage such a healthy and productive environment that would eventually help not only women, but the world at large". Their prior work relating datafication to the rights of marginalised communities includes a research project which details the difficulties faced by religious minorities online in Pakistan. This focused, in particular, on the disproportionate 1 9 hostility directed towards groups marginalised on the basis of gender, ethnic and religious minorities. Digital Rights Foundation focus in their proposal on broadening debates on AI which have been dominated by the Global North in order to speak to 'the intersectional needs of communities in contexts like South Asia and beyond'. |
| Open Data China | Asia | Open Data China is a 'social enterprise based in Shanghai, China, focusing on promoting and building up an open digital future'. They focus on three streams of work: data governance, digital rights and social responsibility and have previously conducted work on bottom-up data trusts and on collective digital rights in the gig economy. The contacts which Open Data China will draw on as part of the Advancing Data Justice project will allow us to access the perspectives of a range of developers across large and small-scale technology provides as well as a range of policymakers, both within public-funded institutions under government supervision and at independent think tanks. |
| Digital Natives Academy | Oceania | Digital Natives Academy was founded in 2014 with the aim 'to create career pathways for whānau wanting to be part of digital tech industries'. They have described their approach as deeply rooted in indigenous epistemologies and Te Ao Māori pedagogies. Their proposal for the Advancing Data Justice project focuses on the need for trusted relationships to form an effective basis for stakeholder engagement. Their work engaging with Māori communities, conducting interviews privately and with compassion will provide a valuable contribution to this project. |
| Engage Media | Oceania and Asia | Based in Australia but working across Southeast Asia and Oceania, EngageMedia is a non-profit media, technology, and culture organisation. EngageMedia uses 'the power of video, the Internet and open technologies to create social and environmental change'. Currently, they are running a digital rights campaign in Thailand to raise awareness and enhance democratic agency. As part of the Advancing Data Justice project, Engage Media will make important contributions thanks to extensive networks across a wide area spanning the Asia-Pacific. |